\documentclass[aps,preprintnumbers,superscriptaddress,twocolumn,amsmath,amssymb,floatfix,pra]{revtex4-2} 
\pdfoutput=1
\usepackage{graphicx}
\usepackage{bm}
\usepackage{tikz}
\usepackage[T1]{fontenc}
\usepackage[utf8]{inputenc}
\usepackage{braket}
\usepackage{bbold}
\usepackage{enumerate}

\usetikzlibrary{mindmap}

\newcommand{\id}{\mathbb{1}}

\newcommand\jd[1]{\textcolor{orange}{\bf #1}}

\def\bra#1{\mathinner{\langle{#1}|}}
\def\ket#1{\mathinner{|{#1}\rangle}}

\begin{document}

\title{Large Random Arrowhead Matrices: Multifractality, Semi-Localization, and Protected Transport in Disordered Quantum Spins Coupled to a Cavity}

\author{J.~Dubail}
\thanks{jerome.dubail@univ-lorraine.fr}
\affiliation{Universit\'{e} de Lorraine and CNRS, LPCT (UMR 7019), 54000 Nancy, France}
\affiliation{Universit\'{e} de Strasbourg and CNRS, ISIS (UMR 7006) and icFRC, 67000 Strasbourg, France}

\author{T. Botzung}
\affiliation{Universit\'{e} de Strasbourg and CNRS, ISIS (UMR 7006) and icFRC, 67000 Strasbourg, France}
\affiliation{Institute for Quantum Information, RWTH Aachen University, D-52056 Aachen, Germany
}
\affiliation{Institute for Theoretical Nanoelectronics (PGI-2), Forschungszentrum J\"{u}lich, 52428 J\"{u}lich, Germany}

\author{J.~Schachenmayer}
\affiliation{Universit\'{e} de Strasbourg and CNRS, ISIS (UMR 7006) and icFRC, 67000 Strasbourg, France}

\author{G.~Pupillo}
\thanks{pupillo@unistra.fr}
\affiliation{Universit\'{e} de Strasbourg and CNRS, ISIS (UMR 7006) and icFRC, 67000 Strasbourg, France}
\affiliation{Institut Universitaire de France (IUF), 75000 Paris, France}

\author{D.~Hagenm\"uller}
\thanks{dhagenmuller@unistra.fr}
\affiliation{Universit\'{e} de Strasbourg and CNRS, ISIS (UMR 7006) and icFRC, 67000 Strasbourg, France}

\begin{abstract}
Large ``arrowhead'' matrices with randomly distributed entries describe a variety of important phenomena where a degree of freedom is non-locally coupled to a disordered continuum of modes, including central-spin problems in condensed-matter, molecular junctions, and quantum emitters in cavity quantum electrodynamics (QED). Here we provide an exact solution of random arrowhead Hamiltonians with diagonal disorder in the thermodynamic limit. For concreteness, we focus on the problem of $N$ emitters homogeneously coupled to a non-local cavity mode, corresponding to the disordered Tavis-Cummings model of cavity-QED in the single excitation limit, for which we provide asymptotically exact formulas for static and dynamical quantities of physical interest. We find that the distribution of energy spacing can be continuously tuned between Poisson statistics and a distribution that is very close to semi-Poisson statistics - the latter statistics being usually associated to the critical point of ``Anderson'' localization-delocalization transitions. We demonstrate that all the eigenstates - including two polaritons and a continuum of dark states - are multifractal, which indicates the existence of a critical ``semi-localized'' phase for all values of the light-matter coupling strength, where dark states are localized over multiple, arbitrarily-distant sites. By computing the time-dependent escape probability from an initial site, we find that the system has a peculiar diffusive-like behavior with an escape probability growing linearly with time for any finite coupling strength, and that the escape rate can be controlled by selecting the energy of the initial site. The escape rate averaged over the disorder configurations is found to exhibit a maximum for intermediate coupling strengths, before saturating at a lower value for sufficiently large $N$ in the collective strong coupling limit - a ``cavity protection'' effect. Surprisingly, we show that the saturation value increases with the disorder, indicating that the cavity does not only protect transport against disorder but can also turn the latter into an ally improving transport. We finally investigate the system in a two-terminal configuration, and show that the steady-state excitation current exhibits similar features as the escape probability, thereby extending our cavity-protected transport scenario to out-of-equilibrium situations. We finally demonstrate that dark states can provide the major contribution to long-distance transport in disordered systems.
\end{abstract}

\date{\today}

\maketitle

\section{Introduction}

Large random Hamiltonians have a long history in physics, usually traced back to Wigner's idea of modeling the spectra of heavy nuclei by the eigenvalues of random hermitian matrices. Typically one considers $N\times N$ hermitian matrices with matrix elements that are independent, identically distributed variables. One calculates their spectrum, with the goal of determining the statistical properties of the distribution of eigenvalues in the large $N$ limit. Random matrices had been introduced previously in the mathematics literature by Wishart, but the field expanded considerably thanks to Wigner's idea, and to the early works of Dyson, Mehta, and many others, see e.g.~\cite{mehta2004random,anderson2010introduction,forrester2010log,akemann2011oxford,tao2012topics} for modern textbooks on random matrix theory. Random matrices have found many applications across various subfields of physics, mathematics, statistics and engineering, ranging from transport in mesoscopic systems~\cite{beenakker1997random}, to characterizations of quantum chaos~\cite{bohigas1984characterization,guhr1998random}, to models of two-dimensional quantum gravity~\cite{knizhnik1988fractal,di19932d}, to the statistics of zeros of the Riemann zeta function~\cite{bogomolny1995random,bourgade2013quantum}, to modeling of channels in wireless communication~\cite{tulino2004random}.

Random ``arrowhead''~\cite{o1990computing,gravvanis1998approximate,najafi2014efficient,stor2015accurate}, or ``bordered diagonal''~\cite{sussman1982reconstruction,pickmann2007extremal}, matrices are particularly simple examples of random matrices having most of their matrix elements equal to zero except the ones along the diagonal, and along the last line and last column.

Large random arrowhead matrices naturally appear in applied mathematics with motivations from electric circuit theory~\cite{sussman1982reconstruction} and control theory~\cite{peng2006two}, and efficient numerical methods for inverting and diagonalizing them have been developed. Such matrices are also of relevance in the common physical situation where a specific degree of freedom is non-locally coupled to a randomly disordered continuum of modes acting as an environment~\cite{Elliott_1974}. Among the broad variety of possible physical systems, this occurs for instance in molecular junctions, where a localized vibrating molecule interacts with the free electron gas of a metallic surface placed in close proximity~\cite{Gadzuk_1981,Vinkler_2014} [see Fig.~\ref{fig_real} (\textbf{a})]. The random arrowhead Hamiltonian model is here highly relevant to describe charge~\cite{Reed_1997,Tao_2006,Solomon_2010} and heat~\cite{Galperin_2007,Cui_2008,Aradhya_2013} currents flowing through the junction. Other related situations are plasmonic nanogap junctions~\cite{Zhu_2019}, semiconductor quantum point contacts~\cite{vanWees_1988,Wharam_1988}, and a single molecule embedded in a plasmonic nanocavity~\cite{Benz_2016,Zhu_2019}. This model is also useful in \textit{central-spin problems}, where a single (central) spin non-locally interacts with a spin bath of localized modes~\cite{gaudin1976diagonalisation,Prokof_ev_2000,Breuer2004}. Recent developments of quantum computation with solid-state qubits such as quantum dots~\cite{Khaetskii_2003,Coish_2004} [see Fig.~\ref{fig_real} (\textbf{b})], superconducting qubits~\cite{Zhukov_2018}, and nitrogen-vacancy centers in diamond~\cite{Hanson352} have revived interest in this model, where proper understanding of the decoherence mechanisms affecting the central spin is of crucial importance.  

\begin{figure}[t]
\includegraphics[width=0.35\textwidth]{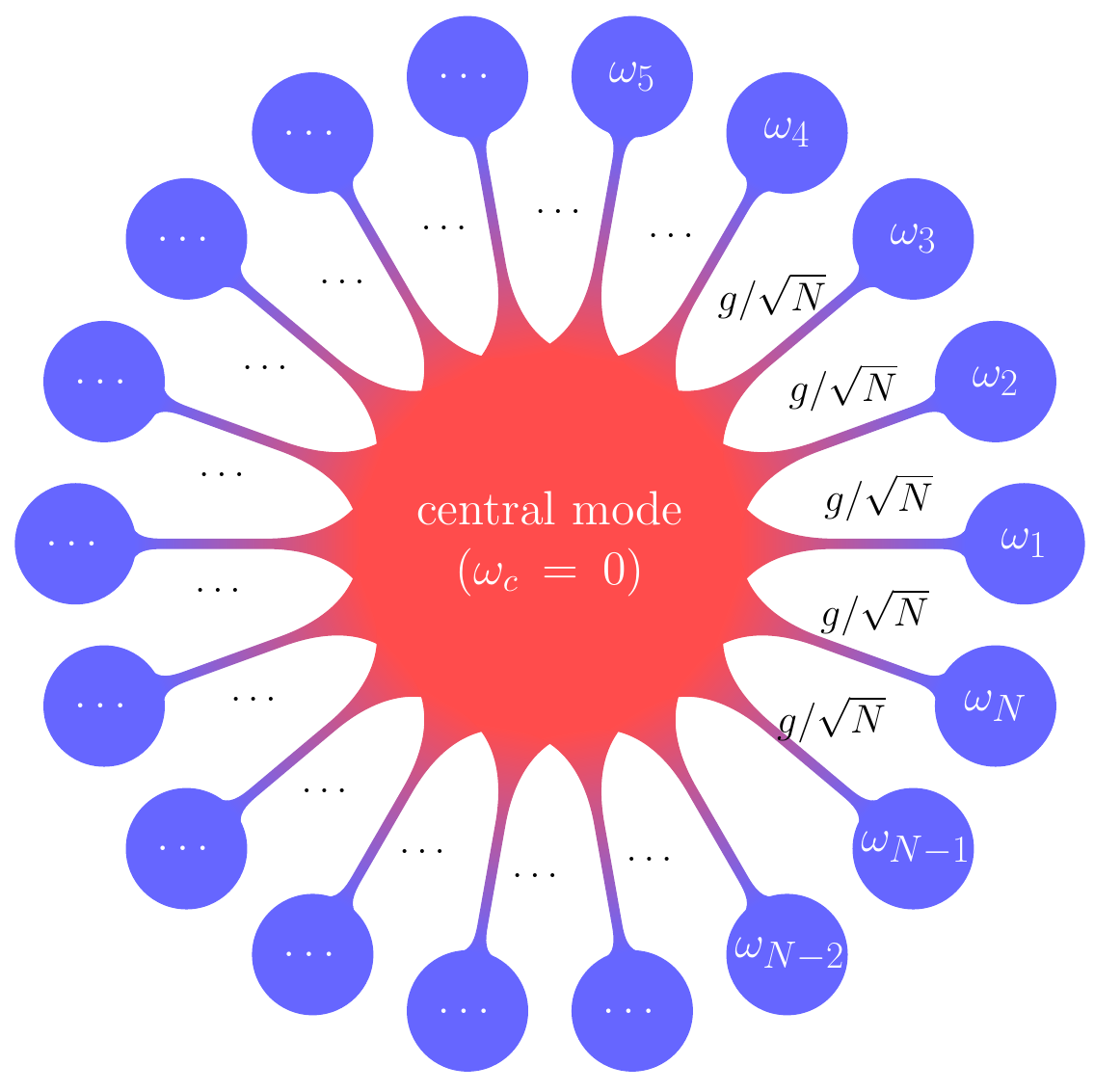}
\caption{Random arrowhead Hamiltonians appear when a specific (central) degree of freedom with energy $\omega_c$ is non-locally coupled to a randomly disordered ensemble of $N$ modes with energies $\omega_1, \omega_2, \dots, \omega_N$, the latter forming a continuum in the thermodynamic limit $N\to \infty$. Without loss of generality the energy of the central mode can be set to zero. In this paper, we focus on the situation with uniform couplings $g/\sqrt{N}$ between the central mode and all the others.}
\label{fig:starshape}
\end{figure}

Other important examples of random arrowhead Hamiltonians arise in cavity-QED. In this case the central site approximates a bosonic mode of the electromagnetic field, which can be either confined in an optical cavity~\cite{Haroche_1983,Carmichael_1989} or a plasmonic structure~\cite{Torma_2014}, and is non-locally coupled to $N$ quantum emitters (two-level systems) described by local pseudo-spin operators [see Fig.~\ref{fig_real} (\textbf{c})]. In the common situation where the transition frequencies of these two-level systems are inhomogeneously broadened, one ends up with a disordered version of the famous Tavis-Cummings (TC) model~\cite{Tavis_Exact_1968}. The TC model has been realized in a variety of physical systems including superconducting qubits~\cite{Wallraff_2009,Shapiro_2015}, nitrogen-vacancy centers in diamond~\cite{Kubo_2010,Astner_2017} and atomic systems~\cite{PhysRevLett.53.1732,Raizen1989,Zhu1990,Dimer_2007,Zhiqiang_2017}.

\begin{figure}[t]
	\includegraphics[width=0.45\textwidth]{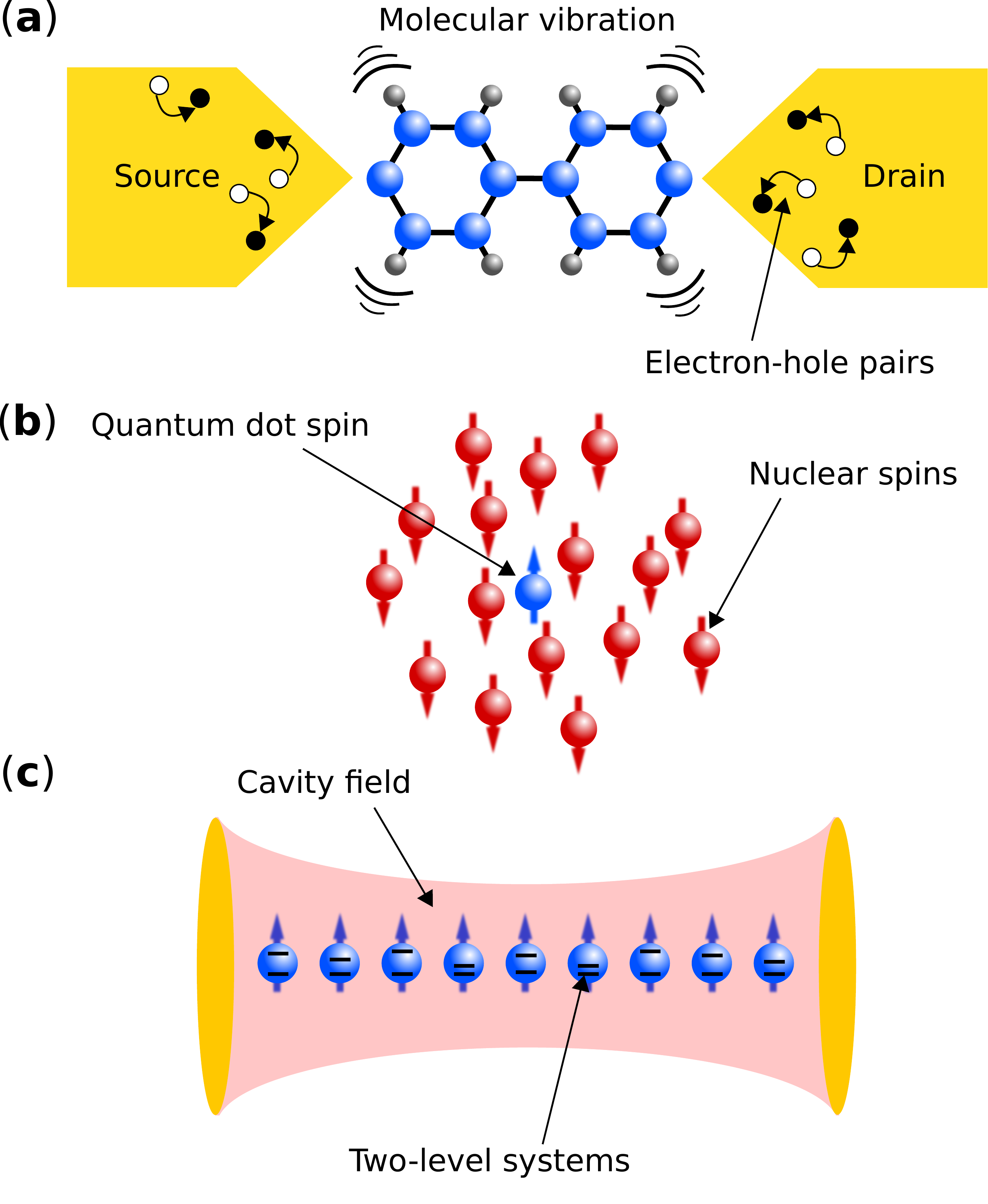} 
	\caption{Physical realizations of random arrowhead Hamiltonians. (\textbf{a}) Molecular junction consisting of a single molecule coupled to metallic leads for charge and thermal transport (Adapted from Ref.~\cite{Quek_2009}). The central (bosonic) mode corresponds to molecular vibrations, which interact with two disordered continua of electron-hole pairs in the leads. (\textbf{b}) A central quantum dot with a single electron of spin $+1/2$ is inhomogeneously coupled to an ensemble of polarized nuclear spins $-1/2$ via hyperfine interaction. (\textbf{c}) Cavity-QED setup consisting of a central cavity electromagnetic (bosonic) mode coupled to an inhomogeneously-broadened ensemble of two-level systems (pseudo-spins), which is the focus of this paper.}
	\label{fig_real}
\end{figure}

The disordered TC model and other closely related models for quantum well exciton-polaritons have been already looked at to study the effect of inhomogeneous broadening on the polariton spectrum~\cite{Houdre_Vacuum_1996,Marchetti2006,Marchetti2007,Kurucz2011,Kirton_Intro_2019} and on superradiance~\cite{Temnov_2005}. Similar models including a hopping term between nearest-neighbor sites were also studied~\cite{Schachenmayer_Cavity_2015,Feist_Extrao_2015} in connection to recent experiments on transport through molecular materials strongly coupled to light~\cite{Orgiu_Conduc_2015,Nagarajan2020,Schwartz2018}. In this case, diagonal disorder (inhomogeneous broadening of molecular excitons), off-diagonal disorder (inhomogeneous hopping)~\cite{doi:10.1021/cr050140x}, and orientational disorder~\cite{PhysRevLett.110.126801} leading to inhomogeneous couplings to the cavity electric field, are all expected to play an important role and cannot \textit{a priori} be neglected. Theoretical investigations on the non-trivial interplay between disorder and strong light-matter coupling are therefore highly desirable for further developments of polaritonic material science.

In this paper we provide an exact solution of random arrowhead Hamiltonians in the thermodynamic limit. For concreteness, we focus on its main cavity-QED application: The TC model in the single excitation limit, with disorder in the spin energies but uniform couplings between the spins and the cavity field. This is described by the Hamiltonian $\hat{H}=\hat{H}_{0}+\hat{V}$, with
\begin{align}
\hat H_{0} &= \sum_{i=1}^{N} \omega_{i} \hat \sigma^{+}_{i} \hat \sigma^{-}_{i} + \omega_{c} \hat a^{\dagger} \hat a \nonumber \\
\hat V &= \frac{g}{\sqrt{N}} \sum_{i=1}^{N} \left(\hat a \hat \sigma^{+}_{i} + \hat \sigma^{-}_{i} \hat a^{\dagger} \right).
\label{eq1}
\end{align}
The first term $\hat{H}_{0}$ provides the energy of the disordered spin ensemble and the cavity mode, with $\hat{\sigma}^{-}_{i}$ ($\hat{a}$) and $\hat{\sigma}^{+}_{i}$ ($\hat{a}^{\dagger}$) the i$^{\rm th}$ spin (cavity) annihilation and creation operators, respectively. The cavity mode energy $\omega_{c}$ is here set to zero without loss of generality. The second term describes the coupling between the spins and the cavity. Since the scaling of the coupling with $N$ comes from its dependence on the cavity-mode volume $V$ as $\sim 1/\sqrt{V}$~\cite{Scully_1997}, the \textit{collective} coupling strength $g$ is the relevant physical parameter in the thermodynamic limit $N\to \infty$ as it remains independent of $N$ for a fixed density $N/V$.

The Hamiltonian Eq.~(\ref{eq1}) commutes with the excitation number operator $\hat N =\sum_{i} \hat \sigma^{+}_{i} \hat \sigma^{-}_{i} + \hat a^{\dagger} \hat a$, so the single-excitation subspace is defined as $\hat N=\id$. This subspace is spanned by the $N+1$ eigenstates of $H_{0}$: $\{\ket{i,0},\ket{G,1}\}$, where $\ket{i,0}$ has the $i^{\rm th}$ spin in its excited state, and $\ket{G,1}$ has all spins in their ground state and one photon in the cavity. In this basis, the Hamiltonian takes the form
\begin{equation}
	\label{eq:arrowhead}
	H = \left( \begin{array}{cccccc|c}
		\omega_1   & 0 & & & \dots & 0 & g/\sqrt{N} \\
		0   & \omega_2 &  0 &  & &  &  \\
		     &  0 & \ddots  & \ddots & &  \vdots &  \\
		\vdots     &  & \ddots   &  & 0 &  & \vdots  \\
		     &  &   &  0 & \omega_{N-1} & 0 & \\
		0   &  & \dots  &  & 0 & \omega_N  & g/\sqrt{N} \\ \hline
		g/\sqrt{N} & &  & \dots && g/\sqrt{N}  & 0
	\end{array} \right),
\end{equation}
where the ``bare energies'' $\omega_j$, $j=1,\dots,N$, are independent, identically distributed (i.i.d.) random variables, drawn from a probability distribution $\rho(\omega)$ (with $\int \rho(\omega) d \omega=1$), such that the density of bare energies
\begin{equation}
	\label{eq:rho}
	\sum_{j=1}^N \delta (\omega-\omega_j)  \; \underset{N \rightarrow \infty}{\simeq } \; N \rho (\omega),
\end{equation} 
in the thermodynamic limit. For simplicity, we assume that $\rho(\omega)$ is supported on an interval $[\omega_{\rm min}, \omega_{\rm max}]$. We write $\mathbb{E}[.]$ for the expectation value taken with respect to the probability distribution $\rho(\omega)$, i.e.
\begin{equation}
    \label{eq:disav}
    \mathbb{E} [O(\{\omega_j \})] \,  = \, \int \left( \prod_{i=1}^N \rho (\omega_i) d\omega_i \right)  \, O(\{ \omega_j \}),
\end{equation}
for any observable $O(\{ \omega_j \})$, and we refer to it as {\it disorder average}. Throughout the paper, we work with a general function $\rho(\omega)$ supported on the interval $[\omega_{\rm min}, \omega_{\rm max}]$. However, for comparison with numerics we find that it is convenient to specialize to the case of a `box' distribution, i.e. the uniform distribution with $\rho(\omega)=1/W$ for $\omega \in [-W/2,W/2]$ and $\rho(\omega)=0$ otherwise. $W$ thus quantifies the \textit{disorder strength}.  

In Ref.~\cite{Botzung_2020}, we studied numerically the disordered TC model in the presence of additional hopping of spin excitations between nearest-neighboring sites. The purpose of including such hopping term was to investigate the effect of the spin-cavity coupling on the metal-insulator Anderson transition~\cite{Anderson_Absenc_1958,Evers_Anders_2008}. It was shown that the cavity mode strongly modifies the localization properties of the eigenstates, which feature a novel character of being localized on multiple, non-contiguous sites, a behavior we dubbed as \textit{semi-localization}. It was also shown that those semi-localized eigenstates can efficiently contribute to coherent energy transport under strong coupling conditions, i.e. when the collective coupling strength $g$ exceeds the bandwidth of the disordered spin ensemble. Such cavity-induced robustness of transport and delocalization against disorder was also highlighted in other very recent works~\cite{PhysRevLett.126.153201,Scholes2020}, showing a growing interest in ``cavity-protected'' transport.

Here we present for the first time a detailed analysis of the properties of the random arrowhead Hamiltonian (\ref{eq:arrowhead}), providing asymptotically exact formulas for the spectrum, average energy shifts induced by the coupling to the cavity mode, and correlation functions for all coupling strengths $g/W$. In particular we show that the distribution of energy spacings can be continuously tuned between Poisson statistics and a distribution very close to semi-Poisson statistics for $g/W \gg 1$ - the latter statistics being usually associated to the critical point of ``Anderson'' localization-delocalization transitions. We also demonstrate that all the eigenstates consisting of polaritons and a continuum of dark states are multifractal. This indicates the existence of a critical ``semi-localized'' phase for all values of $g/W$, where the dark states are localized over multiple, non-contiguous sites. The dark state multifractal spectrum and anomalous dimension are found to be identical to those of the critical point of the Anderson model on the Bethe lattice, which is commonly regarded as a mean-field model (dimension $d\to \infty$) of the localization-delocalization transition. 

In the second part of the paper we explain how the spectral properties of the model can be harnessed to engineer transport through the system of emitters. We show that the escape probability of an excitation from some given site grows linearly with time for any finite coupling strength $g$, similarly to a diffusive motion, and that the escape rate can be fully controlled by tuning the energy of the initial excitation across the disorder distribution. We provide a Fermi's golden rule-based interpretation of these results using second-order perturbation theory, where the non-local coupling of the emitters to the cavity mode leads to an effective long-range hopping with amplitude depending only on the disordered energies of the emitters, but not on the distance between them. We show that the disorder-averaged escape rate exhibits a maximum for intermediate values of the coupling strength $g/W \sim 1$, before saturating to a lower, $g$-independent value for strong couplings $g/W \gg 1$ - a ``cavity protection effect'' of transport that could find important applications in, e.g. optoelectronic devices. Surprisingly, we find that the saturation value increases with the disorder strength, which indicates that the cavity does not only protect transport against disorder, but that the latter can also contributes favourably to transport provided the light-matter coupling is strong enough. We also compute the steady-state excitation current when coupling the system to two external leads, with relevance to mesoscopic physics experiments. In contrast to common expectations from polaritonics, we demonstrate that out-of-equilibrium transport is fully dominated by the contribution of the dark-states. The role of the latter played in influencing dynamics is attracting more and more interest as a possible key aspect of polaritonic material science and polaritonic chemistry~\cite{Gonzalez-Ballestero_Uncoup_2016,Botzung_2020,PhysRevLett.126.153201,Scholes2020,doi:10.1063/1.5136320}.

The remainder of the paper is organized as follows. In Section~\ref{sec:results} we provide a discussion of the main  results of the work. This is followed by in-depth analyses and demonstrations of those results in Sections~\ref{sec:finiteN} to~\ref{sec:trans}. In particular, in Section~\ref{sec:finiteN} we recall  standard formulas for the eigenstates and eigenvalues of arrowhead matrices~\cite{o1990computing}, which are used in subsequent Sections. In Section~\ref{sec:shifts} we discuss the structure of the spectrum and of the $O(1/N)$ energy shifts in the thermodynamic limit. In Section~\ref{sec:IPR}, we analyze the inverse participation ratios of the eigenstates. In Section~\ref{photon_sec}, we analyze the photon Green's function and the photon spectral function. In Section~\ref{sec:propagator}, we derive the large $N$ asymptotics of the real-time Green's function and use it to compute the escape probability. In Section~\ref{sec:trans} we compute the current flowing through the system in a two-terminal configuration. We conclude in Section~\ref{sec:conclusion}.

\subsection{Summary of the results}
\label{sec:results}

Having introduced the model and its underlying physical motivations, we now provide a summary of the main results of this paper.

\subsubsection{Spectrum and energy shifts}

In Sec.~\ref{sec:shifts}, we analyse the spectrum of the Hamiltonian $H$ and derive an asymptotically exact formula at large $N$ for the average energy shift of the disordered energy levels of the emitters induced by the coupling to the cavity. We find that the spectrum consists of two modes with unbounded energies $\varepsilon_{{\rm P}+}$ and $\varepsilon_{{\rm P}-}$ outside the interval $[\omega_{\rm min}, \omega_{\rm max}]$, the support of $\rho(\varepsilon)$. For weak couplings (e.g. $g \ll W$ for a box disorder distribution of width $W$) these modes are parametrically close to the edges of the distribution at $\omega_{\rm min}$ and $\omega_{\rm max}$ and contain a vanishing photon weight. In contrast, for strong couplings ($g \gg W$), they are hybrid modes composed of 50\% cavity photons and 50\% emitters and are well separated from the remaining $N-1$ \textit{dark states} whose energies $\varepsilon_a$ are within the support of $\rho(\varepsilon)$ and close to the bare energies $\omega_j$. For clarity the two modes outside the interval  $[\omega_{\rm min}, \omega_{\rm max}]$ will always be denoted as \textit{polaritons} throughout the paper, even though this term is usually employed only in the strong coupling regime in the literature. 

Upon coupling to the cavity mode, the $a^{\rm th}$ energy of the dark state $\omega_a$ typically gets shifted by an amount of order $1/N$. Defining the energy shift of a dark state $\Delta_a = N [\varepsilon_a - (\omega_{a-1}+ \omega_a)/2] $ for energies sorted in increasing order, we determine the value $\overline{\Delta} (\varepsilon)$ of the shift, averaged over a small energy shell $[\varepsilon - \delta \varepsilon/2 , \varepsilon + \delta \varepsilon/2 ]$ to be
\begin{equation*}
	\overline{\Delta} (\varepsilon) \underset{N \rightarrow \infty}{\simeq} \frac{1}{\pi \rho(\varepsilon)} {\rm arctan} \left[ \frac{\tilde{\rho}(\varepsilon) - \frac{\varepsilon}{\pi g^2} }{\rho(\varepsilon)} \right] ,
\end{equation*}
where $\tilde{\rho}$ is the Hilbert transform of $\rho$ [see Eq.~(\ref{eq:hilbert}) and (\ref{eq:defmomentsdistr})]. This energy shift is closely related to the localization properties of the dark states:
\begin{enumerate}[(i)]
    \item For weak couplings (e.g. for $g/W \ll 1$ for a box distribution of width $W$), the latter are very close to the bare energy levels, i.e. $\overline{\Delta} (\varepsilon)\approx 1/[2 \rho(\varepsilon)]$. The dark states thus follow the same Poissonian statistics as the i.i.d. bare orbitals. The Poisson statistics is a signature of fully localized eigenstates. 
    \item For strong couplings ($g/W\gg 1$), in contrast, the dark states lie roughly at equal distance from the two closest bare energy levels, i.e. $\overline{\Delta} (\varepsilon)\approx 0$, so they follow a statistics that is close to the semi-Poissonian statistics introduced in Ref.~\cite{Bogomolny_1999}. In this regime, the dark states are localized on multiple, non-contiguous sites that are close in energy, a behavior referred to as ``semi-localization''~\cite{Botzung_2020}. 
    \item For intermediate coupling strengths we find that the level spacing statistics depends continuously on $g$ and on the energy $\varepsilon$ (see Fig.~\ref{fig:level_spacings}).
\end{enumerate}

The possibility to interpolate between different statistics is to our knowledge an entirely novel effect.  

\begin{figure*}[t]
	\includegraphics[width=1\textwidth]{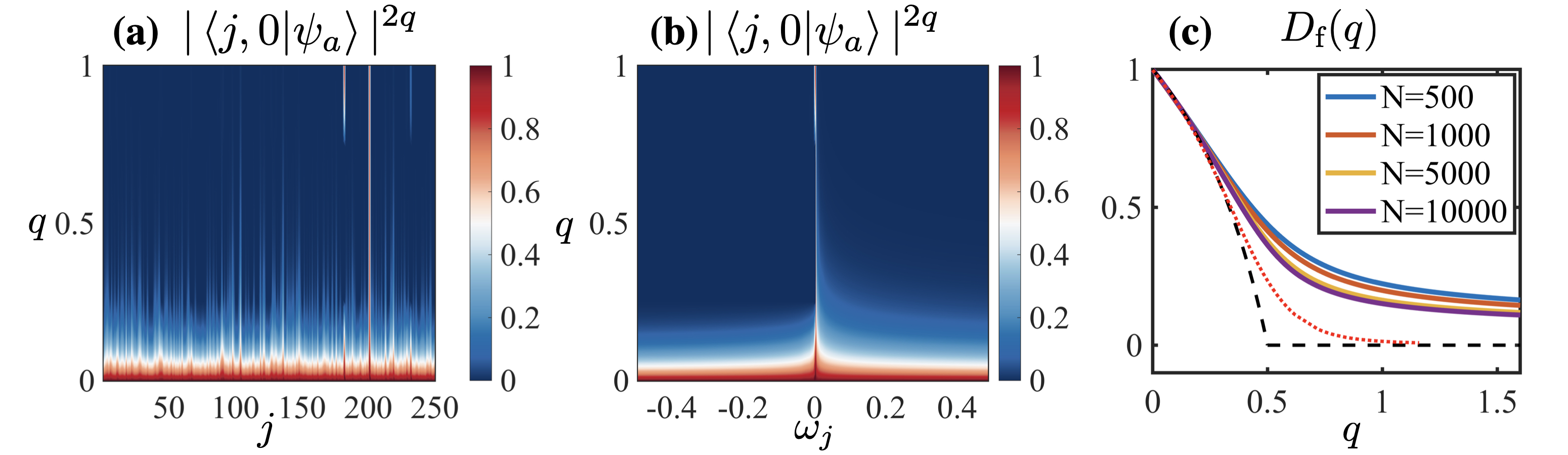}
	\caption{Multifractality of the eigenstates. Components of IPR$_{a} (q)$ for a given eigenstate $\psi_{a}$ in the middle of the spectrum as a function of the site index $j$ (with no ordering with respect to the bare energy $\omega_j$) (\textbf{a}), and as a function of the bare energy $\omega_{j}$ (\textbf{b}). We see that the fractal structure visible in (\textbf{a}) is caused by the random ordering of the sites: when the squared amplitudes are ordered according to the bare energy $\omega_j$, they are a smooth function of $\omega_j$ decaying as $\sim 1/(\varepsilon_a- \omega_j)^2$, see Eq.~(\ref{eq:powerlaw}).
	(\textbf{c}) Fractal dimension $D_{\rm f}(q)$ of the dark states, defined by Eq.~(\ref{eq:fractal_dim}), computed numerically for different system sizes $N$ (lines). The dashed line corresponds to the analytical result $D_{\rm f}(q)/d = {\rm min}(\frac{1-2q}{1-q},0)$. Deviations from the analytical formula are due to finite size effects. The red dotted line shows the $N \to \infty$ extrapolation (see text).}
	\label{fig7}
\end{figure*}
\subsubsection{Inverse participation ratio: Multifractality of the spectrum}

In Sec.~\ref{sec:IPR} we provide an analytical calculation of the so-called inverse participation ratio (IPR) at large $N$, which is a useful quantity to characterize the localization properties of the eigenstates~\cite{edwards1972numerical,wegner1980inverse,evers2000fluctuations,Evers_Anders_2008}. By computing the scaling behavior of the IPR at large $N$, we show that all the eigenstates are {\it multifractal} for all values of $g/W$.

The IPR of the (normalized) eigenstate $\psi_a$ is defined as 
\begin{equation}
{\rm IPR}_a(q) = \sum_{j=1}^{N+1} |\psi_{a,j} |^{2q} \equiv \sum_{j=1}^{N} |\langle j,0 \vert\psi_a \rangle |^{2q} + |\langle G,1 \vert\psi_a \rangle |^{2q}.\nonumber
\end{equation}
For a delocalized eigenstate, all components are of order $\sim 1/N$, and one expects ${\rm IPR}_a(q) = O( N^{1-q})$. In contrast, for a localized eigenstate, a few components are of order $O(1)$, and the others vanish; one thus expects ${\rm IPR}_a(q) = O(1)$. The $q$-dependent IPR can be viewed as an analog of the Renyi entropy for the probability distribution defined by the squared amplitudes $p_j = |\psi_{a,j} |^2$. For small $q$, ${\rm IPR}_a(q)$ is sensitive to the tails of that distribution, while for large $q$ it is most sensitive to the largest components $p_j$. A qualitative change of the behavior of ${\rm IPR}_a(q)$ as a function of $q$ is a signature of multifractality of the wavefunction~\cite{wegner1980inverse,evers2000fluctuations}, see e.g. Section II.C of Ref.~\cite{Evers_Anders_2008} for an introduction. Interestingly, we find that all eigenstates of the TC Hamiltonian (\ref{eq:arrowhead}) exhibit multifractality in that sense.

The scaling behavior of ${\rm IPR}_a(q)$ can be computed analytically at large $N$. In Sec.~\ref{sec:IPR} we show that, remarkably, the ${\rm IPR}_a(q)$ changes as a function of $q$ for all eigenstates of Eq.~(\ref{eq:arrowhead}) with the following different scalings for the two polaritons and for the $N-1$ dark states
\begin{equation}
    \label{eq:res_IPR}
    \begin{array}{l}
    {\rm Polaritons:} \\
    \qquad {\rm IPR}_{\pm}(q) \underset{N\rightarrow \infty}{=} \left\{
        \begin{array}{lcl}
            O(N^{1-q}) & {\rm if} & q<1 \\
            O(1) & {\rm if} & q>1 ,
        \end{array}
    \right. \\ \\
        {\rm Dark \; states:} \\
    \qquad {\rm IPR}_{a}(q) \underset{N\rightarrow \infty}{=} \left\{
        \begin{array}{lcl}
            O(N^{1-2q}) & {\rm if} & q<1/2 \\
            O(1) & {\rm if} & q>1/2 . 
        \end{array}
    \right. \\
    \end{array}
\end{equation}
These scalings are valid for all values of $g \neq 0$ and demonstrate multifractal behavior of all eigenstates of Eq.~(\ref{eq:arrowhead}), for \textit{any finite value} of the light-matter coupling strength. This is different from conventional disordered hopping models where multifractal behavior is usually associated with the single critical point at a localization-delocalization transition~\cite{Evers_Anders_2008}.

In order to get a more precise picture of this multifractal behavior, let us imagine that the $N$ spins lie on a $d$-dimensional lattice of linear size $L$, then $N \sim L^d$, and the IPR behaves as
\begin{equation*}
    {\rm IPR}_{a}(q) \underset{L \rightarrow \infty}{\sim}  L^{-\tau_q} ,
\end{equation*}
which defines a set of multifractal exponents $\tau_q$~\cite{Evers_Anders_2008}. For the dark states in our model, this standard definition gives
\begin{equation}
    \label{eq:exponents}
    \tau_q = \left\{
        \begin{array}{lcl}
            d (2q-1) & {\rm if} & q<1/2 \\
            0 & {\rm if} & q>1/2 . 
        \end{array} \right.
\end{equation}
Then, subtracting the part expected for a normal metal where the eigenstates are delocalized, one can define the anomalous dimension $\Delta_q = \tau_q - d (q-1)$~\cite{Evers_Anders_2008}. Usually, in models of Anderson localization/delocalization transitions, this anomalous dimension is expected to satisfy the exact symmetry relation~\cite{mirlin2006exact}
\begin{equation}
    \label{eq:symmetry_Delta}
    \Delta_q = \Delta_{1-q} .
\end{equation}
Remarkably, in our random arrowhead matrix model, the exponents (\ref{eq:exponents}) for the dark states correspond to the anomalous dimension
\begin{equation}
    \label{eq:anomalousD}
 \Delta_q = \frac{1}{2} - \left| q - \frac{1}{2}\right|,
\end{equation}
which does satisfy the symmetry relation (\ref{eq:symmetry_Delta}). On the other hand, for the two polaritons one gets $\Delta_q = {\rm min}(0,1-q)$ from Eq.~(\ref{eq:res_IPR}), so the symmetry relation is not satisfied by the polariton eigenstates.

This suggests that the dark states (and only the dark states) in our model exhibit some properties that are analogous to the ones of a multifractal wavefunction at an Anderson transition. Moreover, the dark state multifractal spectrum (\ref{eq:exponents}) and anomalous dimension (\ref{eq:anomalousD}) turn out to be exactly the same as the ones for the Anderson model on the Bethe lattice~\cite{abou1973selfconsistent,Evers_Anders_2008}, which is commonly regarded as a mean-field model ($d\rightarrow \infty$) of the localization-delocalization transition. Interestingly, the random arrowhead matrix model thus reproduces some aspects of the physics of the Anderson transition in the $d\rightarrow \infty$ limit.

As illustrated in Fig.~\ref{fig7}, the multifractal behavior of the dark states arises in an \textit{extremely simple way} in the random arrowhead matrix model. We find that for i.i.d. variables $\omega_j$, the probabilities $p_j=|\psi_{a,j}|^2 = |\left< j,0 \left| \psi_a \right> \right. |^2$ look like a random distribution which exhibits a fractal structure [Fig.~\ref{fig7}.(a)]. However, when one sorts the amplitudes in order of increasing $\omega_j$, then the probability $p_j$ turns out to be a smooth function of $\omega_j$ [Fig.~\ref{fig7}.(b)]. In fact it simply decays as a power-law,
\begin{equation}
    \label{eq:powerlaw}
    \psi_{a,j} = |\left< j,0 \left| \psi_a \right> \right. |^2 \sim \frac{1}{(\varepsilon_a - \omega_j)^2} ,
\end{equation}
where $\varepsilon_a$ is the energy of the eigenstate $\psi_a$.
This observation follows directly from the expression of the eigenstates of the matrix (\ref{eq:arrowhead}), see Eq.~(\ref{eq:eigenstates}). It is straightforward to see that the power-law form (\ref{eq:powerlaw}) implies the scaling of the IPR (\ref{eq:res_IPR}), and therefore also the anomalous scaling dimension (\ref{eq:anomalousD}). It is remarkable that the multifractal behavior shown in Fig.~\ref{fig7}(a), which is usually associated to much more complex models in the context of localization-delocalization  transitions~\cite{Evers_Anders_2008}, emerges in such an elementary way in our arrowhead matrix model.

\subsubsection{Photon weight and spectral function}

In Sec.~\ref{photon_sec} we derive asymptotically exact formulas for the photon weight and the closely related photon spectral function, sometimes referred to as photon density of states, at large $N$. The photon weight of an eigenstate $\psi_a$ with energy $\varepsilon_a$ is defined as the weight of its $(N+1)^{\rm th}$ component, i.e. ${\rm PW}_a = | \psi_{a, N+1}|^2\equiv \vert \langle G,1 \vert \psi_{a} \rangle \vert^{2}$. It is an important quantity as it allows to quantify the photon admixture of the eigenstates and therefore to characterize their hybrid nature. The photon weight (or equivalently the photon spectral function) also directly enters the definition of the cavity transmission spectrum~\cite{PhysRevA.69.043805}, and the power spectrum in a resonance fluorescence experiment according to the Wiener-Khintchine theorem~\cite{Scully_1997}.

We find that the {\it photon weight} of the two polaritons ${\rm P}+$ and ${\rm P}-$ is always of order $O(1)$ when $N \rightarrow \infty$, as expected, and that it is given by
\begin{equation}
    \label{eq:pw_pm}
    {\rm PW}_{\pm } \underset{N\rightarrow \infty}{\simeq} \frac{1}{1 - \pi g^2 \tilde{\rho}'(\varepsilon_{{\rm P}\pm} )} .
\end{equation}
[$\tilde{\rho}'$ denotes the derivative of $\tilde{\rho}$, see Eq.~(\ref{eq:defmomentsdistr}).] In contrast, the photon weight of the dark states is of order $O(1/N)$. Its average value over a small energy shell $[\varepsilon - \delta \varepsilon/2 , \varepsilon + \delta \varepsilon/2]$ is
\begin{equation}
    \label{eq:PWdark1}
   \overline{ {\rm PW}} (\varepsilon) \, \underset{N\rightarrow \infty}{\simeq} \, \frac{1}{N} \frac{1/(\pi g)^2}{\rho(\varepsilon)^2 + (\tilde{\rho}(\varepsilon) - \frac{\varepsilon}{\pi g^2})^2 } .
\end{equation}
The photon weight per dark state vanishes in the thermodynamic limit, while the photon weight shared by all dark states does not - it remains of order $O(1)$ - since the number of dark states is $O(N)$. We find that this finite photon weight for $N\to \infty$ is crucial to the existence of semi-localization and cavity-protected transport. 

\begin{figure}[t]
    \centering
    \includegraphics[width=0.5\textwidth]{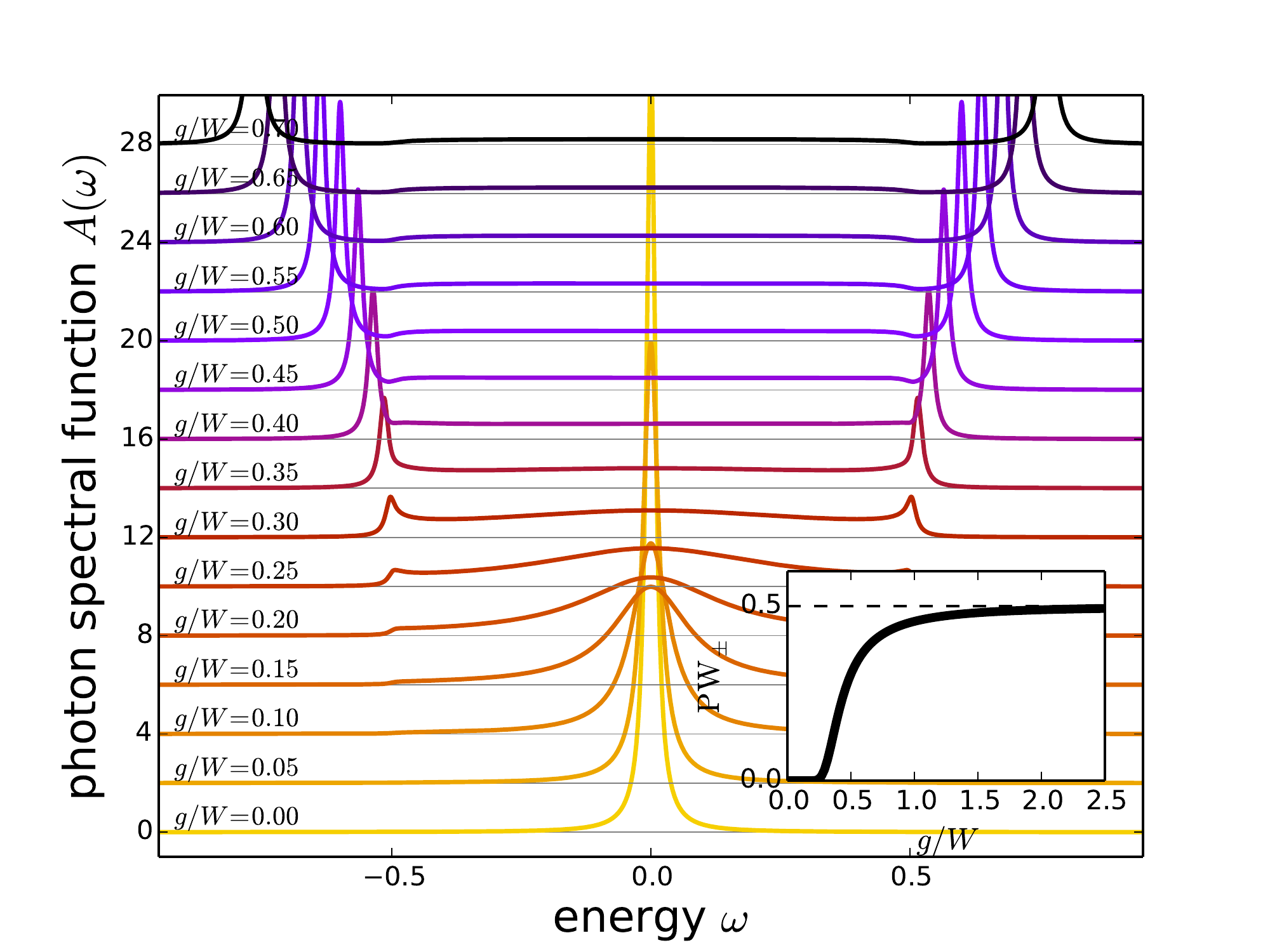}
    \caption{The photon spectral function $A(\omega)$ given by Eq.~(\ref{spectral_pw}), for different values of the collective coupling $g$, and for a box disorder distribution of width $W=1$. For readability, each curve is shifted vertically as $A(\omega) \rightarrow A(\omega) + 40 g$. For the plot we take the convolution of Eq.~(\ref{spectral_pw}) with a Lorentzian of width $\sigma = 0.01$. Inset: the photon weight ${\rm PW}_{\pm}$ given by Eq.~(\ref{eq:pw_pm}), for the box disorder distribution. We see that there are two distinct regimes separated by a smooth crossover. For weak couplings ($g \ll W$) the weight of the two peaks outside the interval $[-W/2,W/2]$ is strongly suppressed, while for strong couplings ($g \gg W$) most of the photonic weight is carried by those two peaks, which correspond to polaritonic modes in that case.}
    \label{fig:spectral_function_g}
\end{figure}

The {\it photon spectral function} is defined as the imaginary part of the photon retarded Green's function, $A(\varepsilon) = -\frac{1}{\pi} {\rm Im }\left(\left[ (\varepsilon - H + i 0^+)^{-1} \right]_{N+1,N+1}\right)$. It is equal to the spectral density, weighted by the photon weights of the eigenstates,
\begin{equation}
    \label{eq:APW}
    A(\varepsilon) \, = \, \sum_{a = 1}^{N+1}   {\rm PW}_a \, \delta(\varepsilon - \varepsilon_a) .
\end{equation}
For large $N$, upon averaging over a small energy shell, the dark states contribution becomes (for $\varepsilon \in [\omega_{\rm min}, \omega_{\rm max}]$)
\begin{equation}
    \overline{A} (\varepsilon) \, \underset{N \rightarrow \infty}{\simeq} \, N \rho(\varepsilon)\, \overline{ {\rm PW}}  (\varepsilon) =   \frac{\rho(\varepsilon)/(\pi g)^2}{\rho(\varepsilon)^2 + (\tilde{\rho}(\varepsilon) - \frac{\varepsilon}{\pi g^2})^2 }.
    \label{spectral_pw}
\end{equation}

All these formulas are exact in the large $N$ limit. At the leading order in $N$ the spectral function and the photon weight (averaged over an energy shell) take deterministic values: they are self-averaging quantities. The fluctuations of the spectral function $A(\varepsilon)$ around its mean value $\overline{A}(\varepsilon)$ are of order $O(1/\sqrt{N})$ in the random arrowhead matrix model. We find that, at this order, the fluctuations of the spectral function $A(\varepsilon)$ around its large-$N$ value $\overline{A}(\varepsilon)$ are Gaussian, characterized by a covariance which we compute exactly [see Eq.~(\ref{eq:covA}) for the result, and Fig.~\ref{fig:spectralfunction} for an illustration]. Thus, the random arrowhead matrix model allows us to provide analytic expressions for observables - the mean value and fluctuations of the spectral function $A(\epsilon)$ - that are of direct relevance to spectroscopy experiments.

\jd{}
\begin{figure}[tb]
	\includegraphics[width=0.45\textwidth]{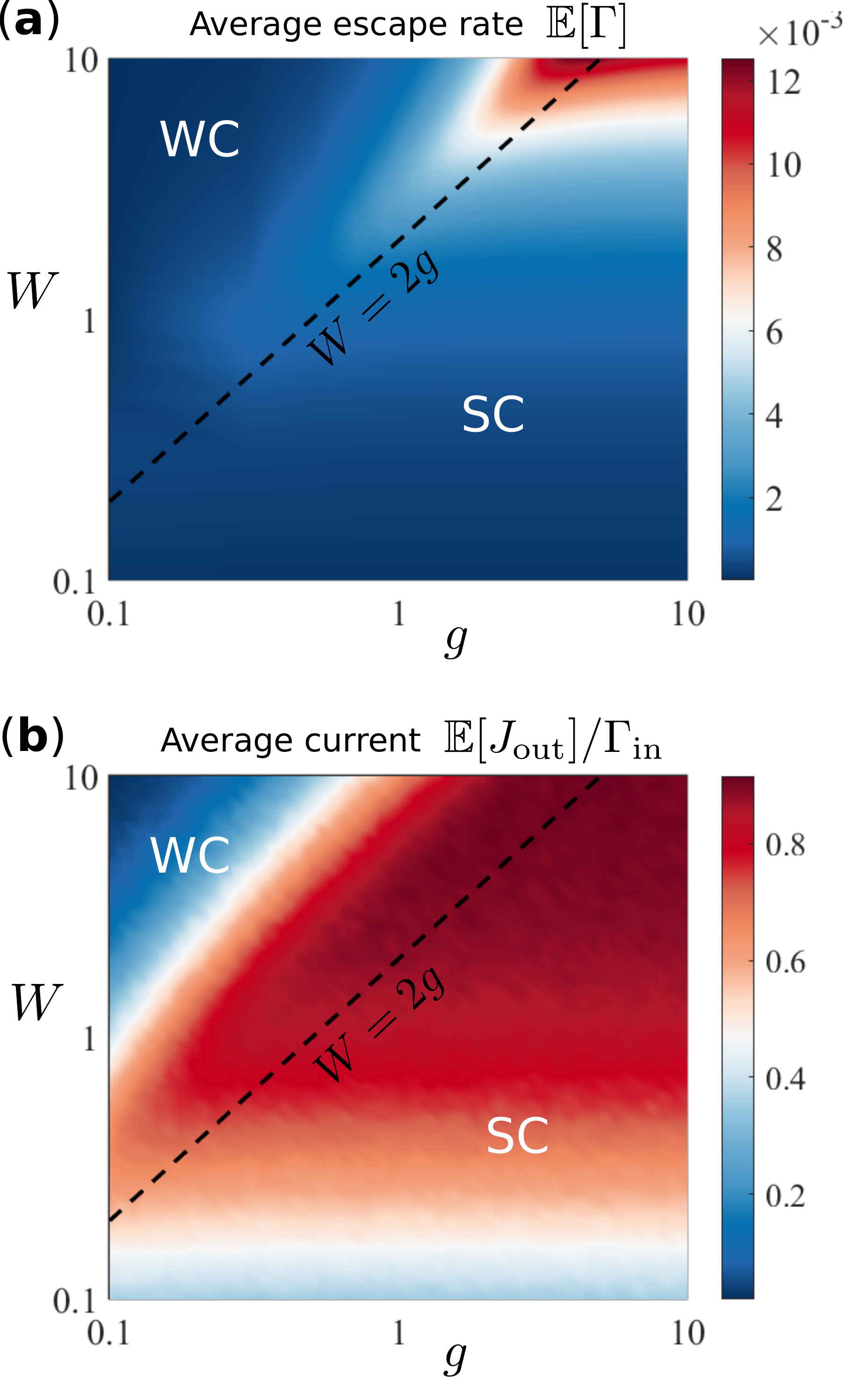}
	\caption{Cavity-protected transport. (\textbf{a}) Disorder-averaged escape rate $\mathbb{E}[\Gamma]$ (\textbf{a}) and normalized output current $\mathbb{E} [J_{\rm out}]/\Gamma_{\rm in}$ (\textbf{b}) as a function of $g$ and $W$ for $N=500$. The disorder-averaged escape rate is given by Eq.~(\ref{esc_av}), with the photon weights PW$_{\pm}=|\psi_{\pm, N+1}|^2\equiv \vert\bra{G,1} \psi_{\pm} \rangle \vert^{2}$ obtained from the eigenstates Eq.~(\ref{eq:eigenstates}). The output current given by Eq.~(\ref{GFs_finn23}), with $\epsilon_{a}$ the eigenvalues of the non-hermitian matrix Eq.~(\ref{eq:arrowhead2}) computed numerically, is averaged over $1000$ disorder configurations. The strong (SC) and weak (WC) coupling regimes are separated by the dotted line $W=2 g$, indicating where the energy splitting between the two polaritons precisely equals the width of the disorder distribution. The contribution of the dark states to $\mathbb{E} [J_{\rm out}]$ is $> 99\%$.}
	\label{Figcurrent}
\end{figure}

\subsubsection{Escape probability}
\label{esc_proba_summary}

In Sec.~\ref{sec:propagator} we derive exact asymptotics of all components of the Green's function, or evolution operator, $\hat{G}(t) = e^{-i \hat{H} t}$, at the leading order in $N$. As an application of that result we provide a solution to the following question: What is the probability $P_j(t)$ that an excitation, initially located on site $j$ with energy $\omega_{j}$, has escaped from this site at time $t$? This quantity is related to the diffusion properties of excitations throughout the system of emitters. We analyze the behavior of the escape probability $P_j(t) = 1- |G_{j,j}(t)|^2$ for times $t$ such that $1 \ll g t \ll N$. In that regime we find a diffusive-like behavior $P_j (t) = \Gamma (\omega_{j}) \, t$, with an exact formula for the escape rate
\begin{equation}
    \label{eq:fermigolden}
    \Gamma (\omega_{j}) = 2\pi \left( g/\sqrt{N}\right)^2 \overline{A} (\omega_j),
\end{equation}
involving the \textit{individual} coupling strength $g/\sqrt{N}$. We also find that the contribution of dark states to this escape dynamics completely dominates over that of polariton states, consistently with the semi-localization properties of the dark states that are localized on multiple, non-contiguous sites. The latter are thus connected to each other via the dark states regardless of the distance separating them, which allows long-range and diffusive-like transfer of excitations throughout the system of emitters. 

The resemblance of Eq.~(\ref{eq:fermigolden}) with the Fermi's golden rule is striking, with the contribution of the dark states to the photon spectral function $\overline{A}(\omega_j)$ [Eq.~(\ref{spectral_pw})] playing the role of the photonic density of states. We present an interpretation of this result using an effective Hamiltonian obtained after integrating out the cavity mode in second-order perturbation. In this effective description the coupling to the cavity provides a long-range hopping term with amplitude depending only on the disordered energies of the emitters, but not on the distance between them. Interestingly, such a diffusive behavior is found for \textit{any} strength of the coupling (in the thermodynamic limit), meaning that arbitrarily small $g/\sqrt{N}$ are sufficient to turn a fully localized phase characterized by the complete absence of diffusion without coupling to the cavity into a new phase with diffusive properties. For any finite coupling strength, we also find that the energy $\omega_{j}$ of the initial site can serve as a knob to control the escape dynamics, which could be exploited in experiments using a narrow band laser for instance:  
\begin{enumerate}[(i)]
    \item For $\omega_{j}=0$, i.e. injection in the middle of the distribution (at resonance with the cavity mode), the escape rate reaches its maximum possible value $2W/(\pi N)$, independently of $g$.  
    \item For $\omega_{j}=\pm W/2$, i.e. injection on the edges of the box distribution, the escape rate vanishes for all values of $g$.
    \item In all other cases, the escape rate increases $\sim g^{4}$ for weak couplings $g \ll W$, reaches its maximum value $2W/(\pi N)$ for intermediate couplings $g\sim W$, and then saturates to a lower, $g$-independent value.    
\end{enumerate}
This can be understood as the photon weight of the dark states PW$(\omega_{j})$ [or equivalently the contribution of the dark states to the photon spectral function $A(\omega_{j})$] governing the escape dynamics reaches its maximum value at the center of the distribution and vanishes (logarithmically for a box distribution) at the edges. Surprisingly, we find that the maximum escape rate $2W/(\pi N)$
increases with the disorder strength, which means that the cavity does not only protect transport against disorder but also turns disorder into an ally that can contribute favourably to transport.  

Another observable relevant to experiments, using e.g. a broad band laser, is the disorder-averaged escape rate $\mathbb{E}[\Gamma (\omega_{j})]$, which exhibits a similar behavior as $\Gamma (\omega_{j})$: It grows as $g^2$ for weak couplings $g \ll W$, reaches a maximum for intermediate couplings $g\sim W$, and then saturates to a lower value for strong couplings $g \gg W$ [Fig.~\ref{Figcurrent}(\textbf{a})]. Moreover, this saturation value is found to grow linearly with $W$, showing that disorder can help transport also after disorder average - provided coupling is strong enough.  

Due to the non-local character of the coupling to the cavity, we note that the escape dynamics does not correspond to a standard diffusive motion as the initial excitation does not propagate from one site to the next. A more accurate picture is that the excitation gets \textit{immediately} delocalized over all the emitters, with occupation rising linearly with time, similarly to a diffusive motion.  

\subsubsection{Out-of-equilibrium transport}

In Sec.~\ref{sec:trans} we investigate out-of-equilibrium transport through the system of emitters in a two-terminal configuration, in relevance to mesoscopic physics~\cite{PhysRevB.41.7411}. Here, two sites (e.g. $j=1$ and $j=N$) are connected to reservoirs respectively injecting and extracting spin excitations with the rates $\Gamma_{\rm in}$ and $\Gamma_{\rm out}$. We provide analytical expressions for the steady-state excitation current flowing through the system of emitters, which depends on a set of non-equilibrium Green's functions for the emitters and cavity photons, simply connected to the evolution operator $e^{-i \hat{H} t}$ introduced in Sec.~\ref{esc_proba_summary}. The former are conventionally denoted as ``lesser'', $\hat{G}^{<} (t)$, ``greater'', $\hat{G}^{>} (t)$, ``retarded'', $\hat{G}^{\rm R} (t)$, and ``advanced'' Green's function, $\hat{G}^{\rm A} (t)$. Similar notations are used for the cavity mode Green's functions. We derive the equations of motion of these Green's functions assuming that the spin and bosonic operators can all be replaced by fermionic ones. This replacement is valid as long as the system remains in the single-particle sector, which we find to hold when rescaling the current injection rate as $\Gamma_{\rm in} \to \widetilde{\Gamma}_{\rm in}/N^{2}$. 

Solving the equations of motion allows us to compute the output current $J_{\rm out}=\Gamma_{\rm out} \, n_{N}$ and the population at each site $j$,
\begin{align*}
n_{j} = \int \! \frac{d\omega}{2\pi} {\rm Im} \left[G^{<}_{j,j} (\omega)\right],
\end{align*}
with $G^{<}_{j,j} (\omega)$ the Fourier transform of the $j^{\rm th}$ matrix element of the lesser Green's function. We derive analytical formulas for the current and populations, which are found to depend on the eigenvalues of the non-hermitian arrowhead matrix
\begin{equation}
\label{eq:arrowhead2}
\left( \begin{array}{ccccc|c}
		\omega_{1} - i \frac{\Gamma_{\rm in}}{2} & 0 & \cdots & 0 & 0 & g/\sqrt{N} \\
		0  & \omega_{2} & \ddots & 0 & 0 & g/\sqrt{N} \\
		\vdots & \ddots & \ddots & \ddots & \vdots & \vdots \\
		0 & 0 & \ddots  & \omega_{N-1} & 0 & g/\sqrt{N} \\
		0 & 0 & \cdots & 0 & \omega_{N}- i \frac{\Gamma_{\rm out}}{2} & g/\sqrt{N} \\ \hline
		g/\sqrt{N} & g/\sqrt{N} & \cdots & g/\sqrt{N} & g/\sqrt{N} & 0
	\end{array} \right).
\end{equation} 
The eigenvalues of this matrix correspond to polaritons and dark states with complex energies due to the in and out couplings to the reservoirs. The latter indeed contribute to the total self-energy through the imaginary numbers $i \Gamma_{\rm in}/2$ and $i \Gamma_{\rm out}/2$. This can be interpreted as saying that the eigenstates have acquired a finite lifetime due to the coupling to the environment, which drives the system out of equilibrium. 

We find that the disorder-averaged current exhibits a similar behavior as the escape rate: It grows as $g^2$ for weak couplings, reaches a maximum at intermediate couplings $g\sim W$, and then saturates to a lower, $g$-independent value for strong couplings [Fig.~\ref{Figcurrent}(\textbf{b})]. While for a given $g \ll W$, the current slightly decreases with the disorder strength $W$, it increases with $W$ for strong enough couplings. This confirms the cavity protection effect observed with the escape dynamics. We also find that the current scales $\sim 1/N^{2}$, consistent with the results of Ref.~\cite{PhysRevLett.126.153201}. While this current is equally carried by polaritons and dark states for small $N$, the contribution of the (broadened) dark states fully dominates in the thermodynamic limit, for all coupling strengths [Fig.~\ref{Figcurrent}(\textbf{c})]. This is to our knowledge the first demonstration of efficient long-range quantum transport solely mediated by the non-local coupling to the cavity, and entirely carried by the dark states.

After this summary of our main results, we turn to the detailed analysis of the spectrum and eigenstates of the arrowhead Hamiltonian (\ref{eq:arrowhead}).

\section{Finite size: Energy spectrum, eigenstates and Green's function}
\label{sec:finiteN}

We start by recalling some simple exact formulas about arrowhead matrices~\cite{o1990computing}. For finite $N$, the eigenstates of the Hamiltonian (\ref{eq:arrowhead}) are, for $1 \leq a \leq N+1$,
\begin{equation}
	\label{eq:eigenstates}
	H \psi_a = \varepsilon_a \psi_a,  \qquad \psi_a=  N_{a} \left( \begin{array}{c}
		\frac{g/\sqrt{N}}{\varepsilon_a - \omega_1} \\
		\frac{g/\sqrt{N}}{\varepsilon_a - \omega_2} \\
		\vdots \\
		\frac{g/\sqrt{N}}{\varepsilon_a - \omega_N} \\
		1
	\end{array} \right) ,
\end{equation}
with a constant $N_{a}= \left(1 + \frac{1}{N} \sum_{j=1}^N \frac{g^2}{(\varepsilon_a - \omega_j)^2} \right)^{-1/2}$ ensuring normalization. The energies $\varepsilon_a$ are the solutions of the equation
\begin{equation}
	\label{eq:defeps}
	\varepsilon_a \, = \, \frac{1}{N} \sum_{j=1}^N \frac{g^2}{\varepsilon_a - \omega_j} .
\end{equation}
Notice that it is a polynomial equation of degree $N+1$ for $\varepsilon_a$.
From now on we work with bare energies $\omega_j$ which are sorted in increasing order: $\omega_1 \leq \omega_2 \leq \dots \leq \omega_N$. Similarly, we label the energies of $H$ in increasing order. Then a very important fact is that the eigenvalues of $H$ are interlaced with the bare energies:
\begin{equation}
	\label{eq:intertwining}
	\varepsilon_1 \leq \omega_1 \leq  \varepsilon_2 \leq \omega_2 \leq \dots  \leq \omega_{N-1} \leq \varepsilon_N \leq \omega_N \leq \varepsilon_{N+1} .
\end{equation}

\begin{figure}[t]
	\includegraphics[width=0.4\textwidth]{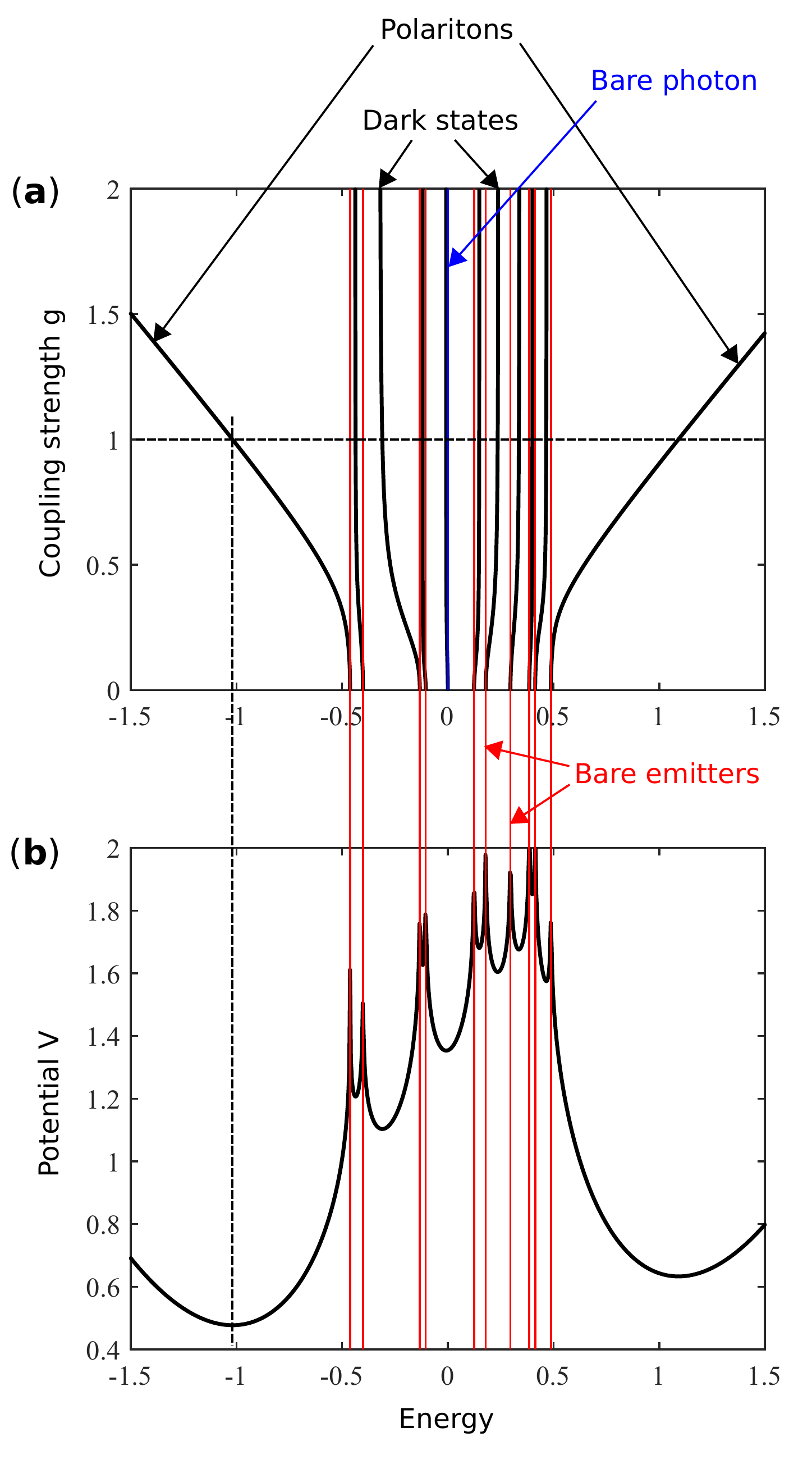} 
	\caption{(\textbf{a}) Energy spectrum (in black) of the Hamiltonian (\ref{eq:arrowhead}) for increasing values of $g$, here for $N=10$ emitters with bare energies (in red) drawn from a uniform distribution in the interval $[-1/2,1/2]$ ($W=1$). The $N-1$ eigenvalues comprised in the interval $[-1/2,1/2]$ are denoted as dark states, while the two unbounded eigenvalues at the edges of the spectrum are referred to as polaritons. The bare photon energy ($\omega=0$) is depicted in blue. (\textbf{b}) The energy spectrum coincides with the $N+1$ minima of the ``potential'' $V(x) = \frac{x^2}{2} - \frac{g^2}{N} \sum_{j=1}^N  \log |x-\omega_j|$, which is plotted with the same bare energies as in (\textbf{a}) and with $g=1$. There is exactly one minimum between each pair of consecutive bare energies.}
	\label{fig:spectrum}
\end{figure}

One elementary way of understanding this interlacing is to observe that the $\varepsilon_a$'s are the $N+1$ minima of the ``potential'' $V(x) = \frac{x^2}{2} - \frac{g^2}{N} \sum_{j=1}^N  \log |x-\omega_j|$, which looks as in Fig.~\ref{fig:spectrum}(b). That ``potential'' diverges when $x \rightarrow - \infty, \omega_1, \omega_2 , \dots, \omega_N, + \infty$, so it is clear that it possesses $N+1$ minima which satisfy (\ref{eq:intertwining}). It is interesting to note that this particular interlacing of the spectrum with the bare energies also appear{\color{blue}s} in other models where long-range interactions generate collective modes on top of a continuum of individual excitations. This occurs for instance in an electron gas with Coulomb interactions, where the normal modes are given by the zeros of the dielectric function computed in, e.g. the random phase approximation. In this case, these normal modes consist of slightly renormalized individual, electron-hole excitations across the Fermi surface lying in between the bare individual excitations, as well as a collective, unbounded plasmon mode with higher energy. Here, the two {\it polariton} states with unbounded energies $\varepsilon_{1}$ and $\varepsilon_{N+1}$ are hybrid light-matter collective modes, while the $N-1$ other individual states with energies $\varepsilon_j$ ($j=2,\cdots,N$) are usually referred to as {\it dark states}. As explained in the following, these dark states acquire a small but finite photonic weight due to the presence of disorder.       

In Sections \ref{sec:propagator} we will be interested in the propagator, or evolution operator, $\hat G(t) = e^{-i \hat H t}$, which governs the time evolution of the system described by the Hamiltonian (\ref{eq:arrowhead}). It is an $(N+1) \times (N+1)$ unitary matrix, $\hat G^\dagger(t) = \hat G^{-1} (t)=\hat G(-t)$. The components of that matrix are the Green's functions
\begin{subequations}
\begin{align}
&G_{i,j}(t) = \bra{i,0} \hat G(t) \ket{j,0}= \langle \hat \sigma^{-}_{i}(t) \hat \sigma^{+}_{j}(0) \rangle \label{gfij} \\ \intertext{for $1 \leq i,j \leq N$,}  
&G_{j,N+1}(t) = \bra{j,0} \hat G(t) \ket{G,1}= \langle \hat \sigma^{-}_{j}(t) \hat a^{\dagger}(0) \rangle \\ \intertext{for $1 \leq j \leq N$,}
&D(t)\equiv G_{N+1,N+1}(t)= \bra{G,1} \hat G(t) \ket{G,1}= \langle \hat a(t) \hat a^{\dagger}(0) \rangle,
\end{align}
\end{subequations}
where $\langle \cdots \rangle \equiv \bra{G,0} \cdots \ket{G,0}$ denotes the expectation value in the ground state $\ket{G,0}$ of both Hamiltonians $\hat H_{0}$ and $\hat H$. The operators $\hat \sigma^{\pm}_{j}(t)=e^{i \hat H t} \hat \sigma^{\pm}_{j} e^{-i \hat H t}$ and $\hat a^{(\dagger)}(t)=e^{i \hat H t} \hat a^{(\dagger)} e^{-i \hat H t}$ are defined in the Heisenberg picture. In order to simplify notations, we call $D(t)$ the cavity photon Green's function, which corresponds to the $N+1^{\rm th}$ element of the matrix $\hat{G} (t)$. We introduce the Fourier transform $\hat G(\omega)=\int dt e^{i\omega t} \hat G(t)$ where $\hat G(t)=i[\hat G^{\rm R}(t)- \hat G^{\rm A}(t)]$ can be expressed as the sum of a ``retarded'' $\hat G^{\rm R}(t)=-i\theta(t)e^{-i \hat H t}$ and an ``advanced'' $\hat G^{\rm A}(t)=i\theta(-t)e^{-i \hat H t}$ propagator, in such a way that $\hat G^{\rm R}(\omega)=(\omega-\hat H + i0^{+})^{-1}$ and $\hat G^{\rm A}(\omega)=(\omega- \hat H + i0^{-})^{-1}$. Since the matrix elements $G^{\rm A}_{i,j}(\omega)=[G^{\rm R}_{i,j}(\omega)]^{*}$, the Green's function $G_{i,j}(\omega)=\bra{i,0} \hat G (\omega) \ket{j,0}$ is a positive definite matrix, which corresponds, up to a factor $2\pi$, to the spectral function
\begin{equation*}
    A_{i,j}(\omega)\equiv -\frac{1}{\pi} {\rm Im}~G^{\rm R}_{i,j}(\omega).
\end{equation*}
One can easily write the Green's functions in closed form using these definitions and Eq.~(\ref{eq:eigenstates}):
\begin{subequations}
\label{eq:prop_finiteN}
\begin{align}
&G_{i,j} (t) =  \frac{g^2}{N} \sum_{a=1}^{N+1} \frac{e^{- i \varepsilon_a t }}{(\varepsilon_a- \omega_i ) (\varepsilon_a- \omega_j ) (1+ \frac{1}{N} \sum_{k} \frac{g^2}{(\omega_k - \varepsilon_a)^2} ) }
 \\ \intertext{for $1 \leq i,j \leq N$,}
&G_{j,N+1} (t) = \frac{g}{\sqrt{N}} \sum_{a=1}^{N+1} \frac{e^{- i \varepsilon_a t }}{(\varepsilon_a- \omega_j ) (1+ \frac{1}{N} \sum_{k} \frac{g^2}{(\omega_k - \varepsilon_a)^2} ) } \\ \intertext{for $1 \leq j \leq N$, and}
&D (t)  = G_{N+1,N+1}(t) = \sum_{a=1}^{N+1} \frac{e^{- i \varepsilon_a t }}{1+ \frac{1}{N} \sum_{k} \frac{g^2}{(\omega_k - \varepsilon_a)^2}  }.
\end{align}
\end{subequations}
That last equation gives the photon Green's function, which plays a distinguished role in the analysis of the model, and for which we derive the large $N$ asymptotics in Section \ref{photon_sec}. In Section \ref{sec:propagator}, we derive the large $N$ asymptotics of the other components (\ref{eq:prop_finiteN}.a,b).

\section{Spectrum and mean energy shift at large $N$}
\label{sec:shifts}

Because of the interlacing Eq.~(\ref{eq:intertwining}), it is clear that the vast majority of energies fall in the same interval $[\omega_{\rm min}, \omega_{\rm max}]$ as the bare energies $\omega_j$. Only the two energy levels $\varepsilon_1$ and $\varepsilon_{N+1}$ fall outside that interval. 

Important insights can be gained by first looking at the energy spectrum in the absence of disorder, when all bare energies are equal to zero: $\rho(\omega) = \delta(\omega)$. In this case, the Hamiltonian only consists of the linear coupling term $\hat V$, which is diagonalized by the collective spin operators $\hat b =\sum_{i} \hat \sigma_{i}^{-}/\sqrt{N}$. In the single-excitation subspace, these operators satisfy the commutation relations
\begin{align*}
&\bra{G,0}[\hat b,\hat b^{\dagger}]\ket{G,0}=1 \\
&\bra{j,0}[\hat b,\hat b^{\dagger}]\ket{j,0}=-\frac{1}{N},
\end{align*}
and are therefore bosonic in the thermodynamic limit $N\to \infty$. The Hamiltonian $\hat V=(g/\sqrt{N}) (\hat a \hat b{^\dagger} + \hat b \hat a^{\dagger})$ features two {\it polariton} eigenmodes $\hat p= (\hat b \pm \hat a)/\sqrt{2}$ with energies $\pm g$. In addition, the spectrum contains $N-1$ {\it dark modes} with energy $\omega_{0}$ (here set to zero) that do not interact with the cavity mode. 

A finite bandwidth for the disorder distribution (e.g. a box distribution of width $W > 0$) changes this picture as the Hamiltonian $\hat H_{0}$ can no longer be diagonalized by the collective operators $\hat b, \hat b^{\dagger}$. The spectrum can be, however, qualitatively understood similarly in terms of individual dark modes and collective polariton modes for $g\gg W$. When $N \rightarrow \infty$, the $N-1$ energies $\varepsilon_2, \dots \varepsilon_N$ form a continuum of dark states with the same density as the bare energies, i.e. $\sum_{a=2}^{N} \delta (\varepsilon - \varepsilon_a) \simeq N \rho(\varepsilon)$, to leading order in $N$ [see Eq.~(\ref{eq:rho}) for the definition of $\rho$]. The remaining two polariton energy levels are equal to $\varepsilon_1 \simeq \varepsilon_{{\rm P}-} < \omega_{\rm min}$ and $\varepsilon_{N+1} \simeq \varepsilon_{{\rm P}+} > \omega_{\rm max}$ where $\varepsilon_{{\rm P}\pm}$ are the two solutions of the equation
\begin{equation}
	\label{eq:gpm}
	\varepsilon_{{\rm P}\pm}  =   \pi g^2 \, \tilde{\rho} (\varepsilon_{{\rm P}\pm}).
\end{equation}
Here $\tilde{\rho}$ is the Hilbert transform of $\rho$,
\begin{equation}
	\label{eq:hilbert}
	 \tilde{\rho}(\varepsilon) = {\rm p.v.} \frac{1}{\pi} \int \frac{\rho(\omega) d\omega}{\varepsilon - \omega},
\end{equation}
where `p.v.' stands for the principal value of the integral. Note that the principal value is needed in the definition when the argument $\varepsilon$ lies in $[\omega_{\rm min},\omega_{\rm max}]$, the support of $\rho$. Equation (\ref{eq:gpm}) is obtained by taking the continuum limit of Eq.~(\ref{eq:defeps}). In the case of a box distribution for the bare orbitals, i.e. $\rho(\omega)=1/W$ for $\omega \in  [-W/2,W/2]$, with $\omega_{\rm min}=-W/2$ and $\omega_{\rm max}=W/2$, Eq.~(\ref{eq:gpm}) becomes 
\begin{equation}
	\label{eq:gpm_mean}
	\varepsilon_{{\rm P}\pm}  = \pi g^2 \, \frac{1}{\pi W}\log \bigg\vert \frac{\varepsilon_{{\rm P}\pm}+(W/2)}{\varepsilon_{{\rm P}\pm}-(W/2)} \bigg\vert.
\end{equation}
For strong couplings $g \gg W$, one recovers $\varepsilon_{\pm} \simeq \pm g$, i.e. the polariton splitting is twice the collective coupling strength and thus barely affected by the disorder~\cite{Houdre_Vacuum_1996}.

A more refined description of the spectrum of dark states is obtained as follows. Since the energy $\varepsilon_a$ ($2 \leq a \leq N$) lies in the interval $[\omega_{a-1}, \omega_a]$, it is typically at a distance of order $1/N$ from the middle of that interval,
\begin{subequations}
\begin{equation*}
	2 \leq a\leq N : \qquad  \varepsilon_a = \frac{\omega_{a-1} + \omega_a}{2} + \frac{1}{N} \Delta_a .
\end{equation*}

\begin{figure*}[ht]
	\includegraphics[width=1.0\textwidth]{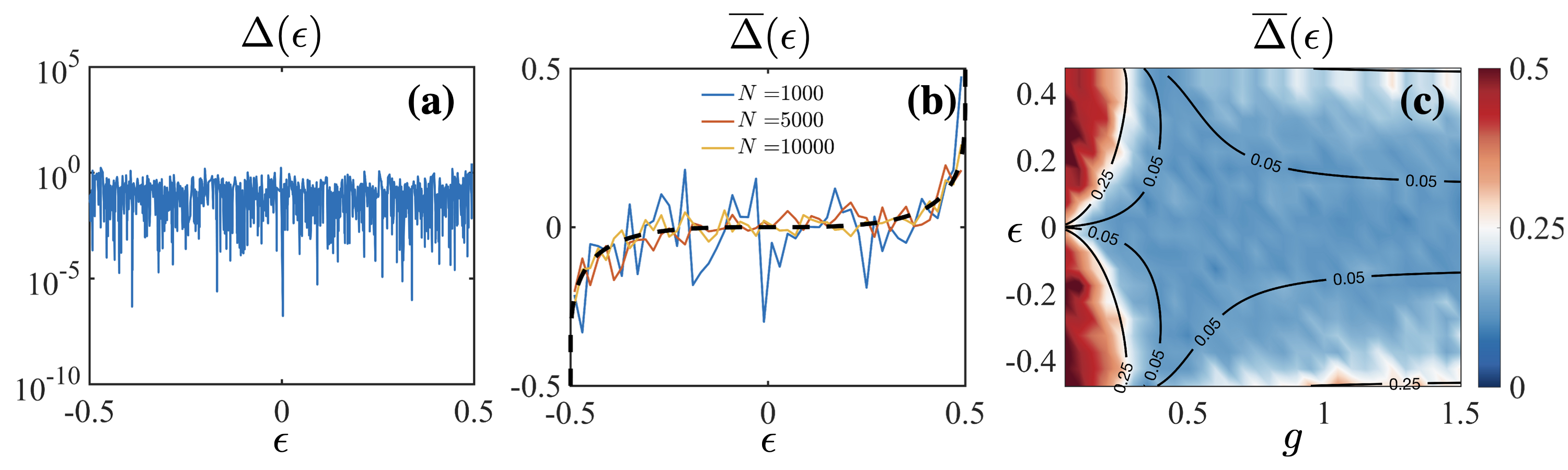}
	\caption{Energy shift computed numerically from the spectrum of the arrowhead matrix Eq.~(\ref{eq:arrowhead}) for a given disorder realization, $g=0.5$, and $W=1$. (\textbf{a}) without energy binning (average over a small energy shell) for $N=1000$, (\textbf{b}) with binning according to Eq.~(\ref{eq:Delta_def}) for different $N$. (\textbf{c}) Binned energy shift versus energy $\varepsilon$ and coupling strength $g$. The energy shift wildly fluctuates, but its mean value after average over a small energy shell converges to the formula (\ref{eq:Delta}) (dotted line) in the thermodynamic limit $N \to \infty$.}
	\label{Fig3}
\end{figure*}

This defines the {\it shift} $\Delta_a$. The shift depends not only
on the intensive energy density $\rho(\omega)$ when $N \rightarrow \infty$, but also on the microscopic details of the distribution of bare energies. In general, $\Delta_a$ is a wildly fluctuating function of its index $a$, and of the bare energies $\omega_j$ [see Fig.~\ref{Fig3} (\textbf{a})].

By averaging over all eigenstates within a small energy shell, $\varepsilon_a \in [\varepsilon - \delta \varepsilon/2, \varepsilon + \delta \varepsilon/2]$, one can define a mean energy shift
\begin{equation}
	\label{eq:Delta_def}
	\overline{\Delta} (\varepsilon) = \frac{1}{N \rho(\varepsilon) \delta \varepsilon} \sum_{| \varepsilon_a - \varepsilon | \leq \delta \varepsilon/2 } \Delta_a .
\end{equation}
For large $N$, this mean energy shift depends only on the intensive distribution $\rho$, and not on the microscopic details or correlations between the bare energies [see Fig.~\ref{Fig3} (\textbf{b})]. We find
\begin{equation}
	\label{eq:Delta}
	\overline{\Delta} (\varepsilon) \underset{N \rightarrow \infty}{=} \frac{1}{\pi \rho(\varepsilon)} {\rm arctan} \left[ \frac{\tilde{\rho}(\varepsilon) - \frac{\varepsilon}{\pi g^2} }{\rho(\varepsilon)} \right] .
\end{equation}
\end{subequations}
Thus, the spectrum of $N-1$ dark states has the same distribution as the one of bare energy levels only at order $O(1)$; to order $O(1/N)$ the energy levels are pushed by an average amount (\ref{eq:Delta}). As seen in Fig.~\ref{Fig3} (\textbf{c}), the energy shift averaged over a small energy shell is close to $0.5$ for weak couplings $g/W \ll 1$, meaning that the dark states are very close to the bare energy levels. On the other hand, the dark states lie roughly at equal distance from the two closest bare energy levels for strong couplings $g/W \gg 1$, i.e. $\overline{\Delta} (\varepsilon) =0$.   

{\it Derivation of 	$\overline{\Delta}$:} To arrive at Eq.~(\ref{eq:Delta}), one can rely on tricks from complex analysis and contour integration. Let $C_{a}$ be a counterclockwise contour in the complex plane which encloses the interval $(\varepsilon_a, \omega_a)$ along the real axis. We assume that $C_a$ does not enclose any of the other points $\varepsilon_{a'}$ or $\omega_{a'}$ for $a'\neq a$. Then
$$
\oint_{C_{a}} \frac{dz}{2\pi i} \, \log \left( \frac{z-\omega_b}{z-\varepsilon_b} \right)  =  ( \varepsilon_a -\omega_a ) \delta_{a,b}.
$$
Summing over $b$ from $1$ to $N$, and adding the term $\log (-1/( z- \varepsilon_{N+1}))$ in the sum (for which the contour integral around $C_a$, $a\leq N$, vanishes), one gets
\begin{align*}
\oint_{C_{a}} \frac{dz}{2\pi i} \, \log \left(- \frac{ \prod_{b=1}^{N} (z- \omega_b)}{ \prod_{c=1}^{N+1} (z- \varepsilon_c)} \right)  =  \varepsilon_a -\omega_a.
\end{align*}
The rational function inside the logarithm is nothing but
\begin{align*}
\frac{1}{\frac{1}{N} \sum_{j=1}^N \frac{g^2}{z- \omega_j} - z},
\end{align*}
as can be checked by inspecting the zeros and poles of the latter expression [see Eq.~(\ref{eq:defeps})]. For a large number of energy levels $\varepsilon_a$ in the small shell $[\varepsilon - \delta \varepsilon/2, \varepsilon+ \delta \varepsilon/2]$, one then finds
\begin{align*}
\overline{\Delta} (\varepsilon) &\simeq  \frac{1}{\rho(\varepsilon) \delta \varepsilon} \left[ \frac{\delta \varepsilon}{2} + \sum_{| \varepsilon_a - \varepsilon | \leq \delta \varepsilon /2} (\varepsilon_a - \omega_a) \right]  \\
&=  \frac{1}{2 \rho(\varepsilon)} + \frac{1}{\rho(\varepsilon) \delta \varepsilon} \\
&\times \sum_{| \varepsilon_a - \varepsilon | \leq \delta \varepsilon /2} \oint_{C_{a}} \frac{dz}{2\pi i} \, \log \left( \frac{1}{ \frac{1}{N}\sum_{j} \frac{g^2}{z- \omega_j} - z } \right) .
\end{align*}
We can rewrite the last expression as a single contour integral, for a contour $C_{\varepsilon,\delta \varepsilon}$ which encloses all the intervals $(\varepsilon_a , \omega_a)$ in the small energy shell $[\varepsilon - \delta \varepsilon/2, \varepsilon + \delta \varepsilon/2]$. That contour can be deformed close to the real axis, so one can think of it as the union of the two intervals $[\varepsilon-\delta \varepsilon/2,\varepsilon+\delta \varepsilon/2] + i 0^+$ and $[\varepsilon-\delta \varepsilon/2,\varepsilon+\delta \varepsilon/2] + i 0^-$. Then one takes the $N \rightarrow \infty$ limit in the integrand, and replaces $\frac{1}{N} \sum_{j=1}^N \frac{g^2}{z- \omega_j} - z$ with $\int \frac{g^2 \rho(\omega) d\omega}{z- \omega} - z$. This leads to
\begin{widetext}
\begin{eqnarray*}
\overline{\Delta} (\varepsilon) & \underset{N \rightarrow \infty}{=} & \frac{1}{2 \rho(\varepsilon)} + \frac{1}{\rho(\varepsilon) \delta \varepsilon} \left( \int_{\varepsilon - \frac{\delta \varepsilon}{2} + i 0^-}^{\varepsilon + \frac{\delta \varepsilon}{2} + i 0^-} -  \int_{\varepsilon - \frac{\delta \varepsilon}{2} + i0^+}^{\varepsilon + \frac{\delta \varepsilon}{2}+i0^+}  \right) \frac{dz}{2\pi i} \, \log \left( \frac{1}{g^2  \int \frac{\rho(\omega) d\omega}{z- \omega} - z} \right) \\
& \simeq & \frac{1}{2 \rho(\varepsilon)}  - \frac{1}{ \rho(\varepsilon)}  \frac{1}{2\pi i}\left[  \log  \left( \int \frac{\rho(\omega) d\omega}{\varepsilon - \omega + i 0^-} - \frac{\varepsilon}{g^2} \right) -  \log  \left(  \int \frac{\rho(\omega) d\omega}{\varepsilon - \omega + i 0^+} - \frac{\varepsilon}{g^2} \right) \right] \\
& = & \frac{1}{\pi \rho(\varepsilon)}  \left( \frac{\pi}{2}  - \frac{1}{2 i}\left[  \log  \left(  i \pi \rho(\varepsilon) +  \pi \tilde{\rho}(\varepsilon) - \frac{\varepsilon}{g^2} \right) -  \log  \left( - i \pi \rho(\varepsilon) +  \pi \tilde{\rho}(\varepsilon) - \frac{\varepsilon}{g^2} \right) \right]  \right) \\
& = & \frac{1}{\pi \rho(\varepsilon)}  {\arg }\left[   \rho(\varepsilon) + i \left(  \tilde{\rho}(\varepsilon) - \frac{\varepsilon}{\pi g^2} \right)\right] ,
\end{eqnarray*}
\end{widetext}
where we have used the Sokhotskyi-Plemelj formula from the second to the third line. The last line is equivalent to Eq.~(\ref{eq:Delta}).

{\it Statistics of nearest neighbor spacings:} To further characterize the distribution of dark states energy levels, we investigate the statistics of nearest-neighbor level spacings, within a small energy window $[\varepsilon - \delta \varepsilon/2,  \varepsilon + \delta \varepsilon/2]$. In that window we define the normalized spacings as $s_a = N \rho(\epsilon_a) (\varepsilon_{a+1}- \varepsilon_a)$, and we numerically evaluate their probability distribution, by averaging over many disorder realizations, for $N$ large enough so that the energy window contains many levels. We find that the resulting distribution of level spacings when $N \rightarrow \infty$ is a one-parameter family of probability distributions $p_\alpha(s)$, parameterized by $\alpha$, related to the normalized mean energy shift $\rho(\varepsilon ) \overline{\Delta} (\varepsilon)$, which itself depends on both the collective coupling $g$ and the energy $\varepsilon$. The parameter $\alpha$ is
\begin{equation*}
    \alpha  =  \, {\rm tan} \left[ \pi \rho(\varepsilon) \overline{\Delta}(\varepsilon)\right] = \frac{  \tilde{\rho}(\varepsilon) - \frac{\varepsilon}{\pi g^2}}{\rho(\varepsilon)} .
\end{equation*}
The probability distribution $p_\alpha(s)$ may alternatively be defined as follows, which is more convenient for numerical purposes. Consider $N$ i.i.d. variables $u_j$ ($j=1,\dots, N$) drawn from a uniform distribution in $[0,1]$, and the set of $N$ solutions of the equation
\begin{equation}
    \label{eq:prob_alpha}
    \alpha \, + \,  \frac{1}{N}\sum_{j=1}^N \frac{1}{\tan (\pi (X - u_j))}  = 0 , \qquad X \in [0,1] .
\end{equation}
One can study the distribution of that set of solutions $\{ X_a \}_{1\leq a \leq N}$ as a function of the parameter $\alpha$. In particular, ordering the solutions as $0 \leq  X_1 < X_2 < \dots < X_N \leq 1$, and defining the normalized level spacings as $s_a = N (X_{a+1}-X_a)$ for $a=1, \dots , N-1$ and $s_a = N (1+X_{1}-X_N)$ for $a=N$, we can study numerically the distribution of level spacings $p_{\alpha}(s)$. Notice that the level spacing statistics is invariant under the transformation $u_j \rightarrow -u_j$ (mod $1$), $\alpha \rightarrow - \alpha$, so we have $p_\alpha(s) = p_{-\alpha}(s)$.

When we replace Eq.~(\ref{eq:defeps}) by Eq.~(\ref{eq:prob_alpha}), the idea is that $\pi/{\tan(\pi (X_a - u_j))} \sim 1/(X_a-u_j)$, so at small distances the statistics of the roots of both equations must be the same. Moreover, because $1/{\tan(\pi (X_a +1 - u_j))} = 1/{\tan(\pi (X_a - u_j))}$, the problem defined by Eq.~(\ref{eq:prob_alpha}) is the analog of the one defined by Eq.~(\ref{eq:defeps}), but with periodic boundary conditions. The problem defined in this way is simpler, because it is translation invariant under $u_j \rightarrow u_j + U \; {\rm mod} \; 1$, $X_a \rightarrow X_a + U \; {\rm mod} \; 1$, for any $U$. This is particularly convenient for numerical purposes, because it allows to study the distribution of energy levels in the entire interval $[0,1]$, without having to restrict to a small energy window as above. 

\begin{figure}[ht]
    \includegraphics[width=0.5\textwidth]{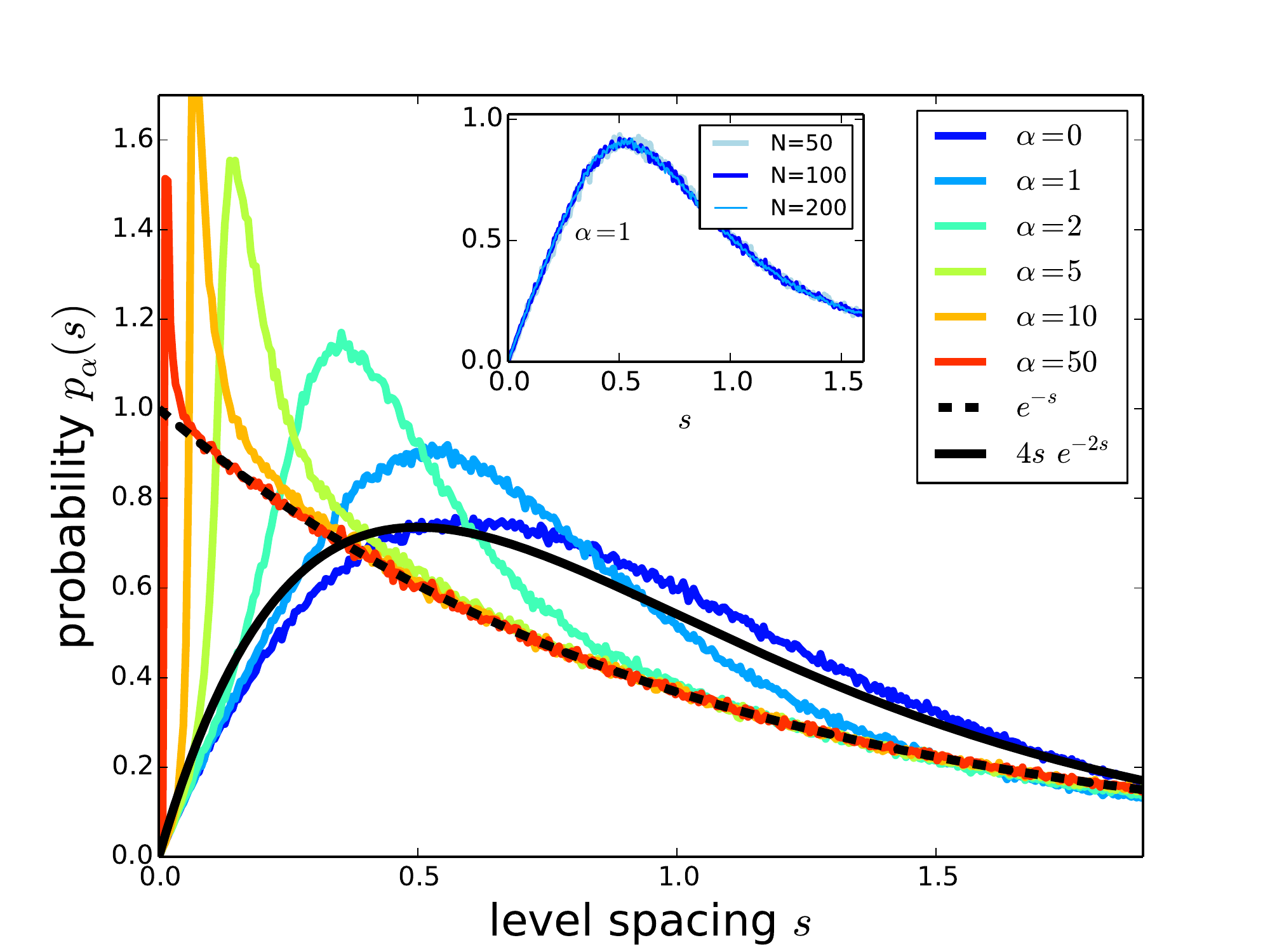}
    \caption{Probability distribution of level spacings $p_\alpha(s)$ in the model defined by Eq.~(\ref{eq:prob_alpha}), evaluated numerically for $N=200$, by averaging over $10^4$ independent disorder realizations. Since $p_\alpha (s) = p_{-\alpha (s)}$, we restrict to $\alpha \geq 0$. When $\alpha \rightarrow \infty$ the distribution is simply the one of independent levels $p_{\rm Poisson} (s) = e^{-s}$, however for finite $\alpha$ it is a different distribution which depends continuously on $\alpha$. When $\alpha = 0$, $p_\alpha(s)$ is close to (but not exactly equal to) the semi-Poisson distribution $p_{\rm semi-Poisson} = 4 s \, e^{-2 s}$ of Ref.~\cite{Bogomolny_1999}. Inset: same quantity evaluated for $N=50$, $N=100$ and $N=200$, showing that the results are converged as a function $N$.}
    \label{fig:level_spacings}
\end{figure}

In Fig.~\ref{fig:level_spacings}, we compute the probability distribution $p_\alpha(s)$ numerically by sampling the level spacings between the solutions of Eq.~(\ref{eq:prob_alpha}), by averaging over $10^4$ realizations for $N=50, 100, 200$. We see in the inset that the results converge quickly to a smooth probability distribution $p_\alpha(s)$ when $N$ increases. $p_\alpha(s)$ coincides with the Poisson distribution when $\alpha \rightarrow \infty$, which simply follows from the fact that $X_j = \omega_j$ in that limit. When $\alpha = 0$, the result is close to a semi-Poisson distribution. For finite values of $\alpha$ the one-parameter family $p_\alpha (s)$ smoothly interpolates between those two limiting distributions.

To summarize, in this section we have characterized the energy spectrum, in particular the spectrum of dark states by computing the average energy shifts (\ref{eq:Delta}) and the statistics of the energy spacings. Notice that, to compute the energy shifts, we have used contour integrals techniques which, in the end, are sensitive only to the thermodynamic density of states $N \rho(\omega)$, as opposed to higher moments of the disorder distribution. In this paper we will encounter other quantities that share that property and are sensitive only to the density of states, in particular the photon Green's function and the photon spectral function. However, other quantities do depend on the higher moments of the disorder distribution: this is the case for the inverse participation ratio of the eigenstates, which we study in the next section.

\section{Inverse Participation Ratio}
\label{sec:IPR}

In this section we turn to the localization properties of the eigenstates of the arrowhead Hamiltonian (\ref{eq:arrowhead}), and set $g=1/\sqrt{\pi}$ to lighten our formulas. We study the inverse participation ratio (IPR) of the eigenstates, which measures their degree of localization, and is defined as [see Eq.~(\ref{eq:eigenstates})]
\begin{equation}
    \label{eq:IPR_def}
	{\rm IPR}_a(q)  \, = \,  \sum_{j=1}^{N+1} | \psi_{a,j} |^{2 q} \, = \, \frac{1+ \frac{1}{(\pi N)^{q}} \sum_{j=1}^N \frac{1}{(\varepsilon_a - \omega_j)^{2q}} }{\left( 1+ \frac{1}{\pi N} \sum_{j=1}^N \frac{1}{(\varepsilon_a - \omega_j)^2} \right)^q} .
\end{equation}

\subsection{IPR of the polaritons}

The IPR of the two polaritons behaves as follows when $N \rightarrow \infty$:
\begin{eqnarray}
    \label{eq:IPR_polariton}
 \nonumber   {\rm IPR}_{\pm} (q) & \underset{N \rightarrow \infty}{\simeq} &
           \frac{1+ \frac{1}{\pi^q N^{q-1}} \int \frac{\rho(\omega) d\omega}{(\varepsilon_{{\rm P} \pm}-\omega)^{2q}}}{ (1- \tilde{\rho}'(\varepsilon_{{\rm P}\pm}))^q } \\
        &=& \left\{ \begin{array}{lcl}
            N^{1-q} \, \frac{\frac{1}{\pi^q} \int \frac{\rho(\omega) d\omega}{(\varepsilon_{{\rm P} \pm}-\omega)^{2q}}}{ (1- \tilde{\rho}'(\varepsilon_{{\rm P}\pm}))^q }  & {\rm if} & q < 1 , \\
           \frac{1}{( 1- \tilde{\rho}'(\varepsilon_{{\rm P}\pm}))^q }   & {\rm if} & q > 1 . 
        \end{array} 
    \right. \qquad 
\end{eqnarray}
Here $\tilde{\rho}'(\omega)$ denotes the derivative of $\tilde{\rho}(\omega)$, the Hilbert transform of $\rho(\omega)$. For instance, for a box distribution of width $W$, i.e. $\rho(\omega)=1/W$, one has
\begin{align}
    \tilde{\rho}(\omega)&=\frac{1}{\pi W}\log \bigg\vert \frac{\omega+(W/2)}{\omega-(W/2)} \bigg\vert \nonumber \\
    \tilde{\rho}'(\omega)&=\frac{1}{\pi} \frac{1}{(W/2)^2- \omega^2}.
    \label{eq:defmomentsdistr}
\end{align}
Interestingly, we see that the polariton IPR exhibits a sharp behavior change as a function of $q$. For $q<1$, it is of order $O( N^{1-q}) $, while it is of order $O(1)$ for $q\geq 1$. This allows to  define the multifractal dimension of the state as~\cite{Evers_Anders_2008,hetterich2017noninteracting}
\begin{equation}
    \label{eq:fractal_dim}
    D_{\rm f}(q) \, = \, d \, \lim_{N \rightarrow \infty} \left( \frac{1}{1-q} \frac{\log {\rm IPR}_{a}(q)}{\log N} \right),
\end{equation}
where $d$ is the spatial dimensionality of the problem. For localized states, $D_{\rm f}(q)= 0$, while for delocalized states $D_{\rm f}(q) = d$. Here we see from (\ref{eq:IPR_polariton}) that, for the polaritons, $D_{\rm f}(q) = d$ for $q<1$ and $D_{\rm f}(q)=0$ for $q>1$. This reflects the fact that the polariton states simultaneously share some properties of both localized and delocalized states: they have one component of order $O(1)$, like a localized state, and an extensive number of components with weights of order $O(1/N)$, like a delocalized state.

\subsection{Scaling of the IPR for the dark states}

The IPR of the dark states behaves differently. For eigenenergies $\varepsilon_a$ and bare energies $\omega_a$ ($2 \leq a \leq N$) sorted in increasing order, the eigenenergy $\varepsilon_a$ lies between $\omega_{a-1}$ and $\omega_a$, at a typical distance of order $O(1/N)$. Then the scaling of the term $\sum_{j=1}^N \frac{1}{(\varepsilon_a - \omega_j)^{2q}}$ is determined as follows. The terms with $j$ such that $|\varepsilon_a - \omega_j |<\delta /N$ for some fixed $\delta >0$ give a contribution of order $O(N^{2q})$, while the contribution of all the other terms can be replaced by the integral $N \int_{|\omega - \varepsilon_a|>\delta/N} \frac{\rho(\omega) d \omega }{ (\varepsilon_a - \omega)^{2q} }$, which is of order $O(N)$. Thus, the numerator of Eq.~(\ref{eq:IPR_def}) is of order $O(N^{{\rm max}(2q,1)-q})$, while the denominator is of order $O(N^q)$. Hence,
\begin{equation*}
    {\rm IPR}_{a} (q) \, =\, \left\{ \begin{array}{lcl}
            O(N^{1-2q})  & {\rm if} & q < 1/2 , \\
           O(1)   & {\rm if} & q > 1/2 . 
        \end{array} 
    \right. 
\end{equation*}
In particular, the definition (\ref{eq:fractal_dim}) gives $D_{\rm f}(q) = \frac{1-2q}{1-q} D$ for $q<1/2$, and $D_{\rm f} (q) = 0$ for $q>1/2$. The multifractality of the spectrum is shown in Fig.~\ref{fig7}, where the components $| \psi_{a,j} |^{2 q}$ of the IPR of the dark states are displayed together with the fractal dimension $D_{\rm f}(q)$ computed numerically via Eq.~\eqref{eq:fractal_dim}. There, we compute $\textrm{IPR}(q) \equiv \sum_a \textrm{IPR}_a(q)/(N-1)$, with $\textrm{IPR}_a(q)$ defined by Eq.~\eqref{eq:IPR_def}. The overlaps $|\psi_{a,j}|^2 = |\left< j,0 \left| \psi_a \right> \right. |^2$ are computed from the eigenstates in Eq.~\eqref{eq:eigenstates}, and $\textrm{IPR}(q)$ is obtained after summation over all sites $j$ and eigenstates $\psi_{a}$, with $2 \leq a \leq N$ (dark states). A single disorder realization is used to compute the fractal dimension in Fig.~\ref{fig7}(\textbf{c}), since the large fluctuations of $|\psi_{a,j} |^{2 q}$ are smoothened by the summation over all eigenstates in the definition above. In addition, we realize a finite-size analysis of $\textrm{IPR}(q)$. We find that the latter scales with the system size as $b/\log(N) + c$, where $b$ and $c$ are free fitting parameters (not shown). The extrapolated thermodynamic limit is shown as a red dotted line in Fig.~\ref{fig7}(\textbf{c}).

In contrast to conventional disordered hopping models where multifractal behavior is usually associated with the critical point at a localization-delocalization transition~\cite{Evers_Anders_2008}, we find that multifractality in our model occurs for any strength of the light-matter coupling, thus signaling the existence of a critical \textit{phase}.

\begin{figure*}[ht]
	\includegraphics[width=1.0\textwidth]{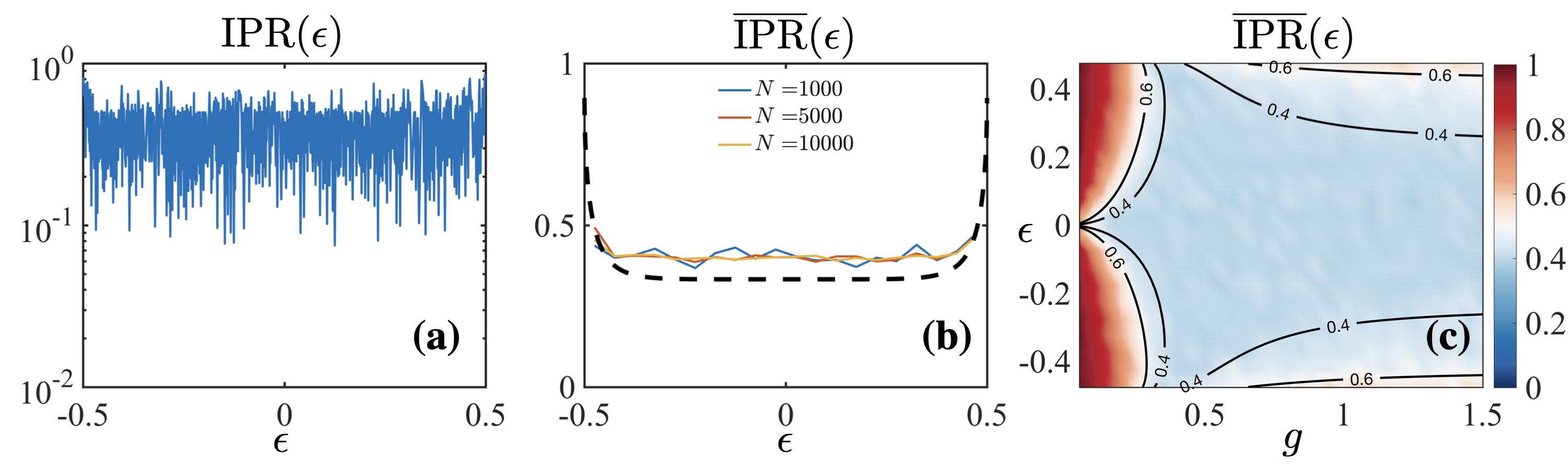}
	\caption{Inverse participation ratio ${\rm IPR}_{a}(q=2)\equiv {\rm IPR}(\epsilon)$ of the dark states ($\epsilon$ denotes the energies of the eigenstates $\epsilon_{a}$) computed numerically from Eq.~(\ref{eq:IPR_def}) for a given disorder realization, $g=0.5$, and $W=1$, (\textbf{a}) without energy binning for $N=1000$, (\textbf{b}) with binning for different $N$. (\textbf{c}) Binned IPR of the dark states versus energy $\varepsilon$ and coupling strength $g$. While the fluctuations of $\overline{\rm IPR}$ are reduced as $N$ is increased, it does not converge to the formula (\ref{eq:IPR}) obtained for an equally spaced distribution. However, formula (\ref{eq:IPR}) still captures its qualitative behavior as a function of $g$ and $\varepsilon$ [contour lines in (\textbf{c})].}
	\label{Fig6}
\end{figure*}

On a $d$-dimensional lattice of size $L\to \infty$, the IPR of the dark states behaves as
${\rm IPR}(q) \sim  L^{-\tau_q}$, with
\begin{equation*}
    \tau_q = \left\{
        \begin{array}{lcl}
            d (2q-1) & {\rm if} & q<1/2 \\
            0 & {\rm if} & q>1/2. 
        \end{array} \right.
\end{equation*}
Therefore the anomalous dimension $\Delta_q = \tau_q - d (q-1)$ for the dark states reads $\Delta_q = \frac{1}{2} - \left| q - \frac{1}{2}\right|$, which satisfy the symmetry relation $\Delta_q = \Delta_{1-q}$ like in models of Anderson localization/delocalization transitions~\cite{Evers_Anders_2008}. Interestingly, the dark state anomalous dimension $\Delta_q$ turns out to be exactly the same as in the Anderson model on the Bethe lattice~\cite{abou1973selfconsistent,Evers_Anders_2008}, which is commonly regarded as a mean-field model ($d\rightarrow \infty$) of the localization-delocalization transition. We emphasize that the emergence of such multifractal behavior is quite elementary in our arrowhead matrix model, as compared to more complex disordered hopping models of Anderson localization-delocalization transitions~\cite{Evers_Anders_2008}.   

\subsection{An exact formula (not valid for the random disorder distribution)}

    We have not been able to find a closed analytical formula for the IPR of an eigenstate $\psi_a$ as a function of its energy $\varepsilon_a$, even after averaging over an energy shell $[\varepsilon-\delta \varepsilon/2, \varepsilon+\delta \varepsilon/2]$. In contrast with the energy shift $\Delta (\varepsilon)$ studied in the previous section, or with the photon spectral function $A(\varepsilon)$ or the photon weight ${\rm PW}(\varepsilon)$ studied below, we find that the IPR (\ref{eq:IPR_def}) cannot be expressed in terms of contour integrals that allow to extract its large $N$ behavior as a functional of the thermodynamic density of states $N \rho(\omega)$ [see Eq.~(\ref{eq:rho})]. Unlike these quantities, the IPR remains sensitive to the details of the microscopic distribution of bare energies in the large $N$ limit.
    
    To illustrate this, we have studied the case of a deterministic distribution of bare energies that are locally equally spaced, with thermodynamic density of states $N \rho(\omega)$. Such a distribution is easily constructed as follows: for a given function $\rho(\omega)$, one asks that the bare energies $\omega_1 < \omega_2 < \dots < \omega_N$ satisfy
    \begin{eqnarray}
        \label{eq:equally_spaced}
        \omega_{a+1} - \omega_a = \frac{1}{N \rho[(\omega_{a+1}+\omega_a)/2]} .
    \end{eqnarray}    
    Importantly, this distribution leads to the correct density of states $N \rho(\omega)$ in the thermodynamic limit, however its correlations between energy levels are clearly very different from the ones obtained from $N$ i.i.d. variables as in the random arrowhead Hamiltonian. In particular, i.i.d. bare energies would lead to ``rare'' pairs of neighboring energies that could be very  close to each other, while the equally spaced distribution (\ref{eq:equally_spaced}) automatically prevent this, and thus exhibits some kind of ``level repulsion''.
    
For the deterministic equally spaced distribution defined by (\ref{eq:equally_spaced}), we have computed the IPR of the dark states ($a = 2, \dots, N$). Using the fact that $$\sum_{j=1}^N \frac{1}{(\varepsilon_a - \omega_j)^4} \underset{N \gg 1}{\simeq} \frac{\pi^{4} N^4}{\rho(\varepsilon_a)^4} \left( \frac{1-\frac{2}{3} \cos^2[\pi \rho(\varepsilon_a) \Delta(\varepsilon_a)]}{\cos^4[\pi \rho(\varepsilon_a) \Delta(\varepsilon_a) ]}\right),$$ we find that the IPR (for $q=2$) has the following large $N$ asymptotics,
\begin{equation}
	\label{eq:IPR}
	{\rm IPR}_a  (q=2) \, \underset{N \rightarrow \infty}{=} \,  \frac{\frac{\rho(\varepsilon_a)^2}{3}+ \left( \tilde{\rho}(\varepsilon_a) - \frac{\varepsilon_a}{\pi g^2} \right)^2}{\rho(\varepsilon_a)^2+\left( \tilde{\rho}(\varepsilon_a)- \frac{\varepsilon_a}{\pi g^2} \right)^2} .
\end{equation}
This analytical formula is compared to the numerically evaluated ${\rm IPR}(q=2)$ for the eigenstates of the random arrowhead Hamiltonian in Fig.~\ref{Fig6}. In Fig.~\ref{Fig6} (\textbf{a}), we compute the $\rm IPR_a  (q=2) \equiv {\rm IPR}(\epsilon)$ as a function of $\epsilon$ for $N=1000$ and a single disorder realization, while in Fig.~\ref{Fig6} (\textbf{b}), ${\rm IPR}(\epsilon)$ is binned into groups of equal energy width and disorder-averaged in each bin for different $N$. While the fluctuations of the IPR for i.i.d. bare energies are reduced upon increasing $N$, we see that the IPR does not converge to formula (\ref{eq:IPR}) (black dotted line) obtained for the deterministic equally spaced distribution. 

This is because, in contrast with the energy shift $\overline{\Delta}(\varepsilon)$ or the photon weight $\overline{{\rm PW}(\varepsilon)}$, the asymptotic value of the IPR at large $N$ is not sensitive only to the density of states $N \rho(\omega)$, it also  remains highly sensitive to the microscopic fluctuations of the bare energy levels.

We note, however, that $\overline{\rm IPR}(q=2)$ is well captured qualitatively by the formula (\ref{eq:IPR}) [see contour lines in Fig.~\ref{Fig6}(\textbf{c})]. For a box disorder distribution of width $W$ and for weak couplings $g/W \ll 1$, the dark states are essentially fully localized with $\overline{\rm IPR} (q=2) \approx 1$, except for the states lying in the band center that can always hybridize together via the cavity mode~\cite{Botzung_2020}. The dark states thus follow the same Poissonian statistics as the i.i.d. bare orbitals. For strong couplings, all dark states can be thought of as semi-localized states, i.e. localized over multiple, non-contiguous sites, since they exhibit an intermediate value $\overline{\rm IPR} (q=2) \approx 0.4$. The dark states lie roughly at equal distance from the two closest bare energy levels, and thus follow a statistics that is close to the semi-Poissonian statistics introduced in Ref.~\cite{Bogomolny_1999}. For intermediate coupling strengths, $g\sim W$, the level spacing statistics depends continuously on $g$ and on the energy $\varepsilon$, and can therefore be \textit{continuously tuned} between different statistics.

Now that we have characterized the localization properties of the eigenstates of the Hamiltonian (\ref{eq:arrowhead}) through their inverse participation ratios, we turn to the large $N$ behavior or the photon Green's function and the photon spectral function.

\section{The photon Green's function}
\label{photon_sec}

As hinted in section~\ref{sec:shifts}, a central role in the analysis of large $N$ properties of arrowhead Hamiltonians is played by the photon Green's function in frequency space,
\begin{equation*}
D(z)  = \left[\frac{1}{z- H}\right]_{N+1,N+1}.
\end{equation*}
Here $z \in \mathbb{C}$, and the limit of $D(z)$ when $z$ approaches a point $\omega$ on the real axis from above/below is the retarded/advanced Green's function  $D(\omega+i0^{\pm})$ at frequency $\omega$. To evaluate $D(z)$, one can take the Fourier transform of Eq.~(\ref{eq:prop_finiteN}c), which leads to
\begin{eqnarray}
    \label{eq:Dz}
\nonumber D(z) &=&  \sum_{a=1}^{N+1} \frac{1}{z-\varepsilon_a} \frac{1}{1+ \frac{1}{\pi N} \sum_{j=1}^N \frac{1}{(\varepsilon_a-\omega_j)^2}  } \\
\nonumber    &=& \frac{1}{z - \frac{1}{\pi N} \sum_{i=1}^N \frac{1}{z-\omega_j}}  \\
     &=&  \frac{\prod_{j=1}^N (z-\omega_j) }{\prod_{a=1}^{N+1} (z-\varepsilon_a)} .
\end{eqnarray}
The equalities in the second and third lines are easily checked by inspecting the zeros and poles of these expressions. The result in the third line makes it clear that the function $D(z)$ is very special.  
It encodes all the spectral properties of our arrowhead Hamiltonian (\ref{eq:arrowhead}): the zeros and poles of $D(z)$ are the bare energies and the eigenvalues of $H$, respectively. 
 
\subsection{Large $N$ limit}
 
The large-$N$ limit of $D(z)$ turns out to be an important tool in the analysis of the properties of large arrowhead Hamiltonians. First, we introduce the photon self-energy $\Pi(z) = \frac{1}{\pi N} \sum_{j=1}^N \frac{1}{z-\omega_j} $, and its large-$N$ limit
 \begin{equation*}
	\overline{\Pi}(z) = \lim_{N \rightarrow \infty} \frac{1}{\pi N} \sum_{j=1}^N \frac{1}{z-\omega_j} = \frac{1}{\pi} \int_{\omega_{\rm min}}^{\omega_{\rm max}} \frac{\rho(\omega) d\omega}{z-\omega} ,
\end{equation*}
where $ [\omega_{\rm min}, \omega_{\rm max}] $ is the support of $\rho$. $\overline{\Pi}(z)$ is analytic in $\mathbb{C} \setminus [\omega_{\rm min}, \omega_{\rm max}]$, and it has a discontinuity along that interval: $\overline{\Pi}(\omega + i 0^\pm) = \tilde{\rho}(\omega) \mp i \rho(\omega)$. Then the large-$N$ limit of the photon Green's function is
\begin{equation*}
	\overline{D} (z) \, = \, \lim_{N \rightarrow \infty} D(z) \, = \,  \frac{1}{z- \overline{\Pi}(z)} .
\end{equation*}
$\overline{D} (z)$ has the following properties:
\begin{enumerate}[(i)]
	\item it is analytic in $\mathbb{C} \setminus ( [\omega_{\rm min}, \omega_{\rm max}] \cup \{ \varepsilon_{{\rm P}+},\varepsilon_{{\rm P}-} \} )$
\item it has two poles at $\varepsilon_{{\rm P}+}$ and $\varepsilon_{{\rm P}-}$: $\overline{D}(z) \underset{z \rightarrow \varepsilon_{{\rm P}\pm}}{\simeq} \frac{1}{ (1 - \tilde{\rho}'(\varepsilon_{{\rm P}\pm})) (z - \varepsilon_{{\rm P}\pm})}$. The residues at these poles are the photon weights of the two polaritons (see next section). 
	\item it has a branch cut along the support of $\rho$, i.e. $[\omega_{\rm min}, \omega_{\rm max}] $. Indeed, for $\omega$ along the real axis, $\overline{D}(\omega + i 0^\pm)  = \frac{1}{ \omega - \tilde{\rho}(\omega)  \pm i  \rho(\omega)}$. Thus, $\overline{D} (z)$ has a discontinuity
	\begin{eqnarray}
    \label{eq:Ddisc}
	\nonumber &&	\overline{D}(\omega + i 0^-) - \overline{D}(\omega + i 0^+) \\
\nonumber	 	&& \quad  = 2\pi i \times \left( \frac{\delta(\omega-\varepsilon_{{\rm P}+})}{1-\tilde{\rho}'(\varepsilon_{{\rm P}+})}  + \frac{\delta(\omega-\varepsilon_{{\rm P}-})}{1-\tilde{\rho}'(\varepsilon_{{\rm P}-})} \right. \\ 
	 	&& \qquad \qquad \qquad  \left. + \, \frac{ \rho(\omega) / \pi}{ \rho(\omega)^2 + (\tilde{\rho}(\omega) - \omega)^2 } \right) .
	\end{eqnarray}
	\item it behaves as $\overline{D}(z) \simeq \frac{1}{z}$ at infinity.
\end{enumerate}
Below we show how physical quantities like the photon weight or the photon spectral function are fixed by the analyticity properties of $\overline{D}(z)$, to leading order in $N$ when $N$ is large. Before that, we discuss the fluctuations of $D(z)$ around its asymptotic value $\overline{D}(z)$.

\subsection{Fluctuations of the photon Green's function}

When the bare energies $\omega_j$ are i.i.d random variables, the function $D(z)$ also has a non-deterministic subleading part. Our goal is to characterize the fluctuations of $D(z)$ around its mean value. Let us start by introducing the following random function,
\begin{equation}
	\label{eq:phi1}
	\phi(z) =  \sqrt{N} \left( \frac{1}{\pi N} \sum_{j=1}^N \frac{1}{z-\omega_j} - \overline{\Pi}(z) \right) ,
\end{equation}
which measures the fluctuations of the photon self-energy $\Pi (z)$ around its large-$N$ limit. Notice that $\phi(z)$ has zero mean value for the disorder average Eq.~(\ref{eq:disav}),
$$\mathbb{E}[ \phi(z) ] = 0.$$ The factor $\sqrt{N}$ in Eq.~(\ref{eq:phi1}) is introduced such that the covariance of $\phi(z)$ is of order one: 
\begin{eqnarray}
    \label{eq:Gz1z2}
\nonumber	&& {\rm cov}[\phi(z_1) \phi(z_2)] = \mathbb{E}[ \phi(z_1) \phi(z_2) ] \\
\nonumber &&= \frac{1}{N} \sum_{i,j} \mathbb{E} \left[ ( \frac{1}{\pi} \frac{1}{z_1-\omega_i} - \overline{\Pi}(z_1) ) ( \frac{1}{\pi} \frac{1}{z_2-\omega_j} - \overline{\Pi}(z_2) ) \right] \\
\nonumber &&= \frac{1}{\pi^2 N} \sum_{j=1}^N \mathbb{E} \left[ \frac{1}{z_1-\omega_j} \frac{1}{z_2-\omega_j}  \right] - \overline{\Pi}(z_1) \overline{\Pi}(z_2)   \\
\nonumber &&= \frac{1}{\pi^2} \int \frac{\rho(\omega) d\omega}{(z_1-\omega) (z_2-\omega)} - \overline{\Pi}(z_1) \overline{\Pi}(z_2)   \\
	&&= - \frac{\overline{\Pi}(z_1) - \overline{\Pi}(z_2)}{\pi (z_1-z_2)}  - \overline{\Pi}(z_1) \overline{\Pi}(z_2).
\end{eqnarray}

More generally, the connected part of the $p^{\rm th}$-order correlation of  $\phi(z)$ scales as $\sim N^{-p/2}$, so at large $N$ it vanishes for all $p \geq 3$,
\begin{equation*}
	\mathbb{E}[ \phi(z_1) \phi(z_2) \dots \phi(z_p) ]_{\rm conn.} \underset{N \rightarrow \infty}{\longrightarrow} 0, \quad  \qquad {\rm if} \quad p \geq 3 .
\end{equation*}
Thus, the random function $\phi(z)$ is Gaussian, with mean value zero, and covariance Eq.~(\ref{eq:Gz1z2}). The fluctuations of the photon Green's function are then obtained as follows,
\begin{eqnarray}
    \label{eq:covD}
\nonumber    D(z ) &=& \frac{1}{z - \overline{\Pi}(z) - \frac{1}{\sqrt{N}} \phi(z) } \\
 \nonumber   & = & \frac{1}{z - \overline{\Pi}(z)} + \frac{1}{(z-\overline{\Pi}(z))^2} \frac{1}{\sqrt{N}} \phi(z) + \dots \\
    &=& \overline{D}(z) + \frac{1}{\sqrt{N}} \overline{D}^2(z) \,\phi(z) + O(1/N) .
\end{eqnarray}
Thus, the fluctuations of $D(z)$ around its large $N$ value $\overline{D}(z)$ are also Gaussian and of order $1/\sqrt{N}$, and they are directly determined by the two-point function Eq.~(\ref{eq:Gz1z2}).

\subsection{Photon spectral function: mean value and fluctuation statistics}
The photon spectral function is proportional to the cavity transmission and fluorescence emission spectra, which can be directly accessed by, e.g.~, Fourier transform spectroscopy~\cite{PhysRevA.69.043805,Scully_1997}. It is obtained from the imaginary part of the retarded Green's function,
\begin{eqnarray}
    \label{eq:defA}
    A(\omega) & = & - \frac{1}{\pi} \, {\rm Im} \, D(\omega + i 0^+) \\
\nonumber    & = & - \frac{D(\omega + i 0^+)  - D(\omega + i 0^-)}{2 \pi i}.
\end{eqnarray}
In the large $N$ limit this is equal to [using the poles of the function $\overline{D}(z)$ and Eq.~(\ref{eq:Ddisc})]
\begin{eqnarray}
    \label{eq:ANinf}
\nonumber \overline{A}(\omega) & \underset{N \rightarrow \infty}{\simeq} & \frac{\delta(\omega- \varepsilon_{\rm P} +)}{1- \tilde{\rho}' (\varepsilon_{\rm P} +) }  + \frac{\delta(\omega- \varepsilon_{\rm P} -)}{1- \tilde{\rho}' (\varepsilon_{\rm P} -) }   \\
    && \quad  + \,   \frac{   \rho(\omega)/\pi}{ \rho(\omega)^2 + (\tilde{\rho}(\omega) - \omega)^2 } .
\end{eqnarray}
Notice that the spectral function should satisfy the sum rule
$\int A(\omega) d\omega \, = \, 1 $, which is a consequence of Eq.~(\ref{eq:APW}) and of the normalization of the eigenstates. The fact that the sum rule is satisfied by (\ref{eq:ANinf}) follows from contour integration. Integrating the function $\overline{D}(z)$ along a counterclockwise contour $C$ that encloses both the interval $[\omega_{\rm min}, \omega_{\rm max}]$ and the two points $\varepsilon_{{\rm P}\pm}$, one gets $\frac{1}{2\pi i} \oint_C \overline{D}(z) dz = \int \overline{A}(\omega) d\omega $. The contour $C$ can then be deformed to infinity, and one concludes that $\frac{1}{2\pi i} \oint_C \overline{D}(z) dz = 1$ using the above property 4 of the function $\overline{D}(z)$.

With the formalism of the previous section, we can go beyond the leading order (\ref{eq:ANinf}) and calculate the $O(1/\sqrt{N})$ fluctuations of $A(\omega)$ around its mean value $\mathbb{E}[A(\omega)] = \overline{A}(\omega)$. Combining Eq.~(\ref{eq:defA}) and (\ref{eq:covD}), one sees that the connected correlations of $A(\omega)$ can be expressed in terms of those of $\phi(\omega)$. In particular, the covariance of $A(\omega)$ can be expressed in terms of the known covariance of $\phi(\omega)$ through
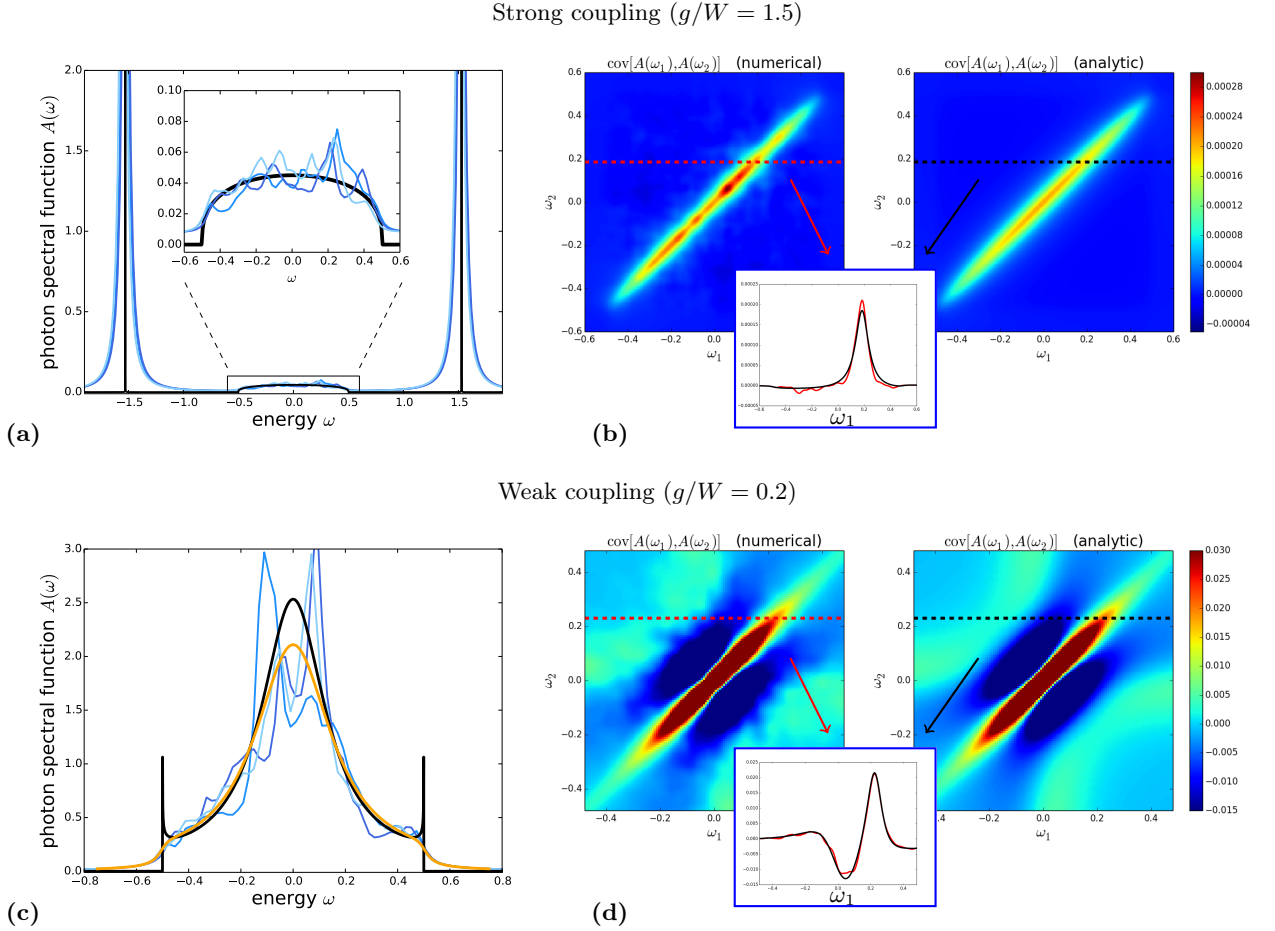
\begin{figure*}[ht]
    \centering
    \begin{tikzpicture}
        \draw (1.8,2.9) node{Strong coupling ($g/W = 1.5$)};
	    \draw (-3,0) node {\includegraphics[width=0.4\textwidth]{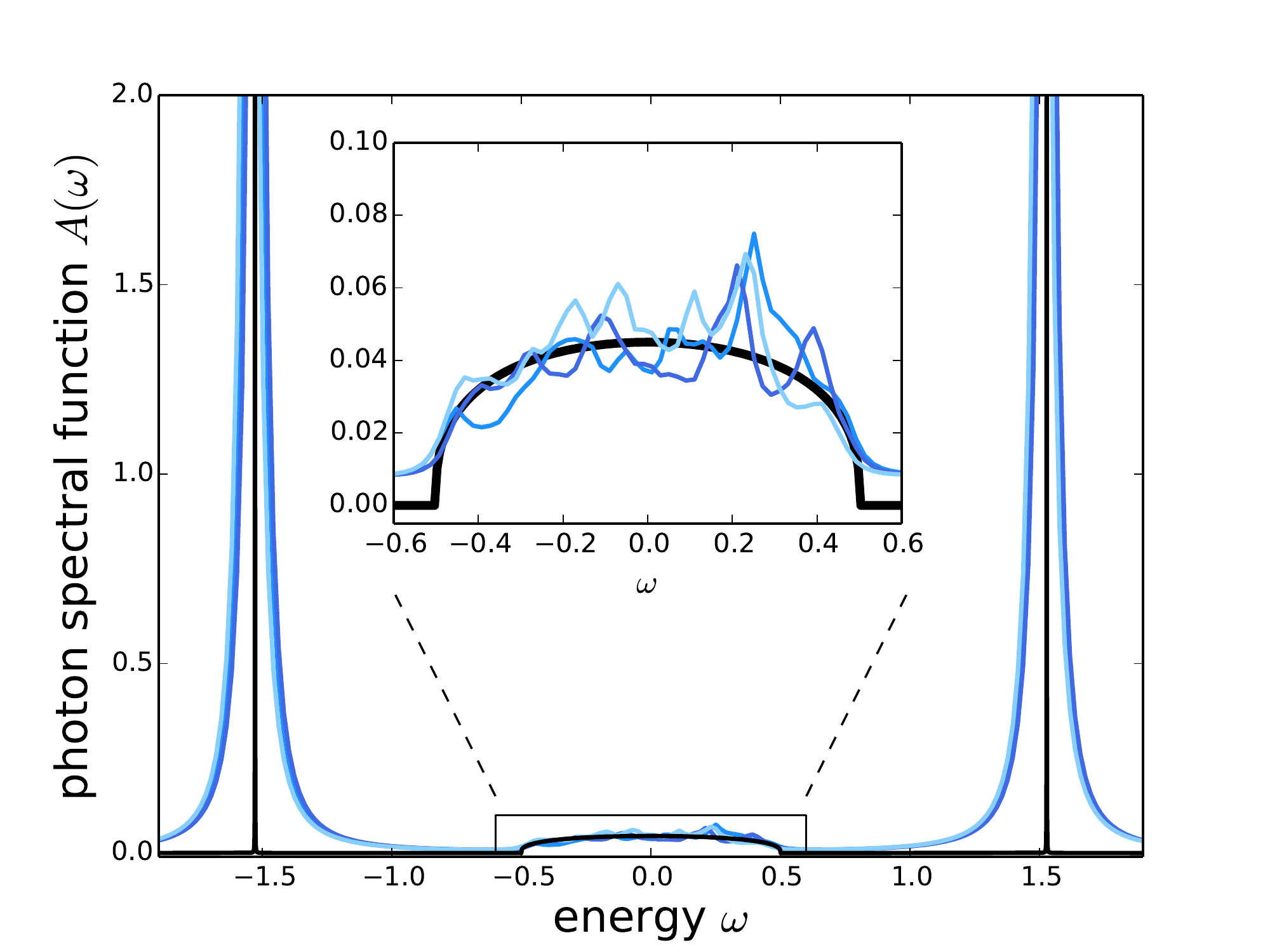}};
	    \draw (3,0.4) node {\includegraphics[width=0.32\textwidth]{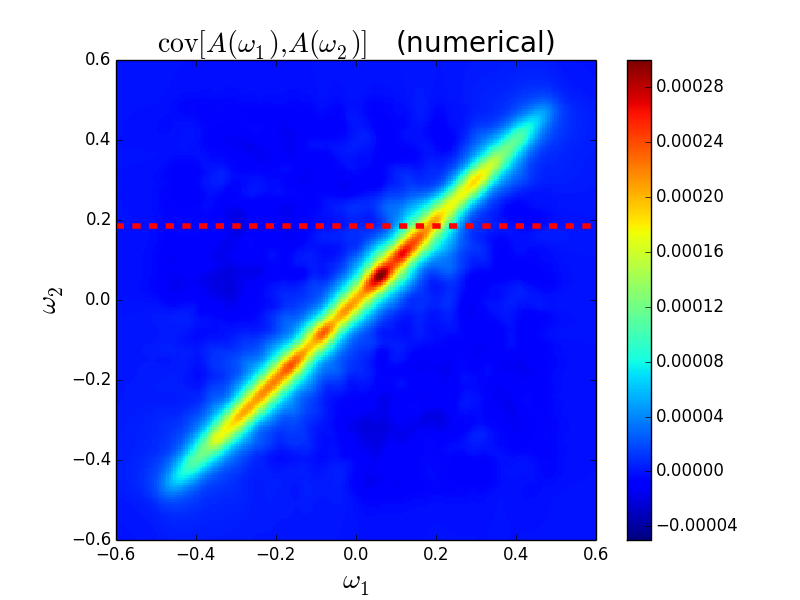}};
	    \draw (7.38,0.4) node {\includegraphics[width=0.32\textwidth]{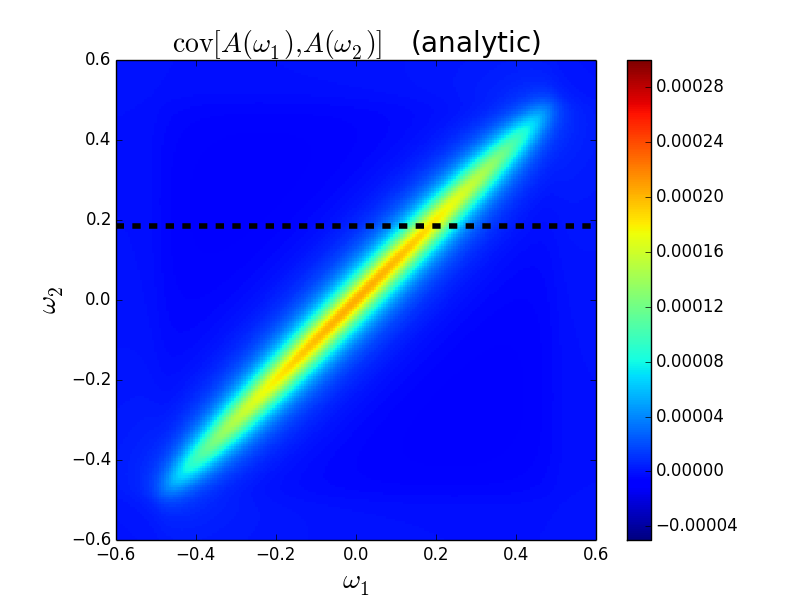}};
	    \draw (4.3,-1.5) node {\includegraphics[width=0.15\textwidth]{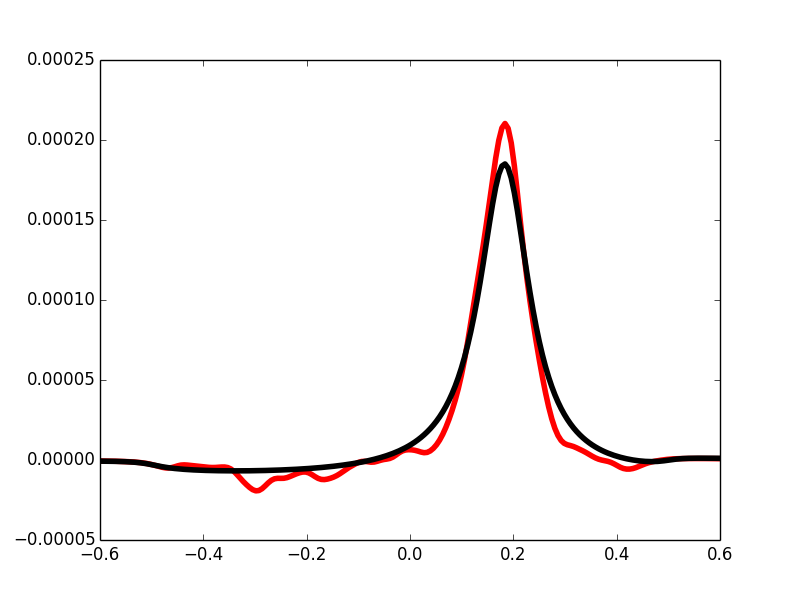}};
	    \draw[thick,blue] (2.96,-2.6) rectangle (5.63,-0.5);
	    \draw[thick,red,->] (3.7,0.7) -- ++(0.5,-1);
	    \draw[thick,black,->] (6.2,0.7) -- ++(-0.7,-1);
	    \draw (-6.5,-2.7) node{{\bf (a)}};
	    \draw (1.3,-2.7) node{{\bf (b)}};
	    \draw (4.4,-2.5) node{\small $\omega_1$};
	\end{tikzpicture} \\ \vspace{0.2cm}
	\begin{tikzpicture}
	   \draw (1.8,2.9) node{Weak coupling ($g/W = 0.2$)};
	    \draw (-3,0) node {\includegraphics[width=0.4\textwidth]{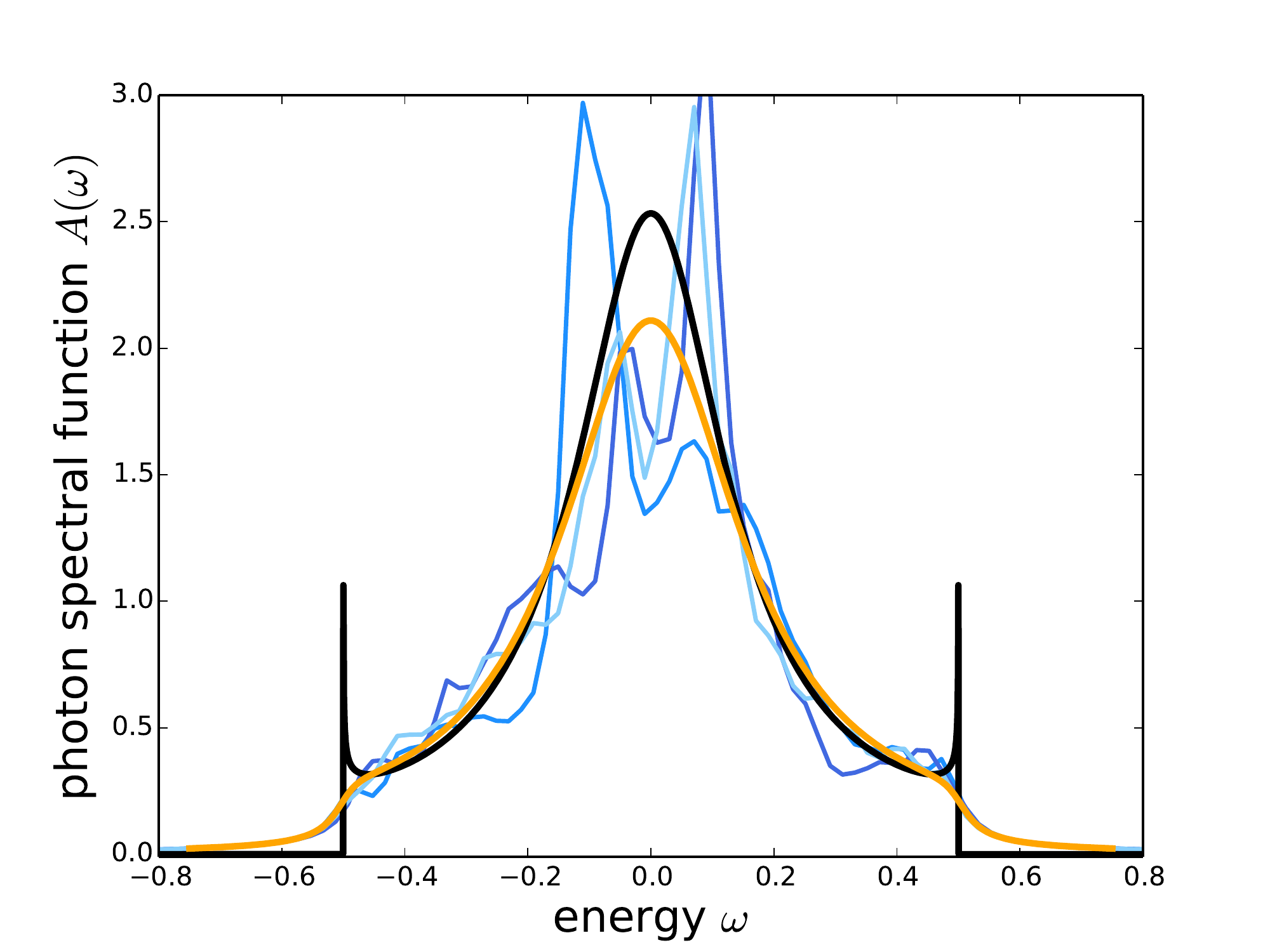}};
	    \draw (3,0.4) node {\includegraphics[width=0.32\textwidth]{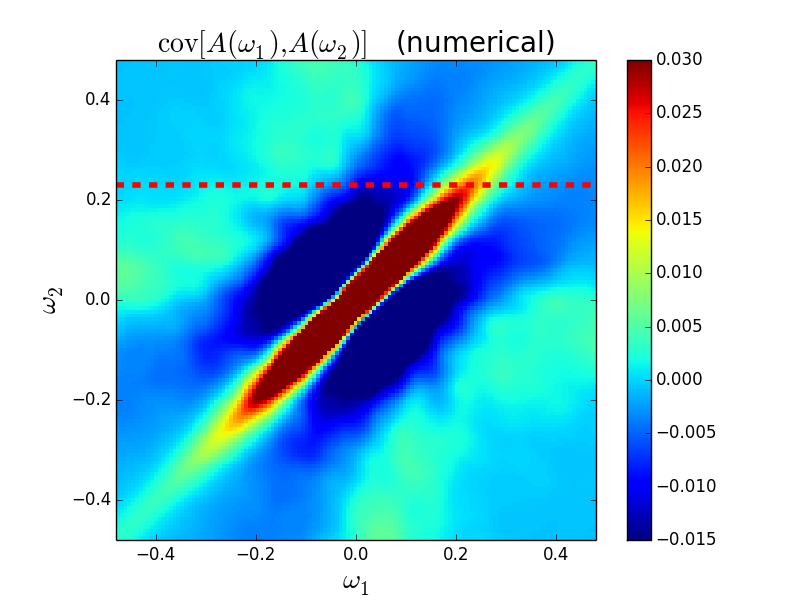}};
	    \draw (7.38,0.4) node {\includegraphics[width=0.32\textwidth]{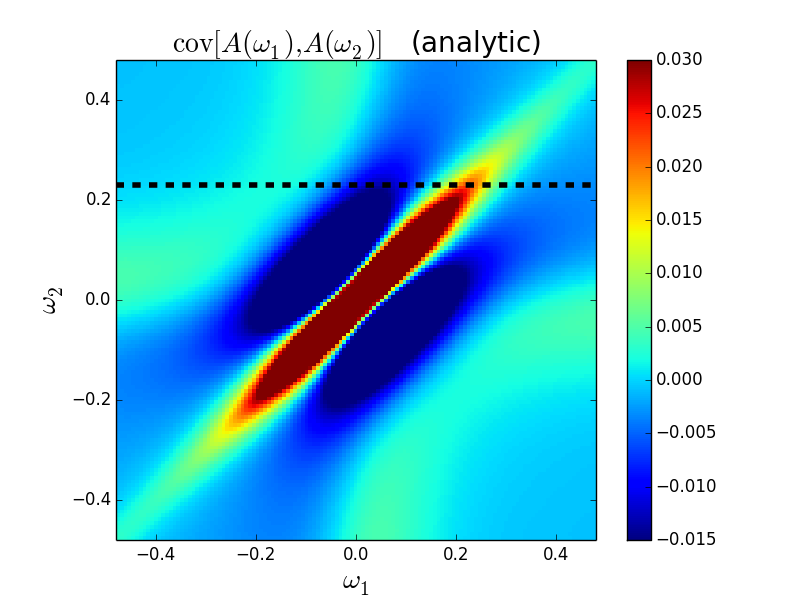}};
	    \draw (4.3,-1.5) node {\includegraphics[width=0.15\textwidth]{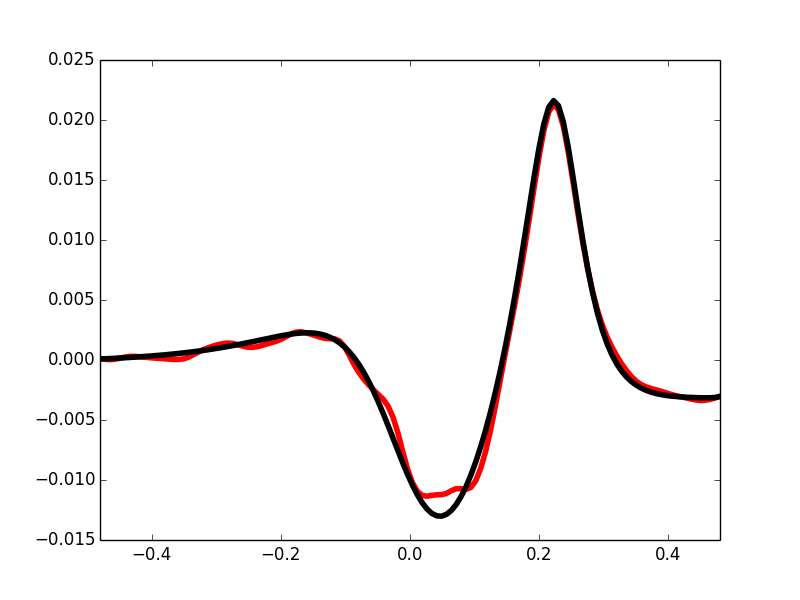}};
	    \draw[thick,blue] (2.96,-2.6) rectangle (5.63,-0.5);
	    \draw[thick,red,->] (3.7,0.7) -- ++(0.5,-1);
	    \draw[thick,black,->] (6.2,0.7) -- ++(-0.7,-1);
	    \draw (-6.5,-2.7) node{{\bf (c)}};
	    \draw (1.3,-2.7) node{{\bf (d)}};
	    \draw (4.4,-2.5) node{\small $\omega_1$};
	\end{tikzpicture}
	\caption{(a) Photon spectral function $A(\omega)$ for $g=1.5$,  $W=1$ and $N=100$, for three different disorder configurations (blue curves), compared to the analytical formula (\ref{eq:ANinf}) valid in the thermodynamic limit (black curve). $A(\omega)$ is computed using Eq.~(\ref{eq:APW2}), where the Dirac delta $\delta (\omega - \varepsilon_a)$ is replaced by the Lorentzian $\sigma/\pi /( (\omega - \varepsilon_a)^2 + \sigma^2)$ of width $\sigma = 0.03$. We clearly see the two large peaks at the lower and upper polariton energies $\varepsilon_{\rm P}^\pm$, see Eq.~(\ref{eq:ANinf}). $A(\omega)$ is displayed for three different disorder realizations (blue curves). The inset is a zoom on the window $\omega \in [\omega_{\rm min}, \omega_{\rm max}] = [-W/2,W/2]$: one sees that the spectral function fluctuates around its mean value $\overline{A}(\omega)$ given by Eq.~(\ref{eq:ANinf}), drawn in black. In the text, we show that the fluctuations of $A(\omega)$ around its mean value are of order $O(1/\sqrt{N})$, and that they are Gaussian at that order. 
	(b) Numerical evaluation of the covariance ${\rm cov}[A(\omega_1) ,A(\omega_2)]$, and comparison with the analytic formula~(\ref{eq:covA}). The numerical estimate of the covariance is obtained by averaging over 1000 independent disorder realizations. Both the numerical and the analytic curves include a convolution with the Lorentzian of width $\sigma = 0.03$.
	(c) Same as in (a), with $g=0.2$ (and $W=1$, $N=100$, $\sigma = 0.03$). The three blue curves are typical results for  different disorder realizations. The black line is the analytic formula (\ref{eq:ANinf}), and the orange line is its convolution with the Lorentzian of width $\sigma = 0.03$. (d) Same as in (b), for $g=0.2$.}
	\label{fig:spectralfunction}
\end{figure*}
\begin{widetext}
\begin{eqnarray}
 \nonumber  && {\rm cov}[A(\omega_1), A(\omega_2)] =  \mathbb{E}[A(\omega_1) A(\omega_2)] -  \mathbb{E}[A(\omega_1)] \mathbb{E}[ A(\omega_2)] \\
\nonumber &&= \frac{-1}{(2\pi)^2} {\rm cov}[ D(\omega_1+i 0^+)-D(\omega_1+i 0^-) , D(\omega_2+i 0^+) - D(\omega_2+i 0^-)] \\
\nonumber && =  \frac{-1}{(2\pi)^2} \frac{1}{N} \left( \overline{D}^2(\omega_1+i 0^+) \overline{D}^2(\omega_2+i 0^+) {\rm cov}[ \phi(\omega_1+i 0^+), \phi(\omega_2+i 0^+)] \right. \\
\nonumber && \qquad -  \overline{D}^2(\omega_1-i 0^+) \overline{D}^2(\omega_2+i 0^+) {\rm cov}[ \phi(\omega_1-i 0^+), \phi(\omega_2+i 0^+)]  \\
\nonumber && \qquad -  \overline{D}^2(\omega_1+i 0^+) \overline{D}^2(\omega_2-i 0^+) {\rm cov}[ \phi(\omega_1+i 0^+), \phi(\omega_2-i 0^+)] \\
\nonumber && \left.  \qquad +  \overline{D}^2(\omega_1-i 0^+) \overline{D}^2(\omega_2-i 0^+) {\rm cov}[ \phi(\omega_1-i 0^+), \phi(\omega_2-i 0^+)] \right) + O(N^{-3/2}).
\end{eqnarray}
Then, using the result (\ref{eq:Gz1z2}) and the Sokhotski-Plemelj formula, and combining all the terms, one arrives at the following result for the covariance of the photon spectral function. For $\omega_1, \omega_2 \in [\omega_{\rm min}, \omega_{\rm max}]$,
\begin{eqnarray}
    \label{eq:covA}
 \frac{{\rm cov}[A(\omega_1), A(\omega_2)]}{\mathbb{E}[A(\omega_1)] \mathbb{E}[A(\omega_2)]}  &=& \frac{1}{N \rho(\omega_1)} \,  \delta(\omega_1-\omega_2)   \\
 \nonumber && + \, \frac{1}{N} \frac{  \frac{2}{\pi}\frac{(\omega_2 - \tilde{\rho}_2) (\rho_1^2 + \tilde{\rho}_1^2 - \omega_1^2) - (\omega_1 - \tilde{\rho}_1) (\rho_2^2 + \tilde{\rho}_2^2 - \omega_2^2)}{ \omega_1 - \omega_2} - (\rho_1^2 + \tilde{\rho}^2_1 - \omega_1^2)(\rho_2^2 + \tilde{\rho}^2_2 - \omega_2^2) }{(\rho_1^2 +(\tilde{\rho}_1 - \omega_1)^2 ) (\rho_2^2 +(\tilde{\rho}_2 - \omega_2)^2 )} + O(N^{-3/2}) ,
\end{eqnarray}
with $\rho_{1,2} = \rho(\omega_{1,2})$ and $\tilde{\rho}_{1,2} = \tilde{\rho}(\omega_{1,2})$.
\end{widetext}
Eq.~(\ref{eq:covA}) entirely characterizes the fluctuations of the photon spectral function around its mean value (\ref{eq:ANinf}), at order $O(1/N)$. The non-Gaussianity of the fluctuations appears only at higher order. In Fig.~\ref{fig:spectralfunction}, we display typical results for $A(\omega)$ for different disorder realizations: the fluctuations around the mean value $\overline{A}(\omega)$ is clearly visible, see the inset of Fig.~\ref{fig:spectralfunction}.(a). We also display the numerical estimate of the covariance ${\rm cov}[A(\omega_1),A(\omega_2)]$, obtained by averaging over 1000 disorder realizations for a system size $N=50$, see Fig.~\ref{fig:spectralfunction}.(b); the result matches Eq.~(\ref{eq:covA}) as expected. We note that the covariance carries distinct features for weak and strong couplings. In the latter case, deviations with respect to the mean value of the spectral function are correlated (same sign) only along the diagonal $\epsilon_{1}=\epsilon_{2}$ [Fig.~\ref{fig:spectralfunction}(\textbf{b})]. For weak couplings, however, there are other regions where both correlations and anticorrelations (opposite sign) are observed.

Now that we have characterized the large $N$ behavior of the photon spectral function $A(\omega)$, including its fluctuations, we turn to the closely related photon weight, i.e. the weight of the photonic component of the eigenstates.

\subsection{Photon weights}
\label{sec:PW}

The photon weight of an eigenstate $\psi_a$ with energy $\varepsilon_a$ is defined as the weight of its $(N+1)^{\rm th}$ component [see Eq.~(\ref{eq:eigenstates})], i.e. ${\rm PW}_a = |\psi_{a, N+1}|^2\equiv \vert\bra{G,1} \psi_{a} \rangle \vert^{2}$. Using the definitions of Sec.~\ref{photon_sec}, one can easily see that the photon weight precisely corresponds to the ``quasiparticle'' weight of the eigenstates
\begin{equation}
    \label{eq:APW2}
    A(\varepsilon) \, = \, \sum_{a = 1}^{N+1}   {\rm PW}_a \, \delta(\varepsilon - \varepsilon_a) .
\end{equation}
The photon weight of the two polaritons ${\rm P}+$ and ${\rm P}-$ is directly obtained from Eq.~(\ref{eq:ANinf}) and reads
\begin{equation*}
    {\rm PW}_{\pm} \underset{N\rightarrow \infty}{\simeq} \frac{1}{1 - \tilde{\rho}'(\varepsilon_{{\rm P}\pm} )}.
\end{equation*}
It is of order $O(1)$ (i.e.~independent of $N$) in the thermodynamic limit, and goes asymptotically to $1/2$ for $g \to \infty$. For instance, for a flat disorder distribution in the interval $[-W/2,W/2]$, one finds ${\rm PW}_{\pm}\approx \frac{1}{2}-\frac{W^{2}}{24 g^{2}}$ for $g/W \gg 1$. In this regime, the polaritons are collective light-matter states half composed of the $N$ excited emitters (each with a weight $O(1/N)$) and the photon. Instead, for weak coupling, polaritons lose their collective nature. For a flat disorder distribution, one can show using Eq.~(\ref{eq:gpm_mean}) that, for $g/W \ll 1$,
\begin{equation*}
{\rm PW}_{\pm} \underset{N\rightarrow \infty}{\simeq} \left(\frac{W}{g}\right)^{2} e^{-\frac{W^{2}}{2g^{2}}} \underset{g/W\rightarrow 0}{\to} 0.
\end{equation*}
The photon weight of the dark states ($2 \leq a \leq N$) is more complicated. Like the mean energy shift, it fluctuates wildly as a function of the index $a$ and of the bare energy levels. However, its average value over eigenstates within a small energy shell $[\varepsilon- \delta \varepsilon/2,\varepsilon+\delta \varepsilon/2]$,
\begin{equation}
	\overline{{\rm PW} } (\varepsilon ) \, = \,  \frac{1}{N \rho(\epsilon)\,\delta \varepsilon} \sum_{|\varepsilon_a - \varepsilon| \leq \delta \varepsilon /2 } {\rm PW}_a ,
	\label{photon_weight_binned}
\end{equation}
is independent of the microscopic details of the distribution of bare energies, and takes a simple form in the limit $N \rightarrow \infty$. Using Eqs.~(\ref{eq:ANinf}) and (\ref{eq:APW2}), we find that
\begin{equation*}
    \overline{A} (\varepsilon) \, \underset{N \rightarrow \infty}{\simeq} \, N \rho(\varepsilon)\, \overline{ {\rm PW}}  (\varepsilon),
\end{equation*}
with 
\begin{equation}
	\label{eq:PWdark}
	\overline{{\rm PW}} (\varepsilon)  \, = \,   \frac{1}{N}  \frac{1/\pi}{\rho(\varepsilon)^2 + \left( \tilde{\rho}(\varepsilon) - \varepsilon \right)^2}.
\end{equation}

Thus, one finds that the photon weight of the dark states, which is inherited from the indirect coupling between these dark states and the cavity mode in the presence of disorder, is of order $O(1/N)$. It is interesting to note that for a finite but arbitrarily small $g/W$, the photon weight of the polaritons saturates to a finite value as $N$ grows, which results in the photon weight being mostly concentrated on the two polaritons even for weak couplings, provided $N$ is large enough.

\begin{figure*}[ht]
	\includegraphics[width=1.0\textwidth]{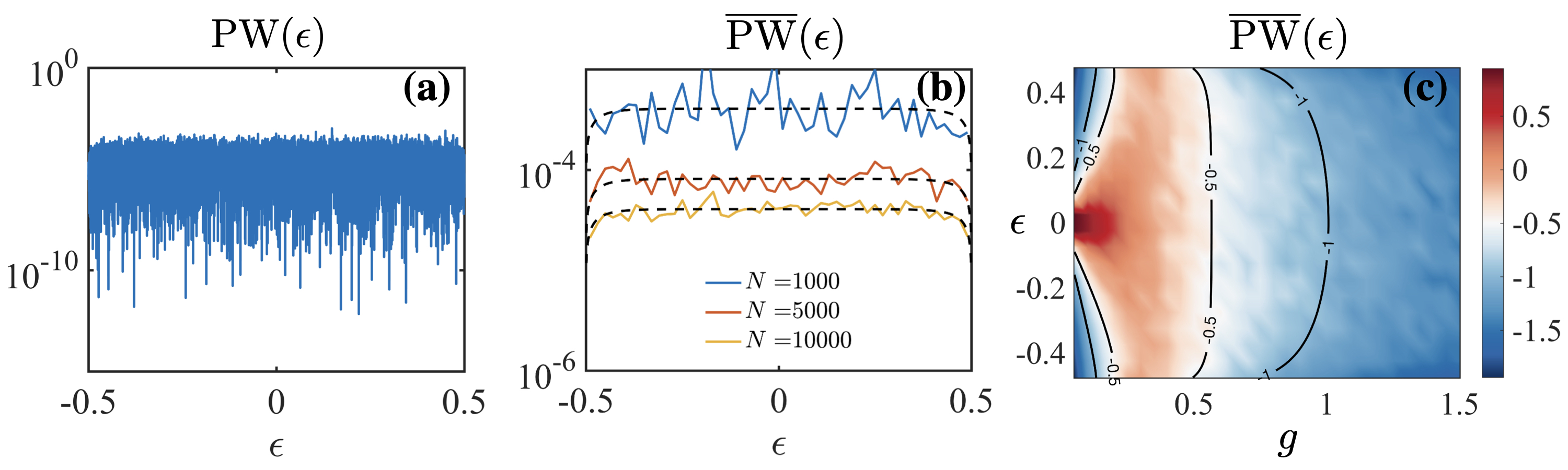}
	\caption{Photon weight of the dark states computed numerically from the eigenvectors Eq.~(\ref{eq:eigenstates}) for a given disorder realization, $g=0.5$ and $W=1$. (\textbf{a}) without energy binning for $N=1000$, (\textbf{b}) with binning according to Eq.~(\ref{photon_weight_binned}) for different $N$. (\textbf{c}) Binned photon weight of the dark states versus energy $\varepsilon$ and coupling strength $g$. The color bar is in log scale. Similarly as for the energy shift, the photon weight wildly fluctuates but its mean value after average over a small energy shell converges to the formula (\ref{eq:PWdark}) (dotted lines) in the thermodynamic limit $N \to \infty$.}
	\label{fig5}
\end{figure*}

The photon weight of the dark states is plotted in Fig.~\ref{fig5}. For a given disorder realization, one finds that PW$(\varepsilon)$ wildly fluctuates in the large $N$ limit [Fig.~\ref{fig5}(\textbf{a})], while the amplitude of fluctuations is reduced upon averaging over a small energy shell (energy binning). Upon increasing $N$, the binned photon weight $\overline{{\rm PW}} (\varepsilon)$ converges to the exact formula (\ref{eq:PWdark}) [Fig.~\ref{fig5}(\textbf{b})]. While being non-zero only at the cavity energy $\varepsilon=0$ for $g=0$, the binned photon weight is spread over the dark states as $g$ is increased for a fixed $N$. For strong couplings $g \gg W$, the photon weight is mostly concentrated in the two polariton states, while the dark states retain a photon weight $\sim O(1/N)$.     

\section{Real-time Green's functions at large $N$, and application to the escape probability}
\label{sec:propagator}

So far we have analyzed the model defined by the arrowhead Hamiltonian (\ref{eq:arrowhead}) from the point of view of its spectral properties: We have focused on its spectrum composed of two polariton modes and of a continuum of dark states in the $N \rightarrow \infty$ limit, and on the photon Green's function in frequency space, closely related to the photon spectral function and to the photon weights of the eigenstates. In this section we turn to the real-time dynamics of the model. We first compute the real-time Green's functions in the large $N$ limit, and provide explicit formulas for those. We then exploit these exact results to analyze the probability that an excitation initially located on site $j$ has escaped from that site after time $t$. We interpret our result for the escape rate in terms of the Fermi's golden rule in an effective model, where the cavity mode has been integrated out in second-order perturbation. We demonstrate the existence of a diffusive dynamics solely mediated by the coupling to the cavity, and a cavity protection effect for strong enough couplings, where transport properties are enhanced with increasing disorder. Cavity-enhanced out-of-equilibrium transport occuring when coupling the system of emitters to two external leads is investigated in the next section.

\subsection{Large $N$ asymptotics of the real-time Green's functions}
\label{green_function_form}

Here we come back to the real-time Green's functions  (\ref{eq:prop_finiteN}.a,b,c) and analyze their large-$N$ asymptotics. We start with the photon Green's function: in real time, we have (see Eq. (\ref{eq:prop_finiteN}), with our convention $\pi g^2 =1$)
\begin{eqnarray}
 \nonumber D(t) \equiv  G_{N+1,N+1} (t) &=& \sum_{a=1}^{N+1} \frac{e^{-i \varepsilon_a t}}{1+ \frac{1}{\pi N } \sum_k \frac{1}{(\omega_k - \varepsilon_a)^2}} \\
 \nonumber   &=& \oint_C \frac{dz}{2\pi i} e^{-i z t} D(z),
\end{eqnarray}
where $D(z)$ is given in (\ref{eq:Dz}), and the counterclockwise contour $C$ encloses all the eigenvalues of $H$. To take the large $N$ limit, we can replace $D(z)$ by $\overline{D}(z)$, and evaluate the contour integral using the analyticity properties of $\overline{D}(z)$, see Sec.~\ref{photon_sec} and especially Eq.~(\ref{eq:Ddisc}). This gives
\begin{eqnarray*}
G_{N+1,N+1} (t)  &\underset{N\rightarrow \infty}{=} & \frac{e^{- i \varepsilon_{{\rm P}+} t}}{ 1- \tilde{\rho}'(\varepsilon_{{\rm P}+})} + \frac{e^{- i \varepsilon_{{\rm P}-} t}}{ 1- \tilde{\rho}'(\varepsilon_{{\rm P}-})} \\
 && +   \frac{1}{ \pi} \int \frac{ e^{-i \omega t} \rho(\omega) d \omega}{\rho(\omega)^2 + \left( \tilde{\rho}(\omega) -\omega \right)^2} . \qquad
\end{eqnarray*}
Importantly, that result does not depend of the microscopic details of the distribution of bare energies. The real-time photon Green's function is not sensitive to fluctuations of that distribution in the large $N$ limit; in other words the real-time Green's function is sensitive only to the average density of states $N \rho(\omega)$ [Eq.~(\ref{eq:rho})].

The other Green's functions (\ref{eq:prop_finiteN}.a,b) share that property, and they can be obtained in a similar way. For the component $G_{j,N+1} (t)$ ($1 \leq j \leq N$), we find
\begin{widetext}
\begin{align*}
G_{j,N+1} (t) &=  \frac{1}{\sqrt{\pi N}} \sum_{a=1}^{N+1} \frac{e^{- i \varepsilon_a t }}{(\varepsilon_a- \omega_j ) (1+ \frac{1}{\pi N} \sum_{k=1}^N \frac{1}{(\omega_k - \varepsilon_a)^2} ) }  = \frac{1}{\sqrt{\pi N}}  \oint_C \frac{dz}{2\pi i} \frac{e^{- i z t }}{z - \omega_j }D(z)  \\
& \underset{N \rightarrow \infty}{=}  \frac{1}{ \sqrt{\pi N}} \bigg[ \frac{e^{- i \varepsilon_{{\rm P}+} t}}{ (\varepsilon_{{\rm P}+}-\omega_j)(1 - \tilde{\rho}'(\varepsilon_{{\rm P}+}) )} +  \frac{e^{- i \varepsilon_{{\rm P}-} t}}{ (\varepsilon_{{\rm P}-}-\omega_j)(1 - \tilde{\rho}'(\varepsilon_{{\rm P}-}) )} \\
& +  {\rm p.v.} \frac{1}{\pi} \int  \frac{ e^{-i \omega t} \rho(\omega) d \omega}{ (\omega - \omega_j) [ \rho(\omega)^2 + \left( \tilde{\rho}(\omega) - \omega \right)^2 ]}  + \frac{1}{\pi} \frac{ e^{-i \omega_j t}   ( \tilde{\rho}(\omega_j) - \omega_j)}{\rho(\omega_j)^2 + \left( \tilde{\rho}(\omega_j) - \omega_j \right)^2} \bigg],
\end{align*}
where `p.v.' denotes the Cauchy principal value of the integral. The large $N$ asymptotics of $G_{i,j}(t)$ with $1 \leq i \neq j \leq N$ also follows 
from that result, because
\begin{eqnarray*}
G_{i,j} (t) & = &   \frac{1}{\pi N} \sum_{a=1}^{N+1} \frac{e^{- i \varepsilon_a t }}{(\varepsilon_a- \omega_i ) (\varepsilon_a- \omega_j ) (1+ \frac{1}{\pi N} \sum_{k=1}^N \frac{1}{(\omega_k - \varepsilon_a)^2} ) } = - \frac{1}{\sqrt{\pi N}} \frac{ G_{i,N+1}(t) - G_{j,N+1}(t)}{\omega_i - \omega_j }.
\end{eqnarray*}

Finally, for $1 \leq j \leq N$, we find
\begin{eqnarray}
G_{j,j} (t) &= &   \frac{1}{\pi N} \sum_{a=1}^{N+1} \frac{e^{- i \varepsilon_a t }}{(\varepsilon_a- \omega_j )^2 (1+ \frac{1}{\pi N} \sum_{k=1}^N \frac{1}{(\omega_k - \varepsilon_a)^2} ) }  \, =\,  \oint_C \frac{dz}{2\pi i}     \frac{e^{-i z t}}{(z-\omega_j)^2}  \left( ( z-\omega_j) + \frac{D(z)}{\pi N}   \right)
\nonumber \\ \nonumber \\
\nonumber	&\underset{N \rightarrow \infty}{=}& e^{- i \omega_j t} + \frac{1}{\pi N}   \left[  \frac{e^{- i \varepsilon_{{\rm P}+} t} }{(\varepsilon_{{\rm P}+}-\omega_j)^2 (1-\tilde{\rho}'(\varepsilon_{{\rm P}+}))}  + \frac{e^{- i \varepsilon_{{\rm P}-} t} }{(\varepsilon_{{\rm P}-}-\omega_j)^2 (1-\tilde{\rho}'(\varepsilon_{{\rm P}-}))}  + i t \frac{ e^{-i \omega_j t}  (\tilde{\rho}(\omega_j) - \omega_j) }{\rho(\omega_j)^2 + \left( \tilde{\rho}(\omega_j) - \omega_j \right)^2  }   \right. \\
 &&  \left. + \, {\rm f.p.} \frac{1}{\pi} \int  \frac{ e^{-i \omega t} \rho(\omega) d \omega}{ (\omega - \omega_j)^2 [ \rho(\omega)^2 + \left( \tilde{\rho}(\omega) - \omega \right)^2 ]}  + e^{-i \omega_j t} \partial_{\omega_j} \left( \frac{\tilde{\rho}(\omega_j) - \omega_j }{\rho(\omega_j)^2 + \left( \tilde{\rho}(\omega_j) - \omega_j\right)^2  } \right)  \right] ,
\label{eq:Gjj}
\end{eqnarray}
where `f.p.' stands for the Hadamard finite part of the integral. We now turn to a concrete example to illustrate how such formulas can be exploited, by evaluating the probability that an excitation, located on site $j$ at time $t=0$, is found elsewhere at time $t>0$.
\end{widetext}

\subsection{Application to the escape probability}
\label{sec:escape}
As a simple application of the above results for the real-time Green's functions, we study the probability that an excitation, initially on a given site $j$ ($j=1,\cdots,N$), has escaped from this site at time $t$:
\begin{equation}
    \label{eq:Pjdef}
	P_j(t) \, = \, 1 - |G_{j,j}(t)|^2 .
\end{equation}
We are interested in the behavior of $P_j(t)$ at large time $t$. More precisely, we want the time to be large, but not too large compared to $N$: $1 \ll t \ll N$. So we take the $N \rightarrow \infty$ limit first, and then consider large time $t$. The $N \rightarrow \infty$ limit of $G_{j,j}(t)$ is given by Eq.~(\ref{eq:Gjj}), so we need to analyze the large $t$ behavior of that expression.

The right-hand-side of Eq.~(\ref{eq:Gjj}) is a sum of six terms; of all these terms, four are clearly bounded as a function of $t$. The two terms that are not bounded are
\begin{align*}
&\frac{i t }{\pi N}  \frac{e^{-i \omega_j t} (\tilde{\rho}(\omega_j) - \omega_j)}{\rho(\omega_j)^2 + \left( \tilde{\rho}(\omega_j) - \omega_j \right)^2  } ,  \nonumber \\ \intertext{and} 
&\frac{1}{\pi N} \, {\rm f.p.} \frac{1}{ \pi} \int  \frac{ e^{-i \omega t} \rho(\omega) d \omega}{ (\omega - \omega_j)^2 [ \rho(\omega)^2 + \left( \tilde{\rho}(\omega) - \omega \right)^2 ]}. 
\end{align*}
While it is not entirely obvious that the latter term diverges with $t$, one can check using the definition of the Hadamard finite part that it behaves as
\begin{align*}
-\frac{t}{\pi N}  \frac{  e^{-i \omega_j t} \rho(\omega_j)}{\rho(\omega_j)^2 + \left( \tilde{\rho}(\omega_j) - \omega_j \right)^2}  
\end{align*}
when $t \rightarrow \infty$. Plugging this into Eq. (\ref{eq:Pjdef}), we find that the escape probability grows linearly with $t$,
\begin{align*}
P_j(t) & \,  \underset{1 \ll t \ll N}{\simeq}  \,   1 -  \left| 1 + \frac{t}{\pi N} \frac{-\rho(\omega_j) + i (\tilde{\rho}(\omega_j) - \omega_j)}{\rho(\omega_j)^2 + (\tilde{\rho}(\omega_j)- \omega_j)^2 }  \right|^2 \\
& \, \simeq \, \Gamma (\omega_j) t ,
\end{align*}
with the following escape rate (here we reinstate the factors $g\sqrt{\pi}$):
\begin{equation}
 \Gamma (\omega)  \, = \,  \frac{2\pi g^2}{N} \frac{\rho(\varepsilon)/(\pi g)^2}{\rho(\varepsilon)^2 + (\tilde{\rho}(\varepsilon) - \frac{\varepsilon}{\pi g^2})^2} \, = \,2\pi  \left( \frac{g}{\sqrt{N}}\right)^2 \, \overline{A} (\omega) . 
\label{FGR}
\end{equation}
It is interesting to note that Eq.~(\ref{FGR}) has a form similar to that given by the Fermi's golden rule, with the photon spectral function playing the role of the usual density of states. Moreover, it is the individual coupling strength $g/\sqrt{N}$ and \textit{not} the collective one that enters the escape rate. We now provide an interpretation of that result based on a second-order perturbative approach. 

\subsection{Fermi's golden rule for an effective long-range hopping Hamiltonian}

To second order, the light-matter coupling term $\hat V$ in the TC Hamiltonian Eq.~(\ref{eq1}) affects the bare orbitals as follows. The eigenstates $\ket{j,0}^{(2)}$ read
\begin{align}
\ket{j,0}^{(2)} =\ket{j,0} + \frac{g}{\omega_{j}\sqrt{N}} \ket{G,1}+ \sum_{i\neq j} \frac{g^{2}}{N(\omega_{j}-\omega_{i})\omega_{j}} \ket{i,0}.
\label{ffu}
\end{align} 
The photon weight of the perturbed eigenstate $\ket{j,0}^{(2)}$ is PW$_{j}=\big\vert\langle G,1 \ket{j,0}^{(2)} \big\vert^{2}=g^{2}/(\omega_{j}^{2} N)$. The perturbative expansion is valid for $g \ll \omega_{j}\sqrt{N}$, which is satisfied in the thermodynamic limit $N\to \infty$ for a fixed $g$ and $\omega_{j}\neq 0$. This condition ensures that the photon weight remains small. Starting with an excitation localized on site $j$ at time $t=0$, we would now like to compute the escape probability to other sites $i$ at time $t$. From energy conservation, one can already expect that these processes imply $\omega_{i}=\omega_{j}$, and therefore the perturbative expansion Eq.~(\ref{ffu}) does not appear to be well suited as the third term in the right-hand-side diverges. It is instead convenient to use a Schrieffer-Wolff transformation on the Hamiltonian Eq.~(\ref{eq1}), which results in a disentanglement of light and matter degrees of freedom~\cite{Zhu_2013}. The new Hamiltonian is written as $\hat H'=e^{\hat S}\hat H e^{-\hat S}$. Under the assumption that the eigenvalues of the generator $\hat S$ remain small (see below), one can expand $\hat H'$ in series as $\hat H'=\hat H+[\hat S,\hat H]+\frac{1}{2}[\hat S,[\hat S,\hat H]]+ \cdots$. The linear coupling term $\hat V$ can be removed from the expansion with the choice $$\hat S = \sum_{j} \frac{g}{\omega_{j}\sqrt{N}} (\hat a \hat \sigma^{+}_{j} - \hat \sigma^{-}_{j} \hat a^{\dagger}),$$
which provides $[\hat S,\hat H_{0}]=-\hat V$. The new Hamiltonian takes the form  
\begin{align}
\hat H' = \hat H_{0}+\frac{1}{2} [\hat S, \hat V] + \mathcal{O}(\hat V^{3}),
\label{ffu2}
\end{align}
and the condition to be satisfied if one is to keep only the first two terms in the right-hand-side of Eq.~(\ref{ffu2}) is therefore $g \ll \omega_{j}\sqrt{N}$, which is the same as the one mentioned above. Calculating the commutator $[\hat S, \hat V]$, we obtain
\begin{align}
\nonumber \hat H' &= \sum_{i} \omega_{i} \hat \sigma^{+}_{i} \hat \sigma^{-}_{i} + \frac{g^{2}}{2N}\sum_{i,j} \left(\frac{1}{\omega_{i}} + \frac{1}{\omega_{j}} \right) \hat \sigma^{+}_{i} \hat \sigma^{-}_{j} \\
&+ \left( \omega_{c}+ \frac{1}{N}\sum_{i} \frac{4g^{2}}{\omega_{i}} \hat \sigma^{+}_{i} \hat \sigma^{-}_{i} \right) \hat a^{\dagger} \hat a,  
\label{ffu3}
\end{align}
up to a constant term. The second term corresponds to an effective hopping between arbitrarily distant sites, while the third term results in a renormalization of the cavity frequency depending on the two-level emitter states. Equivalently, the emitter energies get shifted by cavity photons, which is usually referred to as ``dispersive Stark shift''. This term does not contribute to transitions between states with one excited emitter and zero photon, and can therefore be dropped out of the calculation. As eigenstates of $\hat H_{0}$, these states (denoted as $\ket{j}$ for convenience) satisfy $\hat H_{0}\ket{j}=\omega_{j}\ket{j}$ and $\ket{j (t)}=e^{-i \omega_{j}t}\ket{j}$. Calling $\hat V'$ the second term in the right-hand-side of Eq.~(\ref{ffu3}), the Schr\"{o}dinger equation $$i \frac{\partial \ket{\Psi(t)}}{\partial t} = \hat H' \ket{\Psi(t)},$$ with the ansatz $\ket{\Psi (t)}=\sum_{j} c_{j} (t) e^{-i \omega_{j}t}\ket{j}$ provides the probability amplitudes $c_{j} (t)$ as solutions of 
\begin{align}
\frac{\partial c_{j} (t)}{\partial t} =-i \sum_{i} c_{i} (t) \langle j \vert \hat V' \vert i \rangle e^{i (\omega_{j}-\omega_{i})t}.
\label{eqfi}
\end{align}  
Starting with an excitation on site $j$ at time $t=0$ corresponds to $\ket{\Psi(0)}=\ket{j}$, i.e. $c_{j} (0)=1$ and $c_{i\neq j} (0)=0$. One can then solve Eq.~(\ref{eqfi}) with this initial condition, and sum over all possible final states to obtain the escape probability from site $j$ as
\begin{align}
P_{j}(t)=\sum_{i\neq j} \vert c_{i} (t) \vert^{2} =4 \sum_{i\neq j} \vert \langle i \vert \hat V' \vert j \rangle \vert^{2} \frac{\sin^{2} \left[(\omega_{i}-\omega_{j})t/2 \right]}{(\omega_{i}-\omega_{j})^{2}},
\label{proba}
\end{align}
which features a sharp peak of width $\sim 1/t$ about $\omega_{j}=\omega_{i}$. Under the condition that this characteristic width largely exceeds the mean level spacing $\sim W/N$ (for a box distribution), the summation in Eq.~(\ref{proba}) can be replaced by an integral. With the density of final states $N\rho(\omega)$ [see Eq.~(\ref{eq:rho})], we obtain
\begin{align}
P_{j}(t)=2 t N \int_{\omega_{\rm min}}^{\omega_{\rm max}} \! d\omega \rho(\omega) \vert \langle i \vert \hat V' \vert j \rangle \vert^{2} \xi_{j}(t,\omega).
\label{proba2}
\end{align}
We also assume that the time $t$ is large enough for the width of the function $$\xi_{j}(t,\omega)=\frac{\sin^{2} \left[(\omega-\omega_{j})t/2 \right]}{(\omega-\omega_{j})^{2}t/2}$$ to be much smaller than the extent of the integration domain, in which case $\xi_{j}(t,\omega) \to \pi \delta (\omega-\omega_{j})$ can be replaced by a delta function. With the matrix element $$\langle i \vert \hat V' \vert j \rangle =\frac{g^{2}}{2N} \left(\frac{1}{\omega_{j}} + \frac{1}{\omega_{i}} \right),$$ the escape probability Eq.~(\ref{proba2}) can be written as $P_{j}(t)=\Gamma (\omega_{j}) t$, where the escape rate 
\begin{align*}
\Gamma(\omega_{j})=2\pi g^{2} \, \textrm{PW}(\omega_{j}) \rho(\omega_{j}) =2\pi (g/\sqrt{N})^{2} A(\omega_{j})
\end{align*}
has the same form as in Eq.~(\ref{FGR}). Note that the probability $P_{j}(t)$ must remain small compared to one, and the conditions $\rho(\omega) \ll t \ll N \rho(\omega)$ and $P_{j}(t) \ll 1$ ensuring validity of the Fermi's golden rule (for a box distribution) can be satisfied simultaneously in the thermodynamic limit $N\to \infty$.

\subsection{Physical properties of the escape dynamics}
\label{physescdyn}

The $g$-dependence of the escape rate $\Gamma(\omega_{j})$ given by Eq.~(\ref{FGR}) exhibits some interesting features, which we now illustrate using our usual box distribution of width $W$. In this case, we recall that $\rho=1/W$ for $\omega_{j} \in [-W/2,W/2]$, $\rho=0$ otherwise, and the Hilbert transform $\tilde{\rho}$ is given in Eq.~(\ref{eq:defmomentsdistr}). Injecting the excitation in the middle of that distribution, i.e. $\omega_{j}=0$, the escape rate is independent of $g$ and reads $\Gamma(0)=\frac{2W}{\pi N}$, which can be made \textit{arbitrary large} by increasing the disorder strength $W$. Conversely, injecting exactly on the edges of the distribution, $\omega_{j}=\pm W/2$, the escape rate vanishes for all values of $g$, i.e. $\Gamma(\pm W/2) =0$. In all other cases, the escape rate grows $\sim (2\pi g^{4})/(\omega^{2}_{j}W)$ for weak couplings $g \ll W$, reaches a maximum $(2W)/(\pi N)$ for $$g=\sqrt{\frac{\omega_{j}}{\pi \tilde{\rho}(\omega_{j})}},$$ (which typically corresponds to intermediate coupling strengths $g\sim W$), and then saturates to a lower, $g$-independent value $$\frac{2W}{\pi N} \left(\frac{1}{1+W^2 \tilde{\rho}^{2}(\omega_{j})}\right)$$ for strong couplings $g \gg W$. It is remarkable here that both the maximum escape rate and the saturation value increase with the disorder strength $W$, which originates from the enhancement of the photon weight of the dark states [or equivalently the contribution of the dark states to the spectral function $A(\omega)$], as can be seen from Eqs.~(\ref{eq:PWdark1}) and (\ref{spectral_pw}).

The disorder-averaged escape rate
\begin{align}
\mathbb{E}[\Gamma(\omega_{j})]= &= \frac{1}{W} \int_{-W/2}^{W/2} \Gamma(\omega)  d\omega \nonumber \\
    &= \frac{2\pi g^{2}}{N W} \left(1-  {\rm PW}_{+} -  {\rm PW}_{-} \right)
\label{esc_av}
\end{align}
exhibits similar features and is shown in Fig.~\ref{Figcurrent}(\textbf{a}). Note that since the photon weight is a normalized quantity, i.e. $\sum_{a=1}^{N+1} {\rm PW}_{a}=1$, the quantity in the brackets entering the right-hand-side of Eq.~(\ref{esc_av}) is nothing but the photon weight of the dark states [see Eq.~(\ref{eq:PWdark})] integrated over the support of $\rho$. For weak couplings one finds $\mathbb{E}[\Gamma]\sim \frac{2\pi g^{2}}{N W}$ (grows $\sim g^2$), $\mathbb{E}[\Gamma]$ reaches a maximum at intermediate coupling strengths $g\sim W$, and then saturates for strong couplings to a lower, $g$-independent value $\frac{\pi W}{6N}$, which increases with $W$.

\section{Out-of-equilibrium transport}
\label{sec:trans}

Now that we have identified and discussed cavity-protection effects for the escape probability, it is interesting to look at the excitation current flowing through the system of emitters in an out-of-equilibrium situation. Such a situation occurs when connecting two particular emitter sites to Markovian baths: The source injects excitations at site $j=1$, while the drain extracts excitations at site $j=N$. [In this section we do not assume that the bare energies are sorted in increasing order, so the energies of the `in' and `out' sites $\omega_1$ and $\omega_N$ need not be the minimum and maximum bare energies.] Since we are working in the single excitation subspace, we can replace the spin operators by fermionic ones for simplicity, i.e., $\hat{\sigma}^{-}_{i} \to \hat{\sigma}_{i}$ and $\hat{\sigma}^{+}_{i} \to \hat{\sigma}^{\dagger}_{i}$ ($i=1,\cdots,N$), with $\{\hat{\sigma}_{i},\hat{\sigma}^{\dagger}_{j} \}=\delta_{i,j}$. The excitation current and populations across the system of emitters are computed using the non-equilibrium Green's function formalism. The system is described by the Hamiltonian $\hat{H}_{\rm neq}=\hat{H}+\hat{H}_{\rm r}$, with $\hat{H}$ given by Eq.~(\ref{eq1}) and 
\begin{align*}
\hat{H}_{r} &= \sum_{\alpha} \omega_{\alpha} \hat{\sigma}^{\dagger}_{\alpha,{\rm in}} \hat{\sigma}_{\alpha,{\rm in}} + \sum_{\alpha} \omega_{\alpha} \hat{\sigma}^{\dagger}_{\alpha,{\rm out}} \hat{\sigma}_{\alpha,{\rm out}} \\
& + \sum_{\alpha} \lambda^{\rm in}_{\alpha} \left(\hat{\sigma}_{1} \hat{\sigma}^{\dagger}_{\alpha,{\rm in}} + \hat{\sigma}_{\alpha,{\rm in}} \hat{\sigma}^{\dagger}_{1} \right) \\
& + \sum_{\alpha} \lambda^{\rm out}_{\alpha} \left(\hat{\sigma}_{N} \hat{\sigma}^{\dagger}_{\alpha,{\rm out}} + \hat{\sigma}_{\alpha,{\rm out}} \hat{\sigma}^{\dagger}_{N} \right).
\end{align*} 
Here, $\alpha$ denotes some quantum number running over a continuum of states in each reservoir. For a given function $f_{\alpha}$, summations of the type $\sum_{\alpha} f_{\alpha}$ are replaced with $\int d\omega d (\omega) f(\omega)$, with $d (\omega)$ the density of states in the reservoirs and $\omega \equiv \omega_{\alpha}$ the energy of the state $\alpha$. The operator $\hat{\sigma}_{\alpha,l}$ ($\hat{\sigma}^{\dagger}_{\alpha,l}$) annihilates (creates) a fermion in the reservoir $l={\rm in,out}$, and the real parameters $\lambda^{\rm in}_{\alpha}$ and $\lambda^{\rm out}_{\alpha}$ are the coupling strengths between the reservoirs and the two ends of the system. The steady-state excitation current flowing through the emitters can be computed either from the input current: $$J_{{\rm in}}=\langle \frac{\partial \hat{N}_{{\rm in}}}{\partial t} \rangle=i\langle [\hat{H}_{\rm neq},\hat{N}_{{\rm in}}] \rangle,$$ or the output current: $$J_{{\rm out}}=\langle \frac{\partial \hat{N}_{{\rm out}}}{\partial t} \rangle = i\langle [\hat{H}_{\rm neq},\hat{N}_{{\rm out}}] \rangle,$$ since these two are equal in magnitude. Here, $\hat{N}_{l}=\sum_{\alpha} \hat{\sigma}^{\dagger}_{\alpha,l}\hat{\sigma}_{\alpha,l}$ is the number of fermions in the reservoir $l={\rm in},{\rm out}$, and $\langle \cdots\rangle$ is the expectation value in the steady-state~\cite{haug}.

We force injection and extraction of particles at the first and last sites, respectively, assuming $n_{{\rm in}}(\omega)\equiv \langle \hat{\sigma}^{\dagger}_{\alpha,{\rm in}}\hat{\sigma}_{\alpha,{\rm in}} \rangle =1$ and $n_{{\rm out}}(\omega)\equiv \langle \hat{\sigma}^{\dagger}_{\alpha,{\rm out}}\hat{\sigma}_{\alpha,{\rm out}} \rangle =0$ for all energies $\omega$ in the range of interest. Following Ref.~\cite{Hagenmuller_Cavity_2018}, it can be shown that, in this case, the input and output currents take the form $J_{{\rm in}}=-\Gamma_{\rm in} \, (1-n_{1})$ and $J_{{\rm out}}=\Gamma_{\rm out} \, n_{N}$, with $J_{{\rm in}}=-J_{{\rm out}}$, $\Gamma_{\rm in} \equiv 2\pi d (\omega) \lambda^{2}_{\rm in} (\omega)$ and $\Gamma_{\rm out} \equiv 2\pi d (\omega) \lambda^{2}_{\rm out} (\omega)$ the injection and extraction rates assumed to be frequency-independent (Markovian baths), and
\begin{align}
n_{j} = \int \! \frac{d\omega}{2\pi} {\rm Im} \left[G^{<}_{j,j} (\omega)\right]
\label{eqimp1}
\end{align}
the population on site $j$~\cite{pourfath}. Here, $G^{<}_{i,j} (\omega)=\int dt e^{i\omega t} G^{<}_{i,j} (t)$, with $G^{<}_{i,j} (t) = i\langle \hat{\sigma}^{\dagger}_{j} (0) \hat{\sigma}_{i} (t) \rangle$ the ``lesser'' Green's function. Similarly we call $G^{>}_{i,j} (t) = -i\langle \hat{\sigma}_{i} (t) \hat{\sigma}^{\dagger}_{j} (0) \rangle$ the ``greater'' Green's function, which corresponds up to a $-i$ factor to $G_{i,j} (t)$ as defined in Eq.~(\ref{gfij}) at equilibrium (absence of coupling to the reservoirs). In this situation, $G^{<}_{i,j} (t)$ vanishes since $\langle \cdots \rangle$ denotes the expectation value in the ground state $\ket{G,0}$. Note that we have not used these notations in Sec.~\ref{sec:finiteN} for the sake of clarity. 

\subsection{Equations of motion}
\label{eqofmot_res}

The transport properties of the model are thus encoded in the non-equilibrium Green's function $G^{<}_{i,j} (\omega)$. The latter can be computed by first deriving the equations of motion of the retarded and advanced Green's functions, $G^{\rm R}_{i,j} (t)=\theta (t) \left[G^{>}_{i,j}(t) - G^{<}_{i,j}(t)\right]$ and $G^{\rm A}_{i,j} (t)=\theta (-t) \left[G^{<}_{i,j}(t) - G^{>}_{i,j}(t)\right]$, respectively, with $\theta$ the heaviside step function. These equations of motion are obtained by computing the first time derivative of the Green's functions, and therefore the commutator of $\hat{\sigma}_{i}$ with the different parts of the Hamiltonian~\cite{Hagenmuller_Cavity_2018}. In the frequency domain, the equations of motion can be written in the (Dyson) form~\cite{haug} 
\begin{align}
\sum_{k}\Big\{\!\!\left[G^{0\,\beta}_{i,k} (\omega) \right]^{-1} - \Sigma^{\beta}_{i,k} (\omega)\Big\} G^{\beta}_{k,j} (\omega)=\delta_{i,j},
\label{dyso1}
\end{align}
with $\beta={\rm A},{\rm R}$, the non-interacting Green's functions $G^{0\,{\rm R}}_{i,j} (\omega)=\frac{\delta_{i,j}}{\omega - \omega_{i}+i0^{+}}$ and $G^{0\,{\rm A}}_{i,j} (\omega)=\frac{\delta_{i,j}}{\omega - \omega_{i}-i0^{+}}$, and the self-energies 
\begin{align*}
\Sigma^{\rm R}_{j,k} (\omega)&=\frac{g^{2}}{N} D^{\rm R}_{0} (\omega) - \frac{i}{2}\delta_{j,k} \left(\delta_{j,1} \Gamma_{\rm in} + \delta_{j,N} \Gamma_{\rm out} \right)
\end{align*}
and $\Sigma^{\rm A}_{j,k} (\omega)=\left[\Sigma^{\rm R}_{j,k} (\omega)\right]^{*}$. Here, $D^{\rm R}_{0}=\frac{1}{\omega + i0^{+}}$ denotes the retarded, non-interacting cavity Green's function. The lesser Green's function is obtained from the Keldysh equation $$G^{<}_{i,j} (\omega)=\sum_{k,p} G^{\rm R}_{i,k} (\omega) \Sigma^{<}_{k,p} (\omega) G^{\rm A}_{p,j} (\omega),$$ with the lesser self-energy
\begin{equation}
\Sigma^{<}_{k,p} (\omega)=i\Gamma_{\rm in} \delta_{k,p} \delta_{k,1}.
\label{lesserSE}
\end{equation}
Another useful quantity is the cavity photon population given by~\cite{pourfath}
\begin{align*}
n_{c} = \int \! \frac{d\omega}{2\pi} {\rm Im} \left[D^{<} (\omega)\right].
\end{align*}
In order to compute the lesser cavity photon Green's function $D^{<} (\omega)=\int dt e^{i\omega t} D^{<} (t)$, with $D^{<} (t)=i\langle \hat{a}^{\dagger} (0) \hat{a} (t) \rangle$, we proceed as before and derive the equations of motion of the retarded and advanced cavity Green's functions, $D^{\rm R} (t)=\theta (t) \left[D^{>} (t) - D^{<}(t)\right]$ and $D^{\rm A} (t)=\theta (-t) \left[D^{<}(t) - D^{>}(t)\right]$, respectively, with $D^{>} (t)=-i\langle  \hat{a} (t) \hat{a}^{\dagger} (0) \rangle$. In the frequency domain, these equations of motion take the form $$\Big\{\!\!\left[D^{\beta}_{0} (\omega) \right]^{-1} - \Pi^{\beta} (\omega) \Big\} D^{\beta}(\omega)=1,$$ with $\beta={\rm A},{\rm R}$, and the cavity self-energies
\begin{align*}
\Pi^{\rm R} (\omega)=\frac{g^{2}}{N} \sum_{j} \frac{1}{\omega - \omega_{j} + \frac{i}{2} \left(\delta_{j,1}\Gamma_{\rm in} + \delta_{j,N}\Gamma_{\rm out}\right)}
\end{align*}
and $\Pi^{\rm A} (\omega)=\left[\Pi^{\rm R} (\omega) \right]^{*}$. Again, the lesser cavity Green's function is obtained from the Keldysh equation $D^{<}(\omega)=D^{\rm R}(\omega) \Pi^{<} (\omega) D^{\rm A}(\omega)$, with
\begin{align*}
\Pi^{<} (\omega)=\frac{g^{2}}{N} \frac{i\Gamma_{\rm in}}{\left(\omega - \omega_{1} \right)^{2} + (\Gamma_{\rm in}/2)^{2}}. 
\end{align*}

\subsection{Analytical formulas for the current and populations}
\label{anal_res}

\begin{figure*}[t]
	\includegraphics[width=1.0\textwidth]{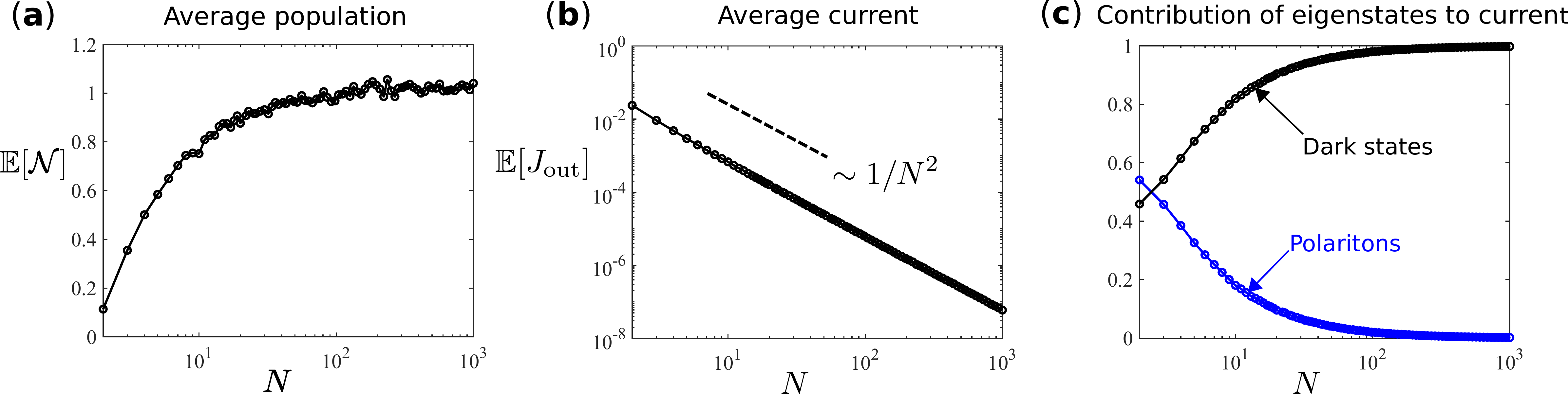}
	\caption{Out-of-equilibrium transport. Disorder-averaged (over $2000$ configurations) total population $\mathbb{E} [\mathcal{N}]$ (\textbf{a}), output current $\mathbb{E} [J_{\rm out}]$ (\textbf{b}), and proportion of the current carried by the different eigenstates (\textbf{c}), as a function of $N$ for $g=1$. Other parameters are $\Gamma_{\rm in} = 0.1/N^2$ and $\Gamma_{\rm out}=1$.}
	\label{Fig8}
\end{figure*}

The Dyson equation Eq.~(\ref{dyso1}) can be solved exactly using matrix inversion carried out with the Sherman-Morrison formula. This leads to 
\begin{align}
G^{\rm R}_{i,j} (\omega) &= \frac{g^{2}}{N} \frac{D^{\rm R}(\omega)}{\left(\omega - \widetilde{\omega}_{i} \right) \left(\omega - \widetilde{\omega}_{j} \right) } \qquad i\neq j \nonumber \\
G^{\rm R}_{i,i} (\omega) &= \frac{1}{\omega - \widetilde{\omega}_{i}} + \frac{g^{2}}{N} \frac{D^{\rm R}(\omega)}{\left(\omega - \widetilde{\omega}_{i} \right)^{2}},
\label{GFs_finn1}
\end{align}
and $G^{\rm A}_{i,j} (\omega)=\left[G^{\rm R}_{i,j} (\omega) \right]^{*}$, with $\widetilde{\omega}_{1}=\omega_{1} - \frac{i\Gamma_{\rm in}}{2}$, $\widetilde{\omega}_{N}=\omega_{N} - \frac{i\Gamma_{\rm out}}{2}$, and $\widetilde{\omega}_{j}=\omega_{j}$ for all $j \neq 1,N$. In order to compute the steady-state populations and current according to Eq.~(\ref{eqimp1}), we use the Keldysh equation $\underline{G}^{<} (\omega)=\underline{G}^{\rm R} (\omega) \underline{\Sigma}^{<} (\omega) \underline{G}^{\rm A} (\omega)$, with $\underline{\Sigma}^{<} (\omega)$ given by Eq.~(\ref{lesserSE}), and $\underline{G}^{\rm R} (\omega)$ given by Eq.~(\ref{GFs_finn1}). The integration in frequency domain is performed using the residue theorem, with the following factorization of the cavity mode Green's functions:
\begin{align*}
D^{\rm R}(z) =\frac{\prod_{j=1}^{N} \left(z - \widetilde{\omega}_{j} \right)}{\prod_{a=1}^{N+1} \left(z- \epsilon_{a} \right)} \quad D^{\rm A}(z)=\left[D^{\rm R}(z)\right]^{*},
\end{align*}
and $\epsilon_{a}$ the $N+1$ solutions of the polynomial equation
\begin{align*}
z \prod_{j=1}^{N} \left(z - \widetilde{\omega}_{j} \right) - \frac{g^{2}}{N} \sum_{j=1}^{N} \prod_{i\neq j} \left(z - \widetilde{\omega}_{j} \right)= 0.
\end{align*}
These solutions all have negative imaginary parts and correspond to the eigenvalues of the complex symmetric (non-hermitian) arrowhead matrix
\begin{equation}
\left( \begin{array}{ccccc|c}
		\widetilde{\omega}_{1} & 0 & \cdots & 0 & 0 & g/\sqrt{N} \\
		0  & \widetilde{\omega}_{2} & \ddots & 0 & 0 & g/\sqrt{N} \\
		\vdots & \ddots & \ddots & \ddots & \vdots & \vdots \\
		0 & 0 & \ddots  & \widetilde{\omega}_{N-1} & 0 & g/\sqrt{N} \\
		0 & 0 & \cdots & 0 & \widetilde{\omega}_{N} & g/\sqrt{N} \\ \hline
		g/\sqrt{N} & g/\sqrt{N} & \cdots & g/\sqrt{N} & g/\sqrt{N} & 0
	\end{array} \right).
	\label{nohermarrow}
\end{equation} 

The real part of these eigenvalues exhibits a similar structure to the equilibrium case (without coupling to the reservoirs): $N-1$ dark states with energies comprised between $-W/2$ and $W/2$, and two polaritons emerging from the continuum of dark states for $g \gtrsim W/2$. In contrast to the equilibrium case, these eigenstates exhibit a finite imaginary part (lifetime) due to the coupling to the two reservoirs. Using the results of Secs.~\ref{eqofmot_res} and \ref{anal_res}, the integration in Eq.~(\ref{eqimp1}) provides the population at site $j\neq 1$: 
\begin{align}
n_{j}=\Gamma_{\rm in} \frac{g^{4}}{N^{2}} \sum_{a} \frac{\tau_{a} \phi_{a}}{\left(\epsilon_{a} - \widetilde{\omega}_{j} \right)\left(\epsilon_{a} - \widetilde{\omega}^{*}_{j} \right)},
\label{GFs_finn23}
\end{align}
the population at the injection site $j=1$:
\begin{align*}
n_{1} & =  1 + \Gamma_{\rm in} \frac{g^{4}}{N^{2}} \sum_{a} \frac{\tau_{a} \phi_{a}}{\left(\epsilon_{a} - \widetilde{\omega}_{1} \right)\left(\epsilon_{a} - \widetilde{\omega}^{*}_{1} \right)}  \\
& + \frac{g^{4}}{N^{2}} \frac{\prod_{k\neq 1} \left(\widetilde{\omega}_{1} - \widetilde{\omega}_{k} \right)\left(\widetilde{\omega}_{1} - \widetilde{\omega}^{*}_{k} \right)}{\prod_{a} \left(\epsilon_{a} - \widetilde{\omega}_{1} \right)\left(\epsilon^{*}_{a} - \widetilde{\omega}_{1} \right)} \\
& +\frac{2 g^{2}}{N} {\rm Re} \bigg\{\frac{\prod_{k\neq 1} \left(\widetilde{\omega}_{1} - \widetilde{\omega}^{*}_{k} \right)}{\prod_{a} \left(\widetilde{\omega}_{1} - \epsilon^{*}_{a} \right)}\bigg\},
\end{align*}
and the population of the cavity mode:
\begin{align*}
&n_{c}=\Gamma_{\rm in} \frac{g^{2}}{N} \sum_{a} \tau_{a} \phi_{a},&
\end{align*}
with $\tau_{a}\equiv -1/[2 \, {\rm Im}(\epsilon_{a})]$ the lifetime of the eigenstates, and $$\phi_{a}=\frac{\prod_{k\neq 1} \left(\epsilon_{a} - \widetilde{\omega}_{k} \right)\left(\epsilon_{a} - \widetilde{\omega}^{*}_{k} \right)}{\prod_{a'\neq a} \left(\epsilon_{a}- \epsilon_{a'} \right)\left(\epsilon_{a}- \epsilon^{*}_{a'} \right)}.$$ 

Importantly, we find that the total population $\mathcal{N}=n_{c}+\sum_{j=1}^{N} n_{j}= O(N)$, which violates the single-excitation assumption. Our fermion model for transport thus does not map onto the disordered TC model. A proper rescaling of the current injection rate $\Gamma_{\rm in} \to \widetilde{\Gamma}_{\rm in}/N^{2}$, (while keeping $\widetilde{\Gamma}_{\rm in}$ fixed) allows to circumvent this issue as it results in a total population $O(1)$ [Fig.~\ref{Fig8}(\textbf{a})]. This ensures that the system remains in the single-excitation subspace, so that the identification of our simple fermionic model for transport with the disorder TC model does hold.

\subsection{Cavity-protected transport}

Using this rescaling of the injection rate and averaging the steady-state output current 
\begin{equation}
J_{\rm out}=\Gamma_{\rm out} n_{N} = \Gamma_{\rm out}\widetilde{\Gamma}_{\rm in} \frac{g^{4}}{N^{4}} \sum_{a} \frac{\tau_{a} \phi_{a}}{\left(\epsilon_{a} - \widetilde{\omega}_{N} \right)\left(\epsilon_{a} - \widetilde{\omega}^{*}_{N} \right)}
\label{eqcurout}
\end{equation}
over disorder, we find that $\mathbb{E}[J_{\rm out}]\sim 1/N^{2}$ [Fig.~\ref{Fig8}(\textbf{b})]. This can be understood from Eq.~(\ref{eqcurout}) by observing numerically that the lifetime of the eigenstates $\tau_{a}\sim N^2$. Moreover, since Eq.~(\ref{eqcurout}) involves a summation over the eigenstates of the matrix (\ref{nohermarrow}), one can easily extract the contribution of these eigenstates to the disorder-averaged current. While for small $N$, the latter is equally carried by polaritons and dark states, we find that it is fully dominated by the dark states in the thermodynamic limit [Fig.~\ref{Fig8}(\textbf{c})]. For a single disorder realization, fixed $N$, and weak coupling $g \ll W$, the current scales $\sim g^{4}$, as can be seen from Eq.~(\ref{eqcurout}). Interestingly, after disorder-averaging we find that $\mathbb{E}[J_{\rm out}]\sim g^{2}$, similarly as for the escape rate discussed in Sec.~\ref{physescdyn}. Output current and escape rate share common features also beyond weak couplings [see Fig.~\ref{Figcurrent}(\textbf{a}) and (\textbf{b})]: $\mathbb{E}[J_{\rm out}]$ reaches a maximum for intermediate coupling strengths $g\sim W$, and then saturates to a slightly lower value for strong couplings. Importantly, the value of the plateau is found to increase with $W$, which means that the out-of-equilibrium current does not only exhibit some robustness against disorder, but also that the latter can be used to enhance transport for large couplings $g \gtrsim W$.   

\section{Conclusion}
\label{sec:conclusion}

In conclusion, we have investigated in detail a class of large arrowhead matrices that are relevant to many physical systems, ranging from molecular junctions, to central-spin problems and cavity QED. We have derived asymptotically exact formulas in the thermodynamic limit for different quantities of physical interest such as the spectrum, average energy shifts, inverse participation ratio, and correlation functions. We have shown that the spectrum and the distribution of energy spacing exhibit characteristics usually associated to the critical point of disordered hopping models for “Anderson” localization-delocalization transitions. We have studied how those peculiar spectral properties are connected to dynamical quantities such as the escape probability and the out-of-equilibrium current, showing that the latter can be efficiently ``protected'' by the cavity. Even more significantly, it was shown that disorder can help transport for strong enough couplings. Interesting perspectives include the effect of dissipation in the model, which would typically be included by adding finite imaginary parts to all diagonal elements of the arrowhead matrix Hamiltonian, as well as many-body effects occuring for larger excitation densities. It is another exciting prospect to investigate whether the combined effects of disorder, light-matter coupling, and particle statistics could lead to non-classical states of light when driving the cavity with a laser field, or upon injecting and removing spin excitations in a transport situation. 

\section*{Acknowledgements}
We are grateful to Fausto Borgonovi, Giuseppe Luca Celardo, Thibault Chervy, Alexandre Faribault, Francesco Mattiotti, Puneet Murthy, and Deepankur Thureja for stimulating discussions, and to Andrea Maroncelli for joint work on closely related projects and comments on the manuscript. This work was supported by the ANR - ``ERA-NET QuantERA'' - Projet ``RouTe'' (ANR-18-QUAN-0005-01), and LabEx NIE (``Nanostructures in Interaction with their Environment'') under contract ANR-11-LABX0058 NIE with funding managed by the French National Research Agency as part of the ``Investments for the future program'' (ANR-10-IDEX-0002-02).  G. P. acknowledges support from the Institut Universitaire de France (IUF) and the University of Strasbourg Institute of Advanced Studies (USIAS). Research was carried out using computational resources of the Centre de calcul de l'Universit\'e de Strasbourg. 

\medskip

\bibliography{biblio}

\end{document}